%% file: main.tex
\documentclass[journal]{IEEEtran}
\pdfoutput=1
\ifCLASSINFOpdf
\else
\fi
\usepackage{cite}
\usepackage{amsmath,amssymb,amsfonts}
\usepackage{algorithmic}
\usepackage{graphicx}
\usepackage{textcomp}
\usepackage{comment}
\usepackage{color, soul}
\usepackage{enumitem}
\setlist[itemize]{leftmargin=0pt}
\usepackage{comment}
\usepackage{hyperref}
\usepackage{import}
\usepackage{caption}
\usepackage{subcaption}
\usepackage{listings}
\usepackage[ruled,vlined]{algorithm2e}
\usepackage{setspace}
\usepackage{tabularx}
\usepackage{epsf}
\usepackage{acronym} 
\usepackage{nomencl}
\usepackage{multirow}

\def\BibTeX{{\rm B\kern-.05em{\sc i\kern-.025em b}\kern-.08em
    T\kern-.1667em\lower.7ex\hbox{E}\kern-.125emX}}

\begin{document}

\title{Cyber-Physical Energy Systems Security:\\Threat Modeling, Risk Assessment, Resources, Metrics, and Case Studies}

\author{Ioannis Zografopoulos,~\IEEEmembership{Graduate Student~Member,~IEEE},  Juan Ospina,~\IEEEmembership{Member,~IEEE}, XiaoRui  Liu,~\IEEEmembership{Student~Member,~IEEE}, and   Charalambos~Konstantinou,~\IEEEmembership{Senior~Member,~IEEE}
}


\IEEEaftertitletext{\vspace{-2\baselineskip}}

\maketitle

\begin{abstract}

Cyber-physical systems (CPS) are interconnected architectures that employ analog, digital, and communication resources for their interaction with the physical environment. CPS are the backbone of enterprise, industrial, and critical infrastructure. Thus, their vital importance makes them prominent targets for malicious attacks aiming to disrupt their operations. Attacks targeting cyber-physical energy systems (CPES), given their mission-critical nature, can have disastrous consequences. The security of CPES can be enhanced leveraging testbed capabilities to replicate power system operations, discover vulnerabilities, develop security countermeasures, and evaluate grid operation under fault-induced or maliciously constructed scenarios. In this paper, we provide a comprehensive overview of the CPS security landscape with emphasis on CPES. Specifically, we demonstrate a threat modeling methodology to accurately represent the CPS elements, their interdependencies, as well as the possible attack entry points and system vulnerabilities. Leveraging the threat model formulation, we present a CPS framework designed to delineate the hardware, software, and modeling resources required to simulate the CPS and construct high-fidelity models which can be used to evaluate the system's performance under adverse scenarios. The system performance is assessed using scenario-specific metrics, while risk assessment enables system vulnerability prioritization factoring the impact on the system operation. The overarching framework for modeling, simulating, assessing, and mitigating attacks in a CPS is illustrated using four representative attack scenarios targeting CPES. The key objective of this paper is to demonstrate a step-by-step process that can be used to enact in-depth cybersecurity analyses, thus leading to more resilient and secure CPS.

\end{abstract}



\maketitle

\vspace{-5mm}
\makenomenclature
\mbox{}

\nomenclature{EPS}{electric power systems}
\nomenclature{CPS}{cyber-physical systems}
\nomenclature{CPES}{cyber-physical energy systems}
\nomenclature{DER}{distributed energy resources}
\nomenclature{BESS}{battery energy storage system}
\nomenclature{TESS}{thermal energy storage system}
\nomenclature{ICT}{information and communication technologies}
\nomenclature{IoT}{internet-of-things}
\nomenclature{RTU}{remote terminal units}
\nomenclature{PLC}{programmable logic controllers}
\nomenclature{IED}{intelligent electronic devices}
\nomenclature{DoS}{denial-of-service}
\nomenclature{HIL}{hardware-in-the-loop}
\nomenclature{ICS}{industrial control systems}
\nomenclature{CHIL}{controller hardware-in-the-loop}
\nomenclature{PMU}{phasor measurement units}
\nomenclature{FDIA}{false data injection attack}
\nomenclature{TES}{transactive energy systems}
\nomenclature{DiD}{defense-in-depth}
\nomenclature{TTP}{tactics, techniques, and procedures}
\nomenclature{IT}{information technology}
\nomenclature{HMI}{human-machine interface}
\nomenclature{EMT}{electromagnetic transient}
\nomenclature{TS}{transient stability}
\nomenclature{T\&D}{transmission and distribution}
\nomenclature{NIC}{network interface cards}
\nomenclature{PHIL}{power hardware-in-the-loop}
\nomenclature{RTS}{real-time simulator}
\nomenclature{CORE}{common open research emulator}
\nomenclature{DIA}{data integrity attack}
\nomenclature{DAA}{data availability attack}
\nomenclature{OS}{operating system}
\nomenclature{ESS}{energy storage system}
\nomenclature{EV}{electric vehicle}
\nomenclature{MG}{microgrid}
\nomenclature{MPPT}{maximum power point tracking}
\nomenclature{MTU}{master terminal unit}
\nomenclature{TDA}{time-delay attack}
\nomenclature{FACTS}{flexible AC transmission systems}
\nomenclature{IDS/IPS}{intrusion detection and prevention systems}
\nomenclature{SDN}{software defined network}
\nomenclature{PCC}{point of common coupling}
\nomenclature{DG}{distributed generation}
\nomenclature{SCADA}{supervisory control and data acquisition}
\nomenclature{RES}{renewable energy source}
\nomenclature{AGC}{automatic generation control}
\nomenclature{MitM}{man-in-the-middle}
\nomenclature{IC}{integrated circuit}
\nomenclature{OT}{operational technology}
\nomenclature{QoS}{quality-of-service}
\nomenclature{CB}{circuit breaker}
\printnomenclature[5em]

\import{sections/}{1-IntroductionREV}

\import{sections/}{2-Related_Work}

\import{sections/}{3-Threatmodeling}
\import{sections/}{4-Framework}

\import{sections/}{5-Case_studies}
\import{sections/}{6-Conclusion}

\bibliographystyle{IEEEtran}
\bibliography{IEEEabrv, biblio}
\newpage

\end{document}

%% file: sections/1-IntroductionREV.tex
\section{Introduction} \label{s:introduction}

\subsection{Background and Motivation}

Over the past years, electric power systems (EPS) have diverged from a unidirectional generation and transmission model towards a more distributed architecture that supports traditional generation sources as well as distributed energy resources (DERs) in the form of distributed generation (DG), such as PV and wind, and distributed storage (DS) sources, such as battery energy storage systems (BESS) and thermal energy storage systems (TESS). The transformation of EPS to cyber-physical energy systems (CPES) is primarily enabled due to the introduction of information and communication technologies (ICT), automated control systems, remote sensing, and embedded industrial internet-of-things (IIoT) devices. According to the National Institute of Standards and Technology (NIST) \cite{NIST}, cyber-physical systems (CPS) refer to architectures that incorporate digital, analog, and physical components. The interaction of these components is determined by the dynamics of the system and the rules which orchestrate its operation. 
\textit{CPES are energy-focused engineered systems that are transforming the way traditional EPS operate by seamlessly integrating physical entities with human, digital, and networking components designed to operate through integrated physics and computational logic.} As such, CPES contribute significantly towards the EPS modernization allowing for better planning, more flexible control, cyber-secure operations, system-wide optimization, transactive energy systems (TES), improvements in power quality, system reliability enhancements, resiliency, interoperability, and cleaner energy generation.

The security of CPS presents significant challenges in controlling and maintaining secure access to critical system resources and services (e.g., for CPES: generation reserves, frequency stability controls, power line protection, etc.), as well as ensuring the confidentiality, accessibility, and integrity of the information exchanged (e.g., control signals of supervisory control and data acquisition -- SCADA systems). CPS, being large-scale complex systems of systems, employ numerous computing components such as remote terminal units (RTUs), programmable logic controllers (PLCs), and intelligent electronic devices (IEDs) that are often designed without  security in mind. Typically, the hardware, software, and communication interfaces of these devices are developed utilizing commercial off-the-shelf components \cite{mclaughlin2016cybersecurity}. Thus, vulnerabilities within such components can be ported to the CPS environments creating potential entry points for malicious adversaries\footnote{Throughout the paper, we use the terms \textit{adversary}, \textit{threat actor}, and \textit{attacker} interchangeably.} aiming to disrupt CPS operations. An indicative incident of malicious behavior targeting CPS operation was reported in March 2019. Attackers targeted the United States (U.S.) grid infrastructure and performed a denial-of-service (DoS) attack through the exploitation of a known CPES vulnerability, namely a web interface firewall vulnerability \cite{CVE1, CVE2}. The attack resulted in the loss of communication between the utility's generation assets and the energy management system \cite{NERC}, causing brief interruptions in the utility's service. The number of cyber-attacks where adversaries exploit known and existing vulnerabilities to compromise CPS is increasing. This fact is validated by security reports stating that ``\textit{99\% of the vulnerabilities exploited in 2020 are known to security professionals, while zero-day vulnerabilities only account for the 0.4\% of vulnerabilities exposed during the past decade}'' \cite{Gartner}.

The importance of CPS, and CPES in particular, for economic prosperity and public health at the national, state, and local level can motivate attackers to compromise such systems in order to obtain financial or political gains. Hence, the evaluation of the CPES robustness and resilience against attacks in realistic scenarios is of paramount importance. At the same time, the quantification of cybersecurity risks is becoming more complex and challenging as EPS -- also referred to as the ``largest interconnected machine on earth'' \cite{largestsmartgrid} -- integrate numerous cyber-components at all levels and scales. In the past, the simulation of specific abnormal scenarios (e.g., faults, overvoltage conditions, frequency fluctuations, etc.) was sufficient to provide insights into EPS operations. However, current advances towards intelligent and interconnected CPES require more accurate models and representations capable of capturing the dynamic behavior of these interoperable systems. The enhancement of CPES security and reliability requires constant probing for potential weaknesses \cite{gritzalis2019critical}. Security studies need to reflect the nature of the CPES infrastructure in actual testing environments that support the interfacing of actual hardware devices designed to operate in the `real' system. In this context, hardware-in-the-loop (HIL) testbeds are effective in providing testing capabilities for evaluating the synergistic relationship between physical and virtual components in controlled environments. Security-oriented HIL testbeds are invaluable in performing cybersecurity and risk analyses, identifying system vulnerabilities in various layers (e.g., hardware, firmware, software, protocol, process), implementing intrusion detection and prevention algorithms, and assessing the efficiency of mitigation techniques without inducing excessive economic burdens or safety hazards \cite{mclaughlin2016cybersecurity, keliris2016enabling}.

The primary motivation of this paper is to develop a framework, which bridges theoretical and simulation-based security case studies and evaluates CPS system behavior leveraging testbed environments, leading to more secure CPES architectures. In order for testbeds to reliably capture the characteristics of the cyber-physical environment, testing and experimental case studies need to be described and modeled considering both the cyber and physical domains. The case studies require detailed descriptions of the resources and metrics that will be utilized for evaluating the CPES performance, reliability, and resilience. In addition, the testing setup must also capture the threat modeling characteristics of the adversary and the attack methodology. In terms of a potential adversary, the threat modeling characteristics are adversarial knowledge, resources, access to the system, and specificity. As for the attack methodology, the threat modeling characteristics include the attack frequency, reproducibility, discoverability, target level, attacked asset, attack techniques, and premise. Doing so, in a holistic and step-by-step approach, allows researchers and stakeholders to thoroughly examine and uncover security risks existing in the CPES under evaluation.

\subsection{Research Contribution and Overview}

The underlying goal of this manuscript is to provide a complete and detailed presentation of CPS security research studies by demonstrating a modular framework for assessing CPS security in the context of CPES. To this end, the paper describes all the required components for evaluating the behavior and performance of CPES under diverse and adverse operational scenarios. The framework exhibits the modeling techniques used to represent the cyber and physical domains of the system, considers the resources used to model the CPES, and presents essential evaluation metrics for each corresponding case study. The contributions of this work, \textit{focusing on CPES security}, can be summarized as follows:

\begin{itemize}[itemsep=0pt,parsep=0pt,leftmargin=*, wide=0pt]
    \item A literature review is provided that presents the research efforts in the area of CPS and CPES security, describes cyber-physical testbeds developed by prominent research centers and laboratories around the world, and illustrates current threat and risk modeling approaches widely used in the industry.
    \item A threat modeling methodology is proposed, comprised of two major parts, the adversary model and the attack model, allowing for an inclusive evaluation of malicious attack strategies.
    \item Leveraging our threat modeling approach, a risk assessment process is provided that takes into account risks related to the effectiveness of an attack, the targeted system component, and the criticality of the cyber-physical process being compromised.
    \item A framework is described that elucidates the crucial components and resources needed to accurately characterize CPS, making it essential for evaluating numerous studies (e.g., cyber, control, etc.). It is important to note that the proposed CPS framework can be used to characterize CPS in other sectors such as healthcare and transportation, but in this work, it is evaluated specifically for CPES.
    \item Four illustrative CPES attack case studies are presented, demonstrating the practicality of the CPS framework. For each case study, we provide the corresponding background and mathematical formulation, threat model, attack setup, and risk assessment. We also describe how each stage of the CPS analysis framework is applied to thoroughly model the specific characteristics of each case study.
\end{itemize}

\begin{figure*}[t!]
\centering
  \includegraphics[width=0.90\textwidth]{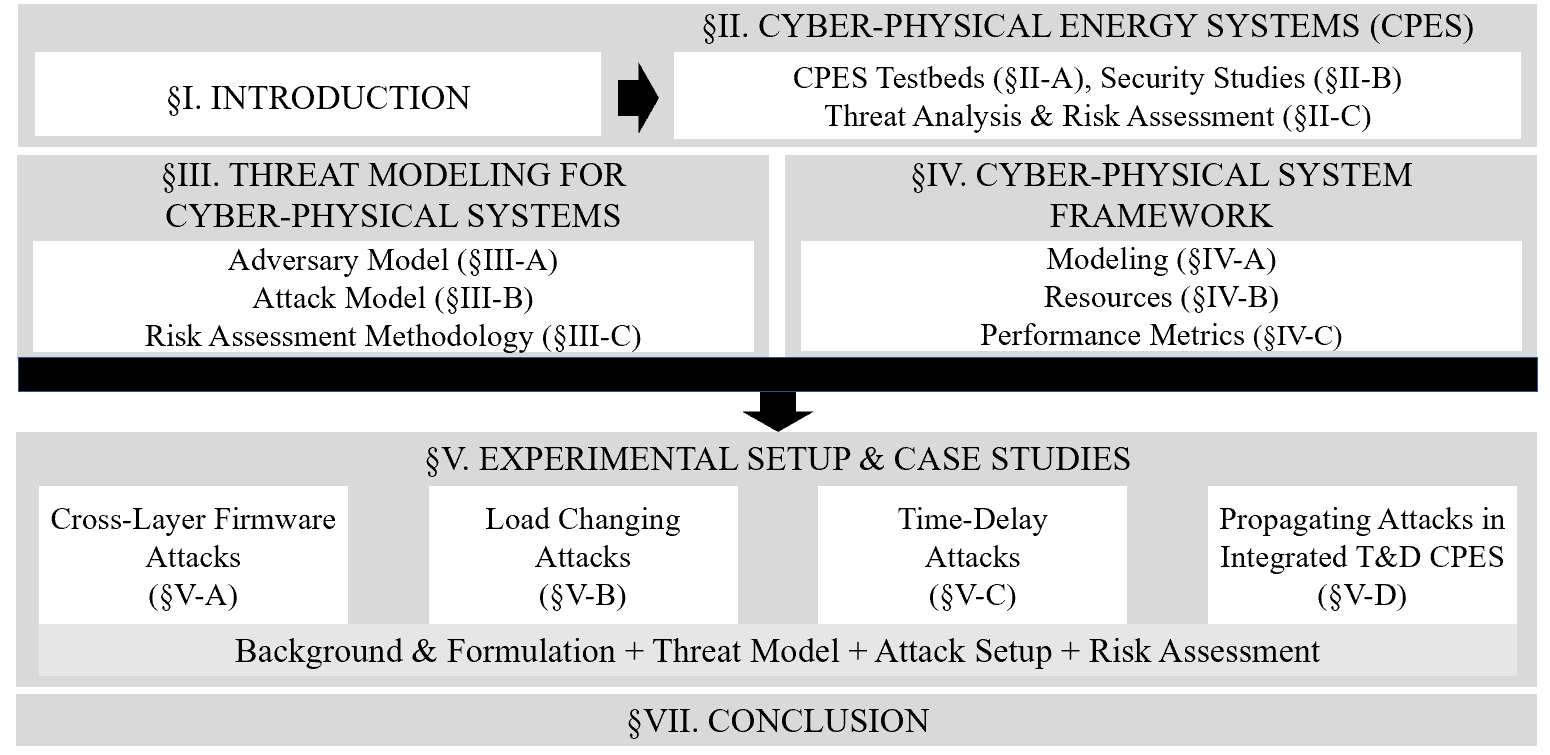}
  \caption{Roadmap of the paper.}
  \label{fig:roadmap}
\end{figure*}

A schematic overview of this paper is illustrated in Fig. \ref{fig:roadmap}. Section \ref{s:related} presents the current state of CPES testbed research, a literature review of CPES security studies, and preliminary information for threat analysis and risk assessment of CPS. Section \ref{s:CPSSecurity} delineates our comprehensive threat modeling and risk assessment methodology. Section \ref{s:cps_framework} provides the description of the proposed CPS framework with details on the modeling, resources, and performance metrics. In Section \ref{s:studies}, we discuss the background information and mathematical formulation for attack cases targeting CPES and present such simulated test case scenarios accompanied by their experimental results implemented using the developed CPS framework. Finally, Section \ref{s:conclusion} concludes this work.

%% file: sections/2-Related_work.tex
\section{Cyber-Physical Energy Systems (CPES):  Testbeds, Studies, and Security Analysis} \label{s:related}

This section provides an overview of different CPES testbeds developed by various research centers and presents their research objectives alongside the equipment used to realize them. We define different classes of CPES security studies from  literature and discuss prominent examples from such categories. Furthermore, we describe threat modeling and risk assessment methodologies and discuss how they can support security studies by defining, preventing, and mitigating threats.

\subsection{CPES Testbeds}\label{s:testbeds}

Throughout the years, EPS were designed and simulated following unidirectional structures in which power is generated at large bulk power generation facilities and then delivered through different stages of transmission and radial distribution systems to consumers. Minimum efforts were exerted to facilitate the integration of renewable energy sources  (RES) and DERs \cite{konstantinou2020towards}. 
However, the increasing penetration of RES and DERs along with the grid modernization efforts through ICT, increase the complexity of EPS \cite{muyeen2017communication}. On the one hand, RES and DERs can be used to meet consumer demands providing reliable, economic, and environmentally friendlier energy. On the other hand, attackers can exploit the fact that these resources are not centrally controlled (i.e., controlled directly by utilities) and stealthily plant their attacks on vulnerable system assets  \cite{ospina2020feasibility, 10.1145/3134600.3134639}. The complex nature of modern EPS introduces a variety of potential entry points for attacks due to the fact that these systems depend on ICT for the communication between system assets  \cite{pbpo095e_ch15}. Although the exigency for secure and resilient EPS is evident, our limited experience with dealing and coordinating such sophisticated architectures exacerbates the situation. We lack mechanisms to detect and mitigate the impact of unexpected adverse events on power system operation. The design of power system monitoring, control, and estimation algorithms, which are inherently secure, regardless of relying on CPES interconnected nature, relies heavily on the existence of representative frameworks where current and future security features and methodologies can be developed and evaluated.  

\begin{figure}[t!]
\centering
\includegraphics[width = 0.30\textwidth]{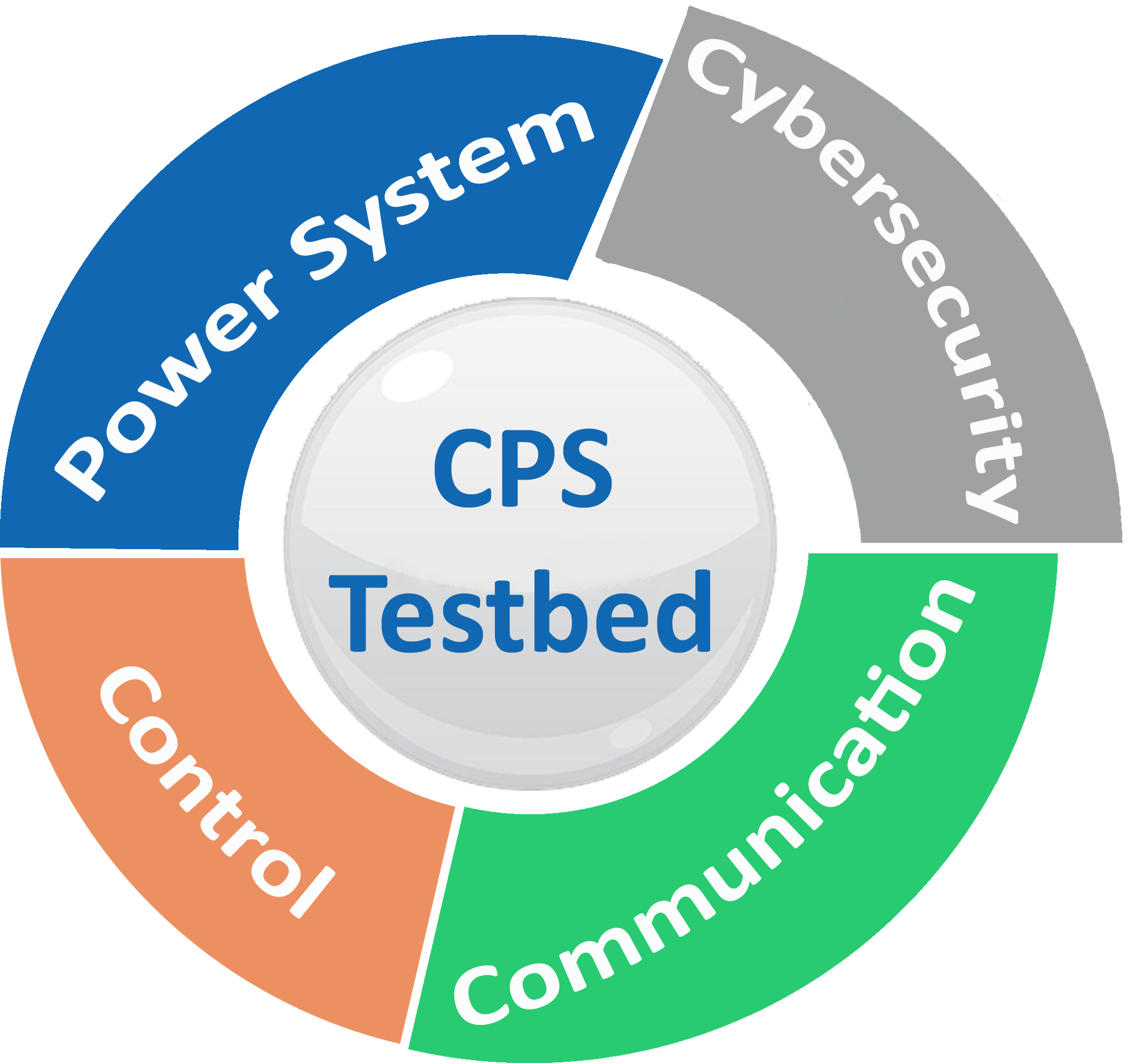}
\caption{Cyber-physical testbed components for EPS research.}
\label{fig:cpstestbed}
\end{figure}

CPES testbeds can provide an ideal environment where thorough system evaluations can be performed without any impact on the actual power system.  The use of testbeds helps de-risk certain procedures before migration to the actual system, and avoid any potential adverse impact they could inflict. Such procedures include the testing and impact evaluation of new EPS equipment (e.g., integration of PV parks, electric vehicles -- EV charging stations, etc.), new control strategies (e.g., power dispatch prioritization between DER, RES, or other power generation resources), and mitigation methodologies for unexpected events (e.g., faults, equipment failures, cyber-attacks, etc.). The main structural components of such cyber-physical testbeds are depicted in Fig. \ref{fig:cpstestbed}. Below, we provide a list of the possible security-related tasks that can be performed on CPES testbeds:

\begin{itemize}[itemsep=0pt,parsep=0pt,leftmargin=*, wide=0pt]
    \item Train users and stakeholders in a simulated/emulated CPES environment. 
    \item Validate interoperable systems' performance holistically, i.e., from the lowest level of operation (e.g., sensor, actuators, process, etc.) to the highest levels including communication between assets, distributed control, and monitoring applications.
    \item Develop and validate cyber-physical metrics and examine system security.
    \item Test novel security mechanisms such as intrusion detection and prevention systems (IDS/IPS), authentication protocols,  and encryption algorithms.
    \item Evaluate the impact of attacks on the cyber and physical domains of the EPS.
    \item Examine the effectiveness of mitigation strategies against adverse cyber-physical events.
\end{itemize}

\begin{table*}
\setlength{\tabcolsep}{1.2pt}
\centering
\caption{Cyber-physical testbed architectures, accuracy, repeatability, cost characteristics, and example testbeds with their simulation resources.}
\label{tab:testbed_types}
\renewcommand{\tabularxcolumn}[1]{m{#1}}

    \begin{tabularx}{\textwidth} { 
      || >{\hsize=.3\hsize\linewidth=\hsize\raggedright\arraybackslash}X 
      | >{\hsize=.20\hsize\linewidth=\hsize\centering\arraybackslash}X
      | >{\hsize=.30\hsize\linewidth=\hsize\centering\arraybackslash}X
      | >{\hsize=.20\hsize\linewidth=\hsize\centering\arraybackslash}X
      | >{\hsize=.20\hsize\linewidth=\hsize\centering\arraybackslash}X
      | >{\hsize=.45\hsize\linewidth=\hsize\centering\arraybackslash}X
      | >{\hsize=.10\hsize\linewidth=\hsize\centering\arraybackslash}X
      | >{\hsize=.25\hsize\linewidth=\hsize\centering\arraybackslash}X || }
      
     \hline \hline
     \textbf{Testbed Architecture} & {\textbf{Accuracy}} & \textbf{Repeatability}  & \textbf{Cost}& \textbf{Example Testbed} & \textbf{Resources*} & \textbf{Year} & \textbf{References} \\ 

     \hline 
     \multirow{2}{2.5cm}{Hardware Assisted} & \multirow{2}{*}{High} & \multirow{2}{*}{Low} & \multirow{2}{*}{High}& INL &PS: \textsl{RTDS} \newline NS: not applicable & 2015 &{\cite{INL, INL1, INL2, INL3, INL4, INL5, INL6}} \\ \cline{5-8}
     &  & &  & NREL & PS: \textsl{RTDS}, \textsl{Opal-RT} \newline NS: not applicable & 2015 &{ \cite{NREL,NREL2, NREL3}} \\
    \hline 
    
    \multirow{2}{2.5cm}{Software Assisted} & \multirow{2}{*}{Medium}  & \multirow{2}{*}{High} & \multirow{2}{*}{Low}& Texas A\&M &PS: \textsl{RTDS} \newline NS: \textsl{OPNET} &2014& { \cite{6965381},  \cite{6009221}}  \\ \cline{5-8}
    & {}  & {} & {}& TU Dortmund &PS: \textsl{Opal-RT} \newline NS: \textsl{OPNET}  &2014& {\cite{dorsch2014software, SDN4SmartGrids} }  \\ 
    \hline 
    
    \multirow{2}{3cm}{Hybrid: Physical Hardware \& Simulated Cyber} & \multirow{2}{*}{Medium}  & \multirow{2}{*}{Medium} & \multirow{2}{*}{Medium}& FSU-CAPS &PS: \textsl{RTDS}, \textsl{Opal-RT} \newline NS: \textsl{OPNET}, \textsl{EXataCPS}  & 2013 &{ \cite{6497874, CAPS_FSU,ogilvie2020modeling,10.1145/3411498.3422926}} \\ \cline{5-8}
     & {}  & {} & {}& PNNL &PS: \textsl{Opal-RT} \newline NS:\textsl{OPNET}, \textsl{ns-2}, \textsl{ns-3} & 2017&{ \cite{osti_1434888, osti_1178900}} \\ 
    \hline 
    
    {Hybrid: Simulated Hardware \& Physical Cyber} & {Medium}  & {Medium} & {Low}& HELICS &PS: \textsl{RTDS}, \textsl{Opal-RT} \newline NS: \textsl{OPNET}, \textsl{OMNeT}  & 2017 &{ \cite{helics_git, helics_tools, le2019smart, duan2019cybersecurity, palmintier2019helics},  \cite{6646193} } \\
     \hline \hline
    \end{tabularx}
    \raggedright\small{\textbf{*}PS, and NS stand for the corresponding CPS tetbeds' power and network simulators. }
    
\end{table*}

The importance of cybersecurity research for CPS and critical CPES infrastructures has led many universities and U.S. national laboratories to develop in-house testbeds, not only for research but also for education and training purposes \cite{9130868}. A variety of testbeds have been designed and implemented based on the application field and the research objectives. In Table \ref{tab:testbed_types}, we provide a summary of some of the existing real-time simulation CPS testbeds along with their inherent resources (i.e., simulation capabilities). We also categorize the cyber-physical testbeds based on their architecture, cost, and accuracy characteristics. Additionally, we present an in-depth overview of the differences between hardware and software-assisted testbeds.

Hardware-assisted testbeds are designed to explicitly study CPS while mostly incorporating several actual physical components encountered in the field. For instance, CPES hardware-assisted testbeds integrate physical equipment such as generators, relays, switchgear, energy storage systems -- ESS, PV panels, wind turbines, etc. By replicating the behavior of the actual system with a considerable amount of physical equipment, these testbeds provide stakeholders the ability to: \textit{i)} make decisions not only based on theoretical analyses but practical studies leveraging the use of hardware resources, \textit{ii)} evaluate the CPS behavior under abnormal operational scenarios without inhibiting the operation of the real system, and \textit{iii)} preemptively assess cyber-attack or fault mitigation and control strategies before the corresponding hardware is deployed to the field, and thus, de-risk this cost-prohibitive and unpredictable process. Hardware-assisted testbeds, however, suffer from three major disadvantages: \textit{i)} they are not cost-effective since they require the testbed components to match the actual equipment deployed in the field, \textit{ii)} once the equipment and testbed configurations are setup in-place, any modification or expansion of the system architecture can be either time-consuming or practically and economically infeasible, and \textit{iii)} scalability issues of representing large-scale EPS due to the requirement of procuring more assets (e.g., generators, inverters, etc.).

A typical example of a hardware-assisted research laboratory that leverages actual operational equipment to perform CPES security research is the  Idaho National Laboratory (INL) of the U.S. Department of Energy (DOE) \cite{INL}. INL's Power and Energy Real-Time Laboratory \cite{INL1, INL2}, alongside their nuclear laboratory \cite{INL3, INL4} and microgrid (MG) testbed \cite{INL5, INL6}, allow the simulation of realistic scenarios supported by actual hardware equipment and data generation routines. The real-time simulation capabilities of INL's testbeds allow researchers to create sophisticated scenarios involving power hardware devices that are interfaced with real-time simulation environments via HIL methodologies such as power hardware-in-the-loop (PHIL) and controller hardware-in-the-loop (CHIL) \cite{INL2}. HIL allows controllers (CHIL) and parts of EPS (PHIL) to be extensively tested before their final integration to the main grid \cite{faruque2009hardware}. The National Renewable Energy Laboratory (NREL) of DOE also includes hardware-assisted testbeds \cite{NREL}. NREL's Flatirons campus specializes in designing, analyzing, and providing accurate simulation models for wind turbines, hydropower, and hydrokinetic generation plants \cite{NREL2}. Their unique facilities drive the improvement of their high-fidelity simulation models, which are cross-referenced to real assets, providing invaluable tools for power engineers performing system analyses incorporating off-shore, or distributed hydro and wind generation \cite{NREL3}. The actual power system assets of wind turbines and hydro-plants, as well as their simulation models, can be leveraged to investigate the potential impact of component failures or cyber-attack incidents with minimum cost, and most importantly, without compromising the actual EPS operation.

Hardware-assisted CPES testbeds do not exclusively utilize physical equipment. In most cases, the conducted research is supported by simulation software enabling the analysis of more complex systems. Since an actual duplicate of an operational CPS in the lab is typically infeasible, in the past years, a high number of software-based CPS testbeds have been developed following, the notion of digital-twin systems \cite{Boschert2016, 8779809}. The main difference between software-assisted testbeds and their hardware-assisted counterparts is that they do not possess any actual field equipment, thus limiting their testing scenarios. Moreover, software-assisted testbeds can be further segmented into sub-categories based on the simulation platform utilized for the system analysis. Some of them utilize widely available software simulators, e.g., \textsl{Matlab/Simulink}, \textsl{PowerWorld}, \textsl{PSSE}, etc., while other rely on real-time simulators (RTS) such as \textsl{Opal-RT}, \textsl{RTDS}, \textsl{Typhoon}, and \textsl{Speedgoat}. The main advantage of software-based CPS testbeds, compared to hardware-based testbeds, is the increased flexibility in designing, modifying, and scaling the systems under test. Also, their cost can be significantly lower for simulating large-scale CPES. However, the validity of the software-based simulated results relies heavily on the fidelity of the models (for emulation, virtualization, etc.) used to represent the corresponding real systems under investigation. 

Examples of testbed environments with extensive CPS simulation capabilities include the ones at Texas A\&M and TU Dortmund. At the Texas A\&M CPS testbed, despite the lack of actual EPS equipment, CPES technologies such as smart grid controllers and RES can be virtualized and evaluated using software-based implementations. The testbed also includes RTS  systems (\textsl{RTDS}) and supports the modeling of communications of CPES components via network simulators (\textsl{OPNET}). Furthermore, it allows researchers to evaluate how communication-enabled devices expand the threat surface \cite{6965381}.

The rapid penetration of ICT technologies in CPS is driving the design and development of large-scale software-defined network (SDN) testbeds  \cite{huang2016survey}. In such SDN-type testbeds, researchers can evaluate novel network technologies, communication protocols, custom data routing algorithms, etc. An example of such an environment is the SDN4SmartGrids CPS testbed at TU Dortmund, where both SDNs and power system RTS are employed for experimentation with ICT-based smart grid applications \cite{dorsch2014software}. In particular, the TU Dortmund's testbed is comprised of a RTS (\textsl{Opal-RT}) responsible for simulating the power system components. The infrastructure emulating the network topology and communication between the simulated grid assets (e.g., EVs, ESS, etc.), management systems, and telemetry units (e.g, phasor measurement units -- PMUs, advanced metering infrastructure -- AMI, etc.) is implemented using the SDN and the \textsl{OPNET} network simulator \cite{SDN4SmartGrids}.

In order to bridge the gap between the hardware and software-assisted CPS testbed methodologies, hybrid testbeds are considered as an effective alternative. As their name implies, hybrid approaches trade-off the utilization of the physical components, that can be found at the transmission and distribution (T\&D) level of CPES, with the utilization of simulators and software suites designed to accurately represent the behavior of real energy systems. Hybrid testbeds enable diverse security investigations that can focus on the physical-system (e.g., programmable controllers, IEDs, grid assets, etc.), the cyber-system (i.e., SCADA communications, telemetry and remote control of assets, monitoring and measurement components, etc.), or any combination of the two. The main advantage of such testbeds is that they provide re-configurable platforms that can scale up, using simulation, to realistic systems' sizes, while also retaining the ability to investigate, with high granularity, the individual security and control properties present in physical devices. Consequently, hybrid CPS testbeds can evaluate holistically the impact of cyber-attacks on CPES, without any of the limitations encountered in hardware-assisted or software-assisted testbeds. 

A prime example of a hybrid CPES testbed framework is HELICS \cite{helics_git, helics_tools}. The HELICS infrastructure enables the integration of different RTS operating at different time-steps as well as the interconnection of T\&D system components. By timely simulating (depending on the temporal constraints) complex T\&D architectures, cybersecurity assessments, including real-time impact analysis and risk mitigation strategies, can be conducted providing meaningful insights regarding the behavior of CPES \cite{le2019smart, duan2019cybersecurity, palmintier2019helics}. The Pacific Northwest National Laboratory (PNNL) also features a hybrid testbed leveraging the aforementioned advantages. The testbed facilitates a variety of cybersecurity studies \cite{osti_1434888}, and provides an effective framework for system vulnerability assessments,  interactive simulations of CPES environments, threat scenario analyses, and risk mitigation strategy evaluations.

The facilities of the Center for Advanced Power Systems (CAPS) of Florida State University (FSU) also include a hybrid testbed setup. The testbed supports the use of RTS, based on the \textsl{RTDS} and \textsl{Opal-RT} platforms, power system simulation software such as \textsl{OpenDSS}, \textsl{PSCAD/EMTDC}, \textsl{Matlab/Simulink}, \textsl{RT-Lab}, \textsl{RSCAD/RTDSphysical}, and EPS components including generators, inverters, and flexible AC transmission systems (FACTS) \cite{6497874}. The center's infrastructure can be segregated into two main subsystems able to perform both real-time and HIL simulations. The first subsystem is composed of $15$ \textsl{RTDS}-enabled racks, each consisting of around $26$-$30$ parallel processors. The subsystem can support real-time simulations comprised of more than $1,000$ electrical nodes (e.g., measurement points) and $5,000$ control units at time-steps in the range of $50\mu$s. It should be noted that for time-critical implementations, such as power electronics converters, the time-step of real-time simulation can be further reduced in the vicinity of $1\mu$s. Fiber-optic networks facilitate the interconnection between the RTS and the physical EPS equipment. Namely, the physical equipment of the testbed includes a $4.16$ kV distribution system, a $7.5$ MVA on-site service transformer, a $5$ MW variable-voltage variable-frequency converter, a $5$ MW dynamometer, and a $1.5$ MVA experimental bus at $480$ V$_{ac}$ \cite{CAPS_FSU}. The second subsystem includes three \textsl{Opal-RT}-enabled racks, supported by multiple processor units along with \textsl{Xilinx field-programmable gate array (FPGA)} computation units. The FPGA hardware accelerators perform the simulation of high-frequency power electronic converters with stringent timing constraints (i.e., in the ns range), while the rest of the EPS is simulated using $\mu$s time-steps. 
Both subsystems have support for multiple industrial protocols utilized for the communications between the physical or simulated EPS assets. Advanced control schemes and experimentation with communication network components are also supported via HIL simulations \cite{ogilvie2020modeling, ospinasampling}. Additionally, the impact of unexpected failures or cyber-attacks targeted at these components can be examined in a controlled environment where minimum risk exists \cite{10.1145/3411498.3422926}.

\begin{table*}[ht!]

\setlength{\tabcolsep}{1.2pt}
\centering
\caption{CPES security study categories and research examples.}
\label{tab:CPESstudies}
\resizebox{0.8\textwidth}{!}{
\begin{tabular}{||l|c|c||}
\hline \hline
\textbf{Security Study Category} & \textbf{Example Literature} & {\textbf{Reference}} \\ \hline 

 \multirow{5}{3.2cm}{Attacks exploiting CPES vulnerabilities} & Low-budget GPS spoofing attacks & \cite{keliris2018low} \\ \cline{2-3}
 & Load redistribution attacks & \cite{xiang2017coordinated}\\ \cline{2-3}
 & Coordinated DoS attacks &\cite{tian2020coordinated}\\ \cline{2-3}
 & Data integrity attacks on state estimation & \cite{tu2020hybrid}\\ \cline{2-3} 
 & FDIAs with limited resources & \cite{pan2017data}\\ \cline{2-3}
 & Hall sensor spoofing attack & \cite{barua2020hall}\\ \hline 
 
 \multirow{7}{3.2cm}{Evaluation of attack impact on CPES} & Graph-theoretic quantitative security assessment & \cite{venkataramanan2019cp} \\ \cline{2-3}
 & Unsupervised learning-based evaluation & \cite{zhang2019cyber} \\ \cline{2-3}
 & Simulation-based impact evaluation & \cite{kuruvila2020hardwareassisted} \\ \cline{2-3}
 & Deep reinforcement learning analysis & \cite{liu2020deep} \\ \cline{2-3}
 & Stochastic modeling assessment& \cite{orojloo2018stochastic} \\ \cline{2-3}
 & Quantitative attack impact evaluation on substations & \cite{fan2019supporting} \\ \cline{2-3}
 & Markov-process based reliability analysis & \cite{yang2018reliability} \\ \hline

 \multirow{8}{3.2cm}{Attack detection algorithms in CPES} &  Unsupervised learning-based FDIA detection & \cite{wei2018false}\\ \cline{2-3}
 & Wavelet transformation-based FDIAs in AC state estimation & \cite{james2018online}\\ \cline{2-3}
 & Historical data-assisted FDIA detection &\cite{ye2019summation}\\ \cline{2-3} 
 & Distributed collaborative FDIA detector & \cite{li2017distributed} \\ \cline{2-3}
 & Machine-learning assisted FDIA detection & \cite{sayghe2020survey}\\ \cline{2-3}
 & Unsupervised learning anomaly detector &\cite{goh2017anomaly}\\ \cline{2-3}
 & Sensor- and process noise-based attack detection & \cite{ahmed2018noiseprint}\\ \cline{2-3}
 & Autoencoder-based anomaly detection &\cite{schneider2018high}  \\ 
    \hline

 \multirow{8}{3.2cm}{Attack mitigations and defenses in CPES} & Semi-supervised method for malware & \cite{huda2017defending}\\ \cline{2-3}
    & Markov-process based reliability analysis & \cite{yang2018reliability}\\ \cline{2-3}
    & Data-driven and compressive sensing resilient state estimator &\cite{8743447}\\ \cline{2-3}
    & Battery-based hardware security authentication & \cite{zografopoulos2020DERauth}\\ \cline{2-3}
    & Battery-power assisted risk mitigation &\cite{kushal2018risk}\\ \cline{2-3}
    & Robust control-based defense mechanism &\cite{kwon2016cyber}\\ \cline{2-3}
    & Control flow integrity validation method & \cite{hao2019integrating}\\ \cline{2-3}
    & Hardware security-based communication protocol extension & \cite{zografopoulos2020harness} \\
 
\hline \hline
\end{tabular}}
\end{table*}

\subsection{CPES Security Studies}\label{s:CPS_Studies}

During the past decade, significant effort has been exerted into CPES security studies with the objective of enhancing CPES resiliency and alleviating cybersecurity vulnerabilities. For instance, a comprehensive work reviewing cybersecurity vulnerabilities and solutions for smart grid deployments is presented in \cite{gunduz2020cyber}. Security solution evaluation, system threat classification, and future cybersecurity research directions are also considered. The authors in \cite{kimani2019cyber}, investigate cyber-attacks on IoT-enabled grid deployments. They discuss how advancements in IoT technologies can drive the power grid modernization process, but at the same time increase the system's threat surface given its interconnected topology encompassing millions of IoT nodes. Researchers in \cite{canaan2020microgrid} examine the security of modern power systems from the viewpoint of interconnection with microgrids. Emphasis is given on the cybersecurity and reliability challenges arising in these architectures. Essential approaches (e.g., testbed-assisted security studies) are discussed to enhance the security of future power systems. In addition, \cite{el2018cyber} provides a complete overview of the cyber-threats encountered on the infrastructure, network protocols, and application levels of power systems. Furthermore, attacks targeting the data availability, integrity, and confidentiality of microgrids are discussed in \cite{nejabatkhah2021cyber}.

In this section, we outline the main topics of existing literature in the area of CPES security. More specifically, the literature work is classified using the following categories: \textit{i)} studies investigating the exploitation of CPES vulnerabilities, \textit{ii)} studies evaluating the impact of cyber-attacks on CPES, \textit{iii)} studies proposing and assessing algorithms (e.g., anomaly detection, IDS/IPS, etc.) for the detection of cyber-attacks, and \textit{iv)} studies focusing on mitigation and defense mechanisms. In Table \ref{tab:CPESstudies} we provide an overview of recent CPES security studies classified under the four aforementioned categories.

\subsubsection{Attacks Exploiting CPES Vulnerabilities}

CPES are advancing towards decentralized interconnected systems in order to support increasing power demand while minimizing transmission losses, leverage MG deployments and their functionalities (e.g., grid-connected or autonomous operations), and incorporate DERs. In addition, to enhance CPES control, reliability, and security, digital ICT equipment such as advanced measuring and monitoring units are being employed in geographically dispersed locations of decentralized CPES. For example, PMUs provide time-synchronized (using GPS) granular measurements for EPS related states including voltage, current, and power magnitudes and phase angles. However, it has been demonstrated that adversaries can leverage open-source public resources to perform GPS spoofing attacks against PMUs \cite{keliris2018low}. By introducing small undetectable timing delays (in the $\mu$s range) in the measurement signals (within the IEEE standard limits for synchrophasors C37.118 \cite{6111222}), the phase differences between actual and measured angles can be significantly altered exceeding allowed limits, tripping circuit breakers (CBs), sectionalizing parts of the EPS, and causing power outages (e.g., brownouts, blackouts) \cite{konstantinou2017gps}.

Moreover, in \cite{xiang2017coordinated}, researchers introduce a coordinated load redistribution attack affecting power dispatch mechanisms. By attacking generators or transmission lines while falsifying load demand and line power flows, system operators are misled into increasing load curtailment. Furthermore, in \cite{tian2020coordinated}, the authors investigate two types of DoS attacks along with their impact on EPS. The first attack is assumed to be a stealthy false data injection attack (FDIA) performed to mask the attack impact from detection algorithms. The second, assumed as a non-stealthy attack, aims to maximize the damage on power system operation by targeting the most vulnerable transmission line, impeding power dispatch, and causing load shedding. In \cite{tu2020hybrid}, the authors propose hybrid data integrity and data availability attacks. They demonstrate how control center measurements can be manipulated leading to undetectable FDIAs. In more detail, by modifying some measurements (i.e., integrity attack) while making some others unavailable to the state estimation algorithm (i.e., availability attack), FDIAs can bypass bad data detection algorithms. 

Ubiquitous power electronics bring new challenges to CPES operation \cite{9324816}. Future CPES are expected to be inverter-dominated systems. As such, vulnerabilities in such components can lead to abnormal system operation. In \cite{barua2020hall}, the authors investigate how stealthy non-invasive attacks on grid-tied inverters can compromise their nominal operation and impact grid operation. Specifically, by spoofing the inverter's hall sensor they demonstrate fluctuations in the output voltage, active and reactive power while also introducing low-frequency harmonics to the grid. Similarly, by exploiting a vulnerability in the authentication mechanism of General Electric Multilin protection and control devices, the authors in \cite{keliris2017ge} show that remote or local attackers can obtain weakly encrypted user passwords, which could then be reversed allowing unauthorized access. Furthermore, the authors in \cite{10.1145/3134600.3134639, ospina2020feasibility} show that by coordinating the power usage of multiple devices, power reserve limits of EPS can be exceeded causing tripping of lines and shedding of loads. A botnet of IoT (internet-of-things)-connected high-wattage loads, such as washing machines, air-conditioning units, dryers,  etc., are coordinated over the network, causing unexpected power usage profiles and pushing the grid to instability limits. Such attacks demonstrate that there is no requirement of strong adversarial knowledge nor considerable attack resources \cite{pan2017data}.

\subsubsection{Evaluation of Attack Impacts on CPES}

Impact evaluation and analysis studies are considered essential for prioritizing and safeguarding critical components in CPES. Such analyses explore the consequences of malicious attacks and can serve to proactively prepare systems for their adverse implications. Impact evaluations can expose critical system components, assist in prioritizing and securing them, and aid in the development of contingency plans in case these vulnerable components get compromised.  For instance, the authors in \cite{venkataramanan2019cp} propose assessment metrics designed to evaluate the resiliency of CPES against adversarial attacks. Different techniques from game theory, graph theory, and probabilistic modeling have been utilized to assess the capability of CPES when supporting critical (or unsheddable) loads after they have been compromised or the system has suffered unexpected disturbances. Other works focus on analyzing the impact of cyber-attacks in transactive energy systems -- TES \cite{zhang2019cyber}. Here, the authors investigate the system operation under two types of attacks that are designed to maliciously affect either the bid prices or the bid quantities. In view of the fact that IEDs, AMI, and smart inverters are penetrating EPS at a rapid pace, the authors in \cite{kuruvila2020hardwareassisted} and \cite{PESGM2021} demonstrate the adverse grid consequences if such devices are compromised. Specifically, the simulated impact of malicious smart inverter firmware modifications in MGs is demonstrated in \cite{kuruvila2020hardwareassisted}. Attacks targeting SCADA-controlled switching devices or monitoring devices impeding situational awareness (in an integrated T\&D system model) are evaluated in \cite{PESGM2021}.

Furthermore, cybersecurity assessment methodologies investigating the impact of RES integration to the grid are also investigated in the literature. For instance, the authors in \cite{liu2020deep} leverage open-source intelligence and contingency analysis methods to discover the most critical system paths. Such transition paths could be utilized by an adversary to maximize the impact of cyber-attacks, leading to disastrous consequences for the EPS. A different approach, which considers intrusion and disruption process modeling, is proposed in \cite{orojloo2018stochastic}, where a stochastic game theory-based CPES security evaluation model is developed. The authors in \cite{fan2019supporting} propose a mathematical framework to estimate the probability and evaluate the impact of malicious attacks on substation automation systems. In \cite{yang2018reliability}, the reliability and security of CPES are analyzed through a communication failure assessment process. Overall, assessment methodologies of attack impacts on CPES are designed with the purpose of aiding CPES evaluation studies. Thus, they should be leveraged as part of a defense-in-depth (DiD) portfolio when assessing potential damages and devising CPES defense strategies.

\subsubsection{Attack Detection Algorithms in CPES}

The severity of the effects of cyber-attacks in CPES underlines the need for accurate and effective attack detection mechanisms that can improve the situational awareness of system operators. Hence, remediation actions can be issued to avoid system and equipment failures, as well as ensure human safety. A plethora of detection schemes have been proposed especially for FDIAs in CPES \cite{wei2018false,james2018online,ye2019summation,li2017distributed}. For instance, in \cite{li2017distributed} researchers develop a distributed host-based collaborative mechanism for detecting false data measurements in PMUs. Each PMU is assigned a host monitor to probe its status (i.e., normal operation or anomalous) by comparing it with predefined nominal values. Then, a majority voting algorithm is executed to decide if the acquired measurements are valid by comparing the status of the under-investigation PMU with the corresponding neighboring PMUs. Unsupervised learning-based anomaly detection methods have also been proposed for cyber-attack detection in CPES \cite{sayghe2020survey}. An example of such anomaly detection scheme is presented in  \cite{goh2017anomaly}, where authors identify suspicious sensor activity using recurrent neural networks (RNNs). Other researchers have also demonstrated how data integrity attacks (DIA) can be identified when sensor and process patterns deviate from a residual-based fingerprinted data  \cite{ahmed2018noiseprint}. Furthermore, given the extensive use of Fieldbus communication devices in CPES, methodologies have been designed to detect anomalous network traffic in a variety of Fieldbus protocols  \cite{schneider2018high}. All of the reviewed detection mechanisms have the objective of notifying system operators once incongruous sensor or monitor behavior is detected in the CPES. As a result, malicious incidents can be effectively handled, minimizing their impact on CPES operations.

\subsubsection{Attack Mitigations and Defenses in CPES} 

The deployment of defense and mitigation mechanisms is critical to enhance the overall CPES security and minimize the adverse impact of cyber-attack scenarios. For example, mitigation strategies can protect CPES against FDIAs which could potentially result in generator equipment damage \cite{kushal2018risk}. Specifically, BESS could be leveraged to assist the generators and reduce the load curtailment inflicted by malicious attacks. A hybrid control-based approach to safeguard systems against cyber-attacks is presented in \cite{kwon2016cyber}. The hybrid controller switches to the most secure controller, from a subset of available controllers, given that some of these controllers might have been compromised by an adversary. In \cite{huda2017defending}, a semi-supervised learning mechanism is utilized to study malware patterns and defend the system from unknown malware targeting the CPES infrastructure. 

Apart from software-based mitigation techniques and defenses, hardware-oriented mechanisms have also been proposed. In \cite{zografopoulos2020DERauth}, the authors propose the use of hardware security primitives leveraging the intrinsic variation of BESS lithium cells to enhance communication protocol security. The practicality of the approach is validated in a simulated testbed environment \cite{zografopoulos2020harness}. Furthermore, in \cite{hao2019integrating}, an instrumentation-based defense technique is presented employing a sub-optimal plan to secure CPES in real-time. Even though the discussed defense and mitigation mechanisms may not be applicable for all cyber-attack scenarios, research and  development in this direction contribute towards understanding attackers' tactics and defending against them, enhancing the security of CPES.

\subsection{Threat Analysis and Risk Assessment} \label{s:threatAssessment}

Precise modeling is essential in order to investigate complex CPES architectures, discover any potential vulnerabilities, and extensively test and evaluate security features. The intricacies of CPS typically consist of multiple interconnected layers bridging assets of varying importance for the system operation, and leveraging ICT and communication protocols. Different methods are being used to review CPS  architectures and assess their cybersecurity. Among them, the DiD and the Purdue models are the most popular ones. The DiD strategy was initially employed in military applications \cite{DefenseInDepth}. It ensures resiliency, redundancy, and the existence of multiple defenses if a vulnerability is exploited, a critical security flaw is identified, or a failure or unintentional fault occurs. Enforcing the DiD multi-layered topology has two main advantages from a security perspective. First, it delays the attack progress in the system since each layer provides an isolated execution environment. Second, it allows system operators to deal with the attack independently on multiple layers, rather than having to rely on a single point-of-defense. Similarly, the Purdue model for industrial control system (ICS) network segmentation \cite{Purdue}, part of the Purdue Enterprise Reference Architecture (PERA), incorporates the DiD concept by demonstrating the interconnections and dependencies between layers and components, allowing for the design of secure CPS \cite{IndustrialCyberSec}. In the following parts (\ref{s:threatRelatedWork} and \ref{s:riskRelatedWork}), we provide the essential information and related work regarding threat modeling and risk assessment methodologies with emphasis on industrial CPS and critical infrastructures.

\subsubsection{Threat modeling} \label{s:threatRelatedWork}

The term `\emph{threat modeling}' refers to the procedure by which potential vulnerabilities are discovered before they can become system threats. This process is crucial for the design of security defenses and mitigation strategies. It is evident that performing threat modeling for CPES is essential since their compromise can have disastrous consequences to the grid operation and the economic and social well-being. However, CPES consist of multiple layers and assets, hence, it can be challenging, due to extensive time, modeling efforts, resources, and cost, to exhaustively examine all the possible scenarios that could arise as system vulnerabilities. To overcome such issues, without compromising the system's reliability, multiple threat modeling approaches have been proposed aiming to prioritize vulnerabilities and assist the implementation of potent security mechanisms. These methodologies provide a holistic view of the system by highlighting the significant assets, commonly referred to as crown-jewels\cite{crownjewels}, and assessing threats based on their potential impact and ease of deployment on the system.

STRIDE\footnote{STRIDE is an acronym for Spoofing, Tampering, Repudiation, Information disclosure, Denial-of-service, and Elevation of privilege.} and DREAD\footnote{DREAD is also an acronym that stands for Damage, Reproducibility, Exploitability, Affected Users, and Discoverability.} are well-established threat modeling frameworks for the security assessment of products and services throughout their life-cycle \cite{STRIDE,DREAD}. For instance, STRIDE uses data flow diagrams for the threat modeling process. The data flow diagrams map system threats to the corresponding vulnerable system components (STRIDE per-element approach). Given the interdependent nature of CPES, an attacker can compromise the system operation by exploiting different component vulnerabilities. Therefore, to guarantee the overall system security,  vulnerabilities need to be addressed both at the component level as well as within the component interrelations (visualized in the data flow diagrams) \cite{khan2017stride}. DREAD can be leveraged to evaluate and rank the severity of threats. A DREAD analysis is comprised of the following six steps: asset identification, system architecture formation, application decomposition, threat identification, threat documentation, and threat impact rating. DREAD and STRIDE methodologies can also be used jointly for comprehensive cybersecurity assessments \cite{DREAD_STRIDE}.

Apart from STRIDE and DREAD, other methodologies for security assessments have been proposed and utilized in the cybersecurity arena. For instance, OCTAVE\footnote{OCTAVE acronym is for Operationally Critical Threat, Asset, and Vulnerability Evaluation.} Allegro is an alternative approach used by organizations when performing mainly information technology (IT) security evaluations and strategic planning for cyber-threats \cite{Allegro}. However, recent works validate the applicability of OCTAVE Allegro for CPS security assessments, both for the enumeration of potential risks as well as the design of countermeasures to maintain nominal system operation \cite{PESGM2021, radanliev2018integration}.
The main steps followed in OCTAVE security assessments include: the development of risk evaluation criteria according to operational constraints, critical asset identification,  critical asset vulnerabilities and corresponding threats discovery, and threat impact assessment. STRIDE, DREAD, and OCTAVE are well-established tools when performing threat modeling analyses and identifying vulnerabilities in the pre-attack context.

The investigation of adversary behavior post-compromise is also important. At this point, the adversary has already overcome the first line of defense and has access to system resources. Notably, there is extensive research on initial exploitation and use of perimeter defenses \cite{stouffer2011guide, 8270432}. However,  there is a knowledge gap of the adversary process after initial access has been gained. To address the aforementioned pitfall and support threat modeling, risk analysis, and mitigation methodologies, pre-and post-compromise events, MITRE developed the ATT\&CK for Enterprise framework \cite{MITREattack}.

MITRE ATT\&CK is an open-source knowledge-base that includes common adversarial attack patterns (e.g., attacks, techniques, and tactics). The ATT\&CK database is constantly being updated with recent attack incidents to enhance enterprise cybersecurity by exposing system vulnerabilities and warrant safer operational environments for businesses and organizations. 
The framework describes the tactics, techniques, and procedures (TTPs) that an adversary could follow in order to make decisions, expand access, and stealthily compromise an organization while residing inside the enterprise network \cite{APT1, APT2}. 
In January 2020, MITRE corporation, realizing that ICS is an essential part of critical CPS infrastructures and with the objective of addressing cybersecurity issues arising by the diverse and interconnected nature of CPS, launched the ATT\&CK for ICS framework \cite{MITRE}. 

The ATT\&CK for ICS framework is also a free community-supported threat knowledge-base that includes information about TTPs that adversaries utilize when targeting ICS (within CPS). The framework assists in understanding the adversarial attack chain and enhance the security standpoint of ICS and related CPS assets. ATT\&CK for ICS is based on MITRE's ATT\&CK for Enterprise framework, i.e., it ports many of the gathered threat intelligence from enterprise networks to ICS since industrial networks often have similarities with enterprise networks. The heterogeneity of ICS, however, with a plethora of operating systems (OS), network devices, and communications protocols co-existing with a variety of field devices (e.g., PLCs, IEDs, PMUs, RTUs, etc.) led to significant revisions from the ATT\&CK for Enterprise to the ATT\&CK for ICS.

The ATT\&CK for ICS framework is designed to support a multi-layer reference approach for adversarial behavior evaluations. The framework is segregated into four core components, making it applicable to a wide spectrum of industrial CPS. The first component category includes \textit{i)} \textit{assets} which consist of control servers, engineering workstations, field controllers, human-machine interface (HMI), among others. All these assets might not be apparent in every system. This is factored in by the ATT\&CK methodology which investigates attacks targeting the respective assets independently as well as their cooperation with other industrial assets. The second core part of ATT\&CK for ICS is the abstraction focusing on the \textit{ii)} \textit{functional levels} of the Purdue architecture. Such levels describe the depth of infiltration that the adversary has achieved. The level ranges from \textit{Level 0}, which corresponds to the physical devices (e.g., sensors and actuators) that orchestrate the industrial process, all the way to \textit{Level 2}, which includes the supervisory control systems, the engineering workstations, and HMIs. These functional levels are depicted in Table \ref{tab:funcLevels}. The last two parts of the framework revolve around the adversarial \textit{iii)} \textit{tactics} and \textit{iv)} \textit{techniques}. The term `\textit{tactics}' refers to the reason why an adversary performs an action, i.e., adversary objective such as disrupting an industrial process control routine.  \textit{Techniques} describe the activities that the adversary uses to achieve the attack goal, i.e., represent ``how'' an attacker accomplishes his/her objectives by taking an action, e.g., through modifying the PLC control logic.

\begin{table}
\caption{ICS functional levels, equipment categories, and their corresponding components\cite{MITRElevels}.}
\label{tab:funcLevels}
\setlength{\tabcolsep}{1.5pt}
\begin{center}

\renewcommand{\tabularxcolumn}[1]{m{#1}}

\begin{tabularx}{0.95\linewidth}
{||>{\hsize=.7\hsize\linewidth=\hsize\raggedright\arraybackslash}X
  |>{\hsize=0.8\hsize\linewidth=\hsize\centering\arraybackslash}X
  |>{\hsize=1.5\hsize\linewidth=\hsize\centering\arraybackslash}X||}
\hline

\hline \hline
Functional Level & Category & Components \\
\hline
\textbf{Level 2} &  Supervisory Equipment &
    Supervisory control functions, site monitoring, local displays \\
\hline

\textbf{Level 1} & Control Equipment &
    Protection devices, local control devices
\\ \hline

\textbf{Level 0} & Process Control &
    Sensors, actuators 
\\ \hline \hline

\end{tabularx}
\end{center}
\end{table}

\subsubsection{Risk Assessment}\label{s:riskRelatedWork}

The term "\emph{risk assessment}" refers to the process of identifying potential risks and their corresponding impact to the system operation as well as determining strategies to mitigate, defer or, accept these risks based on their criticality \cite{Allegro}. Cyber-threat risk assessment is a critical operation that CPES and their ICS need to perform regularly. The introduction of new technologies into CPES (i.e., DERs, EVs, control devices, etc.) along with the interoperable nature of the supported ICT infrastructure increases the risks arising from both the cyber (e.g., measurement, control commands, or communication integrity attacks) and the physical domain (e.g., sensor and actuator compromise).

Typically, risk assessment methodologies rely on probabilistic analyses that leverage Markov-chains \cite{pietre2010attack}, Petri-nets \cite{li2017asset}, Bayesian belief networks \cite{huang2018assessing}, or game theory to estimate the impact of adverse events on system operation \cite{BLOOMFIELD2017198, 5164937}. In \cite{5164937}, for example, researchers model both the attackers and the system's defenses as agents with different action sets and objectives. Due to the contradictory roles of such agents, the corresponding action payoff depends on the ability to compromise the system's assets or the ability to detect the malicious attack from the perspective of attackers or defenders, respectively. Other works have proposed worst-case scenario risk assessment analyses that employ exhaustive Monte Carlo simulations and focus on diverse operation areas of EPS (e.g., automatic generation control --AGC, T\&D system operations, etc.). Then,  the interdependence of such EPS areas with specific risk mitigation mechanisms is analyzed \cite{li2014risk, 7439817}. For instance, the authors in \cite{7439817}, review the impact on buses and transmission lines under abnormal operations caused by cyber-attacks. They also investigate how adverse scenarios can be mitigated if robust protection system strategies, i.e. coordinated bus and transmission line trippings, are correspondingly put in-place. Although probabilistic risk analyses and  worst-case scenario assessments can provide useful results under specific constraints (i.e., if only part of a system is examined), applying such methods to dynamically changing large-scale T\&D integrated models can be a challenging task.
The multitude of T\&D assets expands the search space of exhaustive methods such as Monte Carlo-based risk analyses \cite{WANG201824}. For each asset and every investigated potential attack, the risk analysis process needs to be re-examined and re-computed. The risk calculation overhead is also exacerbated due to the interconnected CPS architecture.

The aforementioned methods, apart from being computationally intensive, can also potentially suffer from poor accuracy. The security risk assessment accuracy of these methods relies on the precise modeling of the CPES physical components (e.g., generators, transmission lines, substations, etc.), their topology, as well as their interconnections with the cyber components (e.g., ICT nodes supporting EPS functions) \cite{giannopoulos2012risk, theocharidou2015risk,6979242}. Failure to properly model CPES can mask interconnection dependencies between components and their layers (cyber or physical), and thus,  perturb the risk score calculation process. 
The presented risk assessment approaches in this section are credible if security assessment is performed partially, i.e., they fail to capture comprehensively system risks as their focus is on specific parts of a CPES ignoring the impact propagation to the rest of the infrastructure. In this work, the threat and system representation is performed meticulously during the threat modeling process (Section \ref{s:CPSSecurity}) and the CPS framework stages (Section \ref{s:cps_framework}), respectively. As a result, our approach determines in advance a detailed system model, overcoming the drawbacks encountered when performing segmented risk evaluations.

In our analysis, system-specific characteristics are formalized and \textit{Risk} scores are calculated by combining the attack \textit{Threat Probability} along with the CPES objective priorities (Section \ref{s:riskAssess}). The proposed methodology expedites the risk assessment analysis of CPS (since the threat modeling, CPS framework analysis, and performance metrics determination have been performed previously), and thus, mitigation policies can be evaluated iteratively until the corresponding \textit{Risk} goals are met. For example, if an EPS asset is compromised, there might be multiple defense mechanisms that could be enforced to mitigate the attack. However, the implementation of some of these mechanisms might result in significant impacts (e.g., uneconomic operation, partial grid disconnections, etc.) or affect other parts of the system due to its interdependent nature. The ability to evaluate, in real-time, the effectiveness of risk mitigation mechanisms provides significant benefits for CPS, aiming to balance security objectives and system performance.

%% file: sections/3-Threatmodeling.tex

\section{Threat Modeling for Cyber-Physical Systems}\label{s:CPSSecurity} 

The fundamental property of any adverse failure is an artifact of the semantics and capabilities of building CPS from a diverse, possibly infinite, set of ways. It is crucial to mitigate any adverse event in CPS, regardless of whether it is accidental or intentional. However, some distinctions need to be made between these two types. For example, there is a high probability that a natural adverse event (e.g., short circuit fault) can be detected by the process, considering a built-in fault detection scheme in the system. In contrast, an intentional fault (possibly caused by an attacker) could alter the results of the system in a congruous way, hence causing the event to go undetected. Traditionally, fault monitoring and detection approaches do not consider the implications that arise due to adversaries and their attack goals. Their aim is solely to recover from transient faults overlooking the actions which trigger this abnormal behavior. Without considering a threat model that includes malicious and motivated adversaries, as well as sophisticated attacks, defense detection schemes can be potentially evaded by attackers entirely, despite the redundancy already built into control processes. A fault can become an exploited vulnerability and the compromised component, if not sanitized properly, can pose a danger to the entire CPS.

The complex nature of CPS, and consequently CPES, urges the identification of attack vectors on both the cyber and the physical domains of the system. Adversaries are constantly improving, adapting, and modifying their attack patterns to evade security mechanisms. As a consequence, security researchers cannot passively await until an asset in the system is compromised to initiate remediation. 
To support the identification, anticipation, and mitigation of cyber-attacks in CPS, we develop a holistic threat model that incorporates the core components of MITRE's ATT\&CK for ICS methodology while providing an additional dimension for security investigations. Specifically, the presented threat modeling approach extends MITRE's methods since: 
\begin{itemize}[itemsep=0pt,parsep=0pt,leftmargin=*, wide=0pt]
    \item We incorporate an \textit{adversary model} to allow for more granular and explicit threat modeling analyses.
    \item  We rigorously define all aspects of potential cyber-attacks so that they can be implemented in CPS testbeds for security evaluations (e.g., evaluate defense mechanisms, mitigation strategy, detection schemes, etc.).
    \item We perform risk assessments considering the actual impact of cyber-attacks on the CPS and leveraging both the threat modeling and CPS framework resource mapping. Hence, every possible attacked CPS component is accounted towards the \textit{Risk} score calculation, aiding threat prioritization, and CPS security posture awareness. 
\end{itemize}

\begin{figure}[t!]
\centering
  \includegraphics[width=0.85\linewidth]{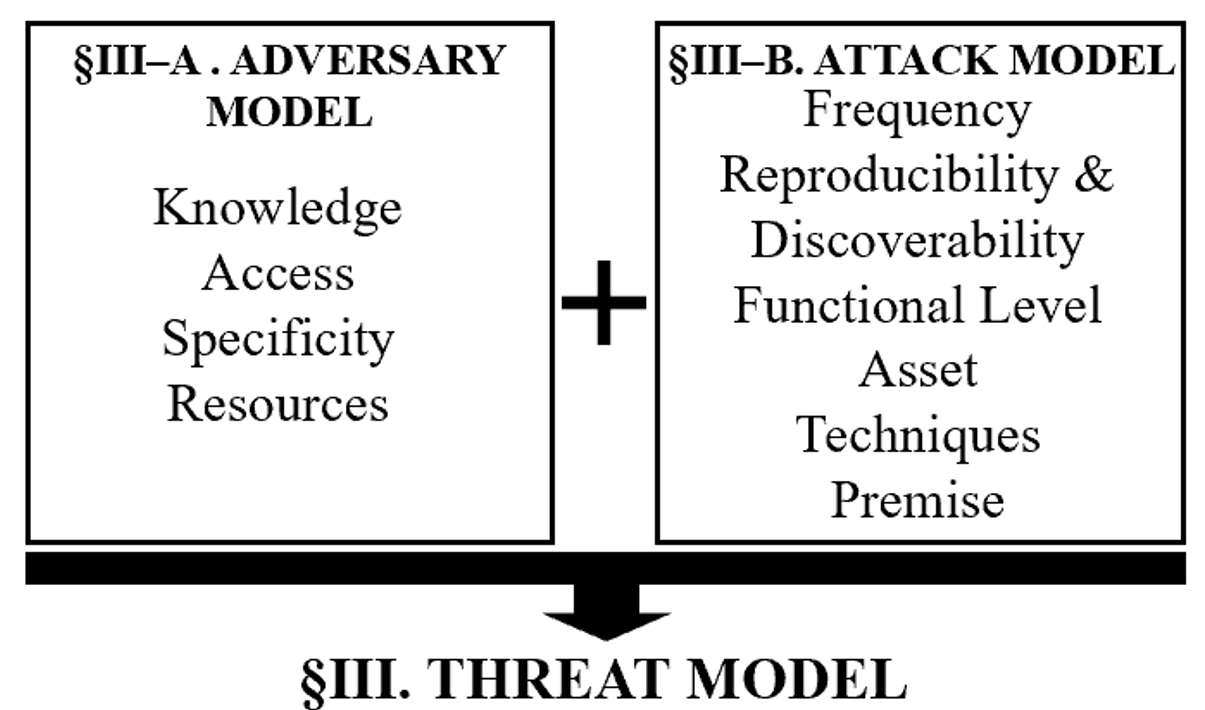}
  \caption{Adversary model and attack model components comprising the comprehensive threat model architecture.}
  \label{fig:threat}
\end{figure}

In the developed threat modeling methodology, we evaluate threats and prioritize them based on the degradation that they can potentially inflict on the CPS. Our threat model consists of two major components, the \textit{adversary model} and the \textit{attack model}, as illustrated in Fig. \ref{fig:threat}. To understand the security implications of threats targeting CPS, the adversary model needs to capture specific information involving the adversary's capabilities, intentions, and objectives. In addition, it is essential to model  attacks based on their specific  methodology, targeted system component, and system impact, as well as define rules that enable multi-layer and severity attack analyses. The adversary and attack models compose the threat score index factored in the threat risk calculation process presented in Section \ref{s:riskAssess}. For instance, the threat score of an attack performed by a stealthy and motivated adversary will be higher than the threat score of the same attack performed by an adversary with limited resources and oblivious knowledge about the system. Our versatile threat modeling approach  can support various types of malicious events and enable end-users to adjust the desired level of threat model granularity.

\subsection{Adversary Model}\label{ss:adversarymodel}
The capabilities of an attacker and the characteristics of the adversary model can be captured by factors such as resources, skills, knowledge of the system, access privileges, and opportunities (i.e., the means to carry out the attack and the number of failed attempts allowed) required to perform the attack. When it comes to system knowledge, distinction has to be made between \textit{white-box} attacks, where an adversary has \textit{complete} information, and \textit{black-box} or \textit{gray-box} attacks, where an adversary has \textit{limited} information about the system \cite{chakraborty2018adversarial}. In the gray-box threat model, the adversarial knowledge is limited to the target model, while in black-box attacks adversaries do not know the target model and can only query to generate adversarial samples \cite{REN2020346}. We port such classification in the context of CPES, in which attackers may have \textit{full, partial, or zero knowledge} of the system model and real-time power grid measurements. Existing work often assumes that adversaries have perfect knowledge of the system model, i.e., the information needed to create the measurement matrix (Jacobian) of the power system that depends on the network topology, the parameters of power lines, and the location of RTUs and PMUs \cite{liang2016review}. However, in realistic attack scenarios, adversaries have limited knowledge of the system due to the dispersed, interconnected, and complex nature of the power grid, the restricted access to CPES control and monitoring functions, and errors in the data collection process \cite{pan2017data, 7741928, rahman2012false}. 

Our adversary model takes into consideration the presented distinctions and defines a hierarchy of the available information to the attackers in order to characterize their knowledge capabilities. At the lowest level of the system, knowledge hierarchy is an adversary that has no information about the system model. At the highest level is an adversary that knows the model characteristics, the algorithmic details, and all the grid measurements. 
In order for the adversary, however, to acquire system measurements or perform reconnaissance and monitoring, he/she should have -- to some extent -- access to the system. As such, our model delineates this access as the \textit{accessibility level} an adversary needs to have in the target CPS. 

In the MITRE ATT\&CK for ICS framework, the term `\textit{attacker tactics}' covers the attackers' access level with their corresponding intentions and objectives. In our modeling approach, however,  the term is captured in two sub-categories, \textit{access} and \textit{specificity}, to allow for a more elaborate adversary classification. The access category defines the degree to which an attacker can interact with a system asset, while the adversarial specificity encapsulates the objectives and goals of the adversary who executes the attack on the CPS. Adversarial objectives are  broadly lumped under \textit{targeted}  and \textit{non-targeted}. In the case of targeted objectives, an adversary's intention is to execute an attack which can result in a specific target output (e.g., the miscalculation of EPS system states, due to topology modifications of wind integrated resources \cite{8214282}, or FDIAs \cite{7313024}). On the other hand, non-targeted attacks can generically maximize malformed outputs of CPS algorithms (with respect to the ground truth) affecting the operational reliability of the system. Finally, our adversary model also captures the adversary \textit{resources}, differentiating between attackers with limited resources and attackers with a variety of intellectual and physical assets at their disposal.

\textit{\underline{Adversary Model Formulation}}: Our attacker model is decomposed into four dimensions: adversarial knowledge, resources, access, and specificity:
\begin{enumerate}[leftmargin=*, labelwidth=0pt, labelindent=0pt, wide=0pt]
    \item \textit{Adversary Knowledge}
    \begin{enumerate}[leftmargin=*, labelwidth=0.1mm, labelindent=0pt, wide=0pt]
        \item \textit{Strong-knowledge adversary}: White-box attacks assume an adversary with \textit{full knowledge} of the system model, parameters, and state vectors. 
        \item \textit{Limited-knowledge adversary}: Gray-box attacks assume an adversary with \textit{some knowledge} of the system's internals with a partial understanding of the network and system model. 
        \item \textit{Oblivious-knowledge adversary}: Black-box attacks assume an adversary with \textit{zero knowledge} about the details of the system and can only estimate the system outputs using confidence scores. In such scenarios, the attacker does not have knowledge in regard to the system model. 
    \end{enumerate}

    \item  \textit{Adversary Access}
    
    \begin{enumerate}[leftmargin=*, labelwidth=0.1mm, labelindent=0pt, wide=0pt]
        \item \textit{Possession}: This type of attack requires the adversary to have physical access to the attacked component (e.g., IED, solar inverter, transformer, etc.) operating either in the digital or analog domain. The access could involve chassis intrusions (e.g., microprobing, memory flashing, circuit bending, etc.), or interface access to the device (e.g., side-channel analysis, power analysis, protocol decode, etc.).
        \item \textit{Non-possession}: In this type of attacks, the adversary cannot physically manipulate the asset under attack. Attacks can be performed leveraging proximity access (e.g., GPS spoofing, side-channel analysis), or by exploiting network interfaces (e.g., replay attacks, rollback attacks, etc.).
    \end{enumerate}

    \item \textit{Adversarial Specificity}
    
    \begin{enumerate}[leftmargin=*, labelwidth=0.1mm, labelindent=0pt, wide =0pt]
        \item \textit{Targeted} attacks occur in multi-class identification, control, and monitoring -based scenarios and misclassify CPS algorithms and operations to a specific malicious result category $x_j \in \mathcal{X}$ from all possible results $\mathcal{X}$. The adversary goal is to maximize the probability of the \textit{targeted} class, i.e., maximize $\mathcal{P}(x_j)$.
        \item \textit{Non-targeted} attacks are similar to \textit{targeted} attacks in terms of misclassification objective, however, the selection of category $x_j$ is relaxed to any arbitrary output category except the correct one $x_i$. 
    \end{enumerate}

    \item \textit{Adversarial Resources}
    
    \begin{enumerate}[leftmargin=*, labelwidth=0.1mm, labelindent=0pt, wide=0pt]
        \item \textit{Class-I} attackers that, despite their adversarial motivation, do not have the financial resources, equipment support, or access privileges, to successfully realize any attack without being detected.
        \item \textit{Class-II} attackers that can be funded individuals, organizations, or nation-state actors with large budgets and substantial access privileges, skills, and tools capable of realizing sophisticated attacks.
    \end{enumerate}

\end{enumerate}

\subsection{Attack Model}\label{ss:attackmodel}

The second part of the proposed threat modeling method focuses on the specific characteristics of malicious attacks (e.g., frequency, reproducibility), the targeted CPS components, and the process aiming to achieve the system compromise. The attack model improves MITRE's taxonomy, which includes concepts such as the attack levels, assets, and techniques, by incorporating supplemental dimensions necessary for holistic security investigations. For instance, particular attention is drawn on aspects like the attack frequency, reproducibility/discoverability, along with the premise of the compromise. The aforementioned features enable the comprehensive characterization of the attacks elucidating all their underlying elements, and as a result, they assist in performing threat and system impact evaluations for CPS environments.

\begin{figure}[t]
\centering
\includegraphics[width = 0.95\linewidth]{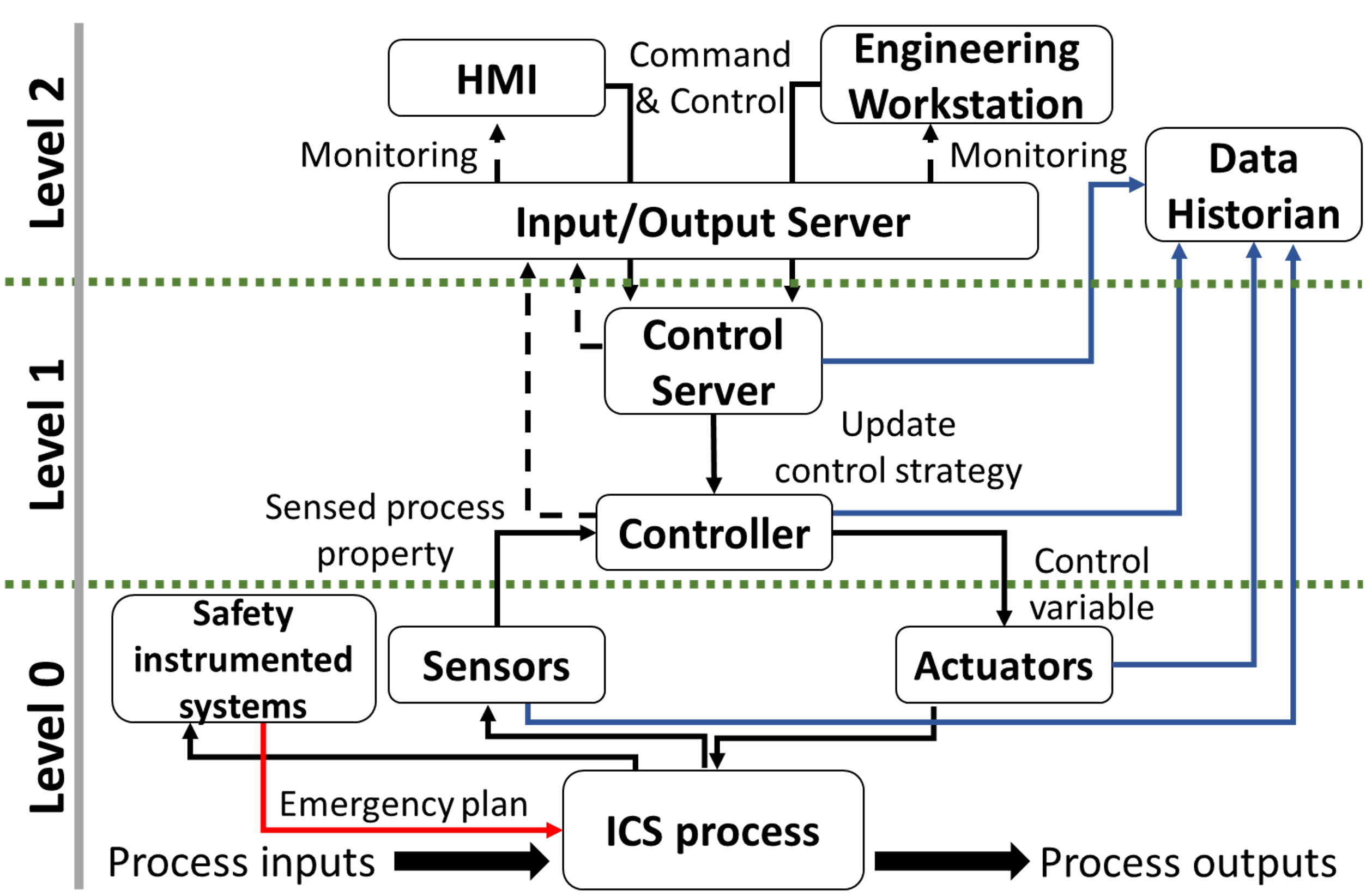}
\caption{CPS functional attack levels and assets overview.}
\label{fig:assets}
\end{figure}

The presented attack model accounts for the CPS structure and interconnections. Given that the same adversarial objective can be achieved following different attack paths, propagation scenarios with diverse attack entry points should be investigated. These attack paths can be initiated from process control devices such as sensors or actuators and propagate to supervisory and control equipment like HMIs. In particular, the attack model considers the attack \textit{frequency}, i.e.,  the number of compromises required to achieve a particular adversarial objective, and the attack \textit{reproducibility and discoverability}. The aspects of reproducibility and discoverability are crucial for CPS risk evaluations. This is attributed to the fact that even catastrophic attacks might not pose any actual danger for the CPS if materializing them is nearly impossible, or they can be easily discovered during their initial stages. The attack \textit{functional level}, attacked \textit{asset}, and attack  \textit{techniques} notions correspond to the definitions introduced in MITRE \cite{MITRE}. The only difference is that the selected attack techniques in our methodology represent some of the most common use cases encountered specifically in CPES. An overview of the CPS functional levels along with the corresponding attacked assets are illustrated in Fig. \ref{fig:assets}. Also, we consider the attack \textit{premise} which indicates whether the attack is targeting the physical or cyber domain of a CPS, trailing the attack path origin and its expected impact. The formulated attack model with the additional aspects of attack frequency, attack reproducibility and discoverability, and attack premise allows to fine-tune each attack case study's model, and overall compose a well-defined CPS threat modeling approach.

\textit{\underline{Attack Model Formulation}}: Our attack model is decomposed into six dimensions: attack frequency, attack reproducibility and  discoverability, attack level, attacked asset, attack techniques, and attack premise.
\begin{enumerate}[leftmargin=*, labelwidth=0pt, labelindent=0pt, wide=0pt]
    
    \item \textit{Attack Frequency}
    \begin{enumerate}[leftmargin=*, labelwidth=0.1mm, labelindent=0pt, wide=0pt]
        \item \textit{Iterative attacks}: attacks that need multiple iterations to achieve the desired malicious output. 
        \item \textit{Non-Iterative attacks}: attacks that only need to be realized once to achieve the desired malicious output.
    \end{enumerate}
    
    \item \textit{Attack Reproducibility and Discoverability}
    \begin{enumerate}[leftmargin=*, labelwidth=0.1mm, labelindent=0pt, wide=0pt]
        \item \textit{One-time attacks}: attacks that can only be realized once since they are detected after the first attempt. 
        \item \textit{Multiple-times attacks}: attacks that can be reproduced multiple-times before they are identified and detected.
    \end{enumerate}

    \item \textit{Attack Functional Level}
    \begin{enumerate}[leftmargin=*, labelwidth=0.1mm, labelindent=0pt, wide=0pt]

        \item \textit{Level 0}: attacks that target CPS processes and their corresponding operational equipment (e.g., sensors, actuators, etc.).

        \item \textit{Level 1}: attacks that target the industrial control network (e.g., PLCs, system controllers, RTUs, etc.) and aim to stealthily manipulate functions that control CPS processes. 
        
        \item \textit{Level 2}: attacks that target the SCADA, and monitoring devices (e.g., HMIs, engineering workstations, data historians, etc.) on the network level (i.e., LAN) overseeing CPS processes.

    \end{enumerate}
    
    \item \textit{Attacked Asset}

    \begin{enumerate}[leftmargin=*, labelwidth=0.1mm, labelindent=0pt, wide=0pt]
        \item \textit{Field controllers:} Such assets are low-level embedded devices (e.g., RTUs, PLCs, IEDs) that enable the control of CPS processes. They typically possess limited computation capabilities and they are in charge of coordinating industrial processes (e.g., generator governors, manufacturing process controllers, etc.).
        
        \item \textit{Control servers:} These devices cover the functionality of both programmable controllers (e.g., PLCs) as well as communication servers (e.g., SCADA master terminal units (MTU), distributed control servers, etc.). Thus, apart from interfacing with low-level CPS devices (e.g., sensors, actuators), they can also support software-based services in industrial environments.  
        
        \item \textit{Safety instrumented systems (SIS):} These systems (e.g., protective relays, recloser controllers) are designed to perform automated remediation actions if an abnormal system behavior is detected (e.g., short-circuit, fault, etc.). The goal of protection systems is to keep the industrial CPS plant online, while avoiding hazard conditions. 
        
        \item \textit{Engineering workstations:} These units are usually powerful and reliable computing configurations used for the monitoring and control of CPS,  processes, and equipment. They are often accompanied by hardware components and software packages that enable CPS supervision.
        
        \item \textit{Data historians:} Such elements are databases used to keep records and store process data. This information is stored in a time-series format that enables the examination, display, and statistical analysis of process control information.

        \item \textit{Human-machine interfaces (HMIs):} A graphical user interface that enables users to monitor system operations, diagnose malfunctioning system behavior, and initiate control and mitigation actions. HMIs can vary between vendors supporting different capabilities, graphical representations, and control interfaces (e.g., web-based, LAN-based, etc.). Additionally, different user groups can have access to different HMIs according to the systems they are monitoring and their clearance level for managing the CPS.

        \item \textit{Input/output (I/O) servers:} Such servers constitute the connecting link between system applications and the field devices which coordinate the ICS equipment under the control subsystems directions. I/O and data acquisition servers (DAS) operate as buffers since they can convert low-level control system data to packets, and forward them to the supervision locations (e.g., HMIs, engineering workstations). Additionally, they serve as intermediate translation units as they collect information from field devices (utilizing diverse communication technologies) and translate them to the predefined formats expected by system applications.

    \end{enumerate}
    
    \item \textit{Attack Techniques}
    \begin{enumerate}[leftmargin=*, labelwidth=0.1mm, labelindent=0pt, wide=0pt]
        \item \textit{Modify control logic:}
        In such attacks, adversaries can cause the CPS to operate abnormally by modifying the code running on the system's control devices (e.g., PLC, RTU, IED). These system devices are orchestrating physical processes via actuators and other field equipment.
 
        \item \textit{Wireless compromise:} 
        In these attack scenarios, adversaries can gain unauthorized remote access to the CPS network by exploiting: the vulnerabilities of devices with wireless connectivity, insecure wireless communication protocols, and/or network connections leaking sensitive information.  

        \item \textit{Engineering workstation compromise:}
        In such attack setups adversaries, after granted access to a CPS engineering workstation, can cause system malfunctions via compromising CPS configurations controlled by engineering workstations, e.g.,  security systems, process controls, ICT infrastructure, etc.
        
        \item \textit{Denial-of-service (DoS):}
        Malicious adversaries performing DoS attacks can compromise a CPS asset by inhibiting its nominal functionality rendering it unresponsive. For instance, overflowing a device with artificial data, blocking its inbound or outbound communications, or even suspending/disrupting its operation can impact time-critical CPS.
        
        \item \textit{Man-in-the-middle (MitM):}
        During MitM attacks adversaries can maliciously intercept, modify, delay, block, and/or inject data streams exchanged between CPS asset communications. Depending on the adversary access level on the CPS networks, numerous attacks (e.g., modify or inject control commands, delay alarm messages, etc.) can be planted affecting CPS operations. 
        
        \item \textit{Spoof reporting messages:}
        Adversaries performing this type of attack can broadcast malicious modified system messages. The attack goal is to either impact CPS operations by limiting the situational awareness (e.g., suppressing critical alarm messages), or misreport information (e.g., sensor measurements), thus, driving systems to unstable and potentially irreversible states.

        \item \textit{Module firmware:}
        In module firmware attack cases,  adversaries can upload maliciously modified code to embedded devices of CPS (e.g., PLCs, smart inverters, etc.). These actions can affect devices operation via modification of their control objectives, and/or insertion of backdoor features (e.g., remote access, exploit system logs, etc.) allowing them to stealthily manipulate CPS assets.

        \item \textit{Rootkits:}
        In this type of attack, adversaries employ rootkits, typically planted in the OS of devices, to disguise malicious software, services, files, network connections ports, etc. Rootkits provide attackers with user or even root-level privileges while hiding their presence from CPS defense mechanisms. 

    \end{enumerate}
    
    \item \textit{Attack Premise}
    \begin{enumerate}[leftmargin=*, labelwidth=0.1mm, labelindent=0pt, wide=0pt]
        \item \textit{Attacks targeting the cyber domain}
            \begin{enumerate}[label=\roman*), leftmargin=0pt, labelwidth=0.1mm, labelindent=0pt, wide=0pt]
                \item \textit{Communications and protocols:} refers to attacks targeting the in-transit CPS data, i.e., exchanged communications data including remote access credentials, measurements, system reports and warnings, etc., with the objective to get unauthorized access (data espionage) or insert malicious modifications (data alteration).
                \item \textit{Asset control commands:} includes attacks targeting the CPS data integrity, i.e., mask counterfeit system data as genuine and unmodified, and trustworthiness (e.g., impersonate authorized user groups and issue, access,  or modify control commands).
                \item \textit{Data storage:} accounts for attacks targeting the accuracy and non-repudiation of CPS data (e.g., logs and historical records of all the performed tasks such as asset setpoint modifications, user sign-ins and action histories, inbound/outbound connections and traffic, etc.).
            \end{enumerate}
            
        \item \textit{Attacks targeting the physical domain}
            \begin{enumerate}[label=\roman*), leftmargin=0pt, labelwidth=0.1mm, labelindent=0pt, wide=0pt]
            \item \textit{Invasive:} attacks that require physical access to the CPS asset (e.g., PLC hardware including micro-controller, memory, integrated circuit -- IC, etc.) in order to manipulate it (e.g., desoldering, depackaging) \cite{konstantinou2016case}. These attacks are time-consuming and require specialized equipment, however, they are difficult to detect.
            \item \textit{Non-Invasive:} attacks that do not require any physical tampering of the ICs residing on the CPS assets, and performing them multiple times can be achieved with minimum effort. No traces are left after the attack is performed rendering them the most difficult type of attacks to detect. Common examples of non-invasive attacks include power analysis attacks, timing attacks, electromagnetic emission attacks, brute force attacks through physical means, hall sensor spoofing, etc. \cite{barua2020hall}.
            \item \textit{Semi-Invasive:} attacks that are a trade-off between invasive and non-invasive attacks, given that they are not as difficult to perform as invasive attacks and can be easily performed multiple times similar to non-invasive ones \cite{sergei}. Common examples of semi-invasive attacks include fault injection, laser scanning,  ultra-violet radiation, or control process tampering \cite{aurora}.
        \end{enumerate}
    \end{enumerate}
\end{enumerate}

\subsection{Risk Assessment Methodology} \label{s:riskAssess}

Risk assessment is a fundamental process in every cybersecurity analysis study. Its importance is further accentuated in the context of mission-critical CPS where operational disruptions can have disastrous impacts. Existing efforts often port IT risk assessment methodologies into operational technology (OT) security evaluations, and consequently, fail to holistically capture CPS constraints and objective \cite{8861487}. 
Some key differences between IT and OT security revolve around the risks associated with loss of operation, asset availability, communication latency, architectural differences, and contingency management strategies \cite{7427514, 8938806, 8586807}.

Qualitative assessments for cybersecurity risks require substantial system knowledge of the CPS structure and experience from the organizations and groups conducting the analysis \cite{rodriguez2015qualitative, 8861487}. On the other hand, quantitative studies calculate exact risk scores aiding the prioritization and mitigation procedures \cite{huang2018assessing, cardenas2011attacks}. Other works employ simulation-assisted investigations in order to evaluate the corresponding impact of cyber-attacks \cite{shi2011security}. Moreover, researchers have also considered dynamically adapting risk assessment models factoring the system and attack impact evolution for the risk score calculations \cite{7366430}. Recent works have proposed combinations of different risk methods harnessing the advantages of more than one strategy and providing more realistic evaluations \cite{7360925, 6846672, ZIO2018176}. These combined approaches are motivated by the fact that in CPS we can have the same impact on system operation using different attack paths. Thus, although the system impact remains the same, the risk scores of these attacks would substantially differ. For example, such scenarios, i.e., following different attack procedures to achieve the same adversarial objective, would be difficult to capture using a qualitative-only risk assessment method.

In this paper,  we utilize a hybrid risk assessment method bridging the advantages of both quantitative and qualitative methods. The hybrid approach adapts to dynamic system operation and adjusts risk scores based on the current system state. Specifically, we assess qualitatively the impact of attacks. To calculate the corresponding attack damage, however, we quantitatively prioritize CPS objectives. The threat probability is also assessed quantitatively to weigh the attack damage and model the risk. It is important to note that both the objective priority as well as the threat probabilities can change during the real-time system operation. 

Such scenarios can be accommodated by our risk model. For instance, the loss of power at a residential area has the same outage impact, regardless if this is due to a natural disaster (e.g., hurricane, thunderstorm), a malicious attack, or EPS electrical faults (e.g., short-circuits). However, the threat probabilities and CPS objective priorities for the three aforementioned scenarios differ significantly. ``People health and personnel safety'' objective during a natural disaster has much higher priority compared to a power outage due to an EPS fault. The latter event, being not a life-threatening situation, would have a higher ``uninterrupted operation and service provision'' priority.

Furthermore, the presented threat modeling methodology of Section \ref{ss:adversarymodel} and \ref{ss:attackmodel},  which enables precise adversary and attack descriptions,  serves as the backbone of our risk assessment method. Specifically, the definitive granularity of threat characterizations, not only exposes the vulnerable system assets but also can infer which CPS objective will be affected the most. The CPS objective is critical for the attack impact evaluations, while vulnerable assets demonstrate the feasibility of an attack. Thus, CPS attack risk scores can be calculated and their prioritization can be performed based on the affected CPS objective. 
The threat \textit{Risk} is defined as:
\begin{equation}
\label{eq:risk}
    Risk = Threat \ Probability \times Damage
\end{equation}
\noindent The \textit{Threat Probability} portion of the risk formula accounts for the threat details, i.e., the adversary and attack models discussed previously, in addition to how likely it is for the investigated threat to materialize in the specific system context. The second component of Eq. \eqref{eq:risk} includes the \textit{Damage}, which assesses the corresponding impact inflicted on the system. The \textit{Damage} is defined as follows:
\begin{eqnarray}
\label{eq:damage}
Damage = \sum_{k=1}^{n}Objective \ Priority \times \nonumber \\ Attack \ Impact
\end{eqnarray}
\noindent where the \textit{Objective Priority} and the \textit{Attack Impact} are used to address the consequences of the attack in the context of the specific CPS objectives. The \textit{Attack Impact} is evaluated qualitatively using a number from $1$ to $3$, reflecting \texttt{Low}, \texttt{Medium}, or \texttt{High} impact, respectively. In addition, for every CPS, the objectives are ranked in order of importance. We utilize four ($n=4$) main objective categories: \textit{i)} people health and personnel safety, \textit{ii)} uninterrupted operation and service provision, \textit{iii)} organization financial profit, and \textit{iv)} equipment damage and legal punishment. Numbers from $1$ to $4$ are used for the objective priorities; $1$ indicates the least significant goal while $4$ stands for the most critical objective. 

In Table \ref{table:damage}, we demonstrate a damage calculation example where we provide a subjective priority ranking as well as the attack impact values. Using Eq. (\ref{eq:damage}), the total potential damage score can be calculated as $\sum(4+9+6+2) = 21$.  Given a specific \textit{Threat Probability} value, we can then assess the total \textit{Risk} for the examined attack scenario. Overall, the presented application-aware risk assessment procedure is taking into consideration all the underlying components of sophisticated and multi-layer threats targeting complex CPS. In addition, it provides a universal method to assess attack risk, regardless of the particular CPS architecture or the corresponding operational objectives. Based on the assessment results, administrative authorities can prioritize which assets need immediate attention and which threats pose the highest risk (if vulnerabilities of the in-scope CPS assets are exploited).

\begin{table}[t]
\setlength{\tabcolsep}{1.2pt}
\centering
\caption{Example of attack damage calculation. }
\label{table:damage}
\renewcommand{\tabularxcolumn}[1]{m{#1}}

\begin{tabularx}{\linewidth} { 
  || >{\hsize=0.85\hsize\linewidth=\hsize\raggedright\arraybackslash}X 
  | >{\hsize=.35\hsize\linewidth=\hsize\centering\arraybackslash}X
  | >{\hsize=.55\hsize\linewidth=\hsize\centering\arraybackslash}X
  | >{\hsize=.25\hsize\linewidth=\hsize\centering\arraybackslash}X || }
  
 \hline
 \hline
 \textbf{Objective} & {\textbf{Priority}} & \textbf{Attack Impact}  & \textbf{Score} \\ 
 \hline
 {People health and personnel safety} & {4}  & \texttt{Low (1)} & \textbf{{4}}  \\
\hline
{Uninterrupted operation and service provision} & {3}  & \texttt{High (3)} & \textbf{{9}}  \\
\hline
{Organization financial profit} & {2}  & \texttt{High (3)} & \textbf{{6}}  \\
\hline
{Equipment damage and legal punishment} & {1}  & \texttt{Medium (2)} & \textbf{{2}}  \\
\hline
\hline
\end{tabularx}
\end{table}

%% file: sections/4-Framework.tex
4\section{Cyber-Physical System Framework: Modeling, Resources, and Metrics for CPES Studies}\label{s:cps_framework}

\begin{figure*}[t!]
\centering
  \noindent\makebox[\textwidth]{\includegraphics[width=15cm]{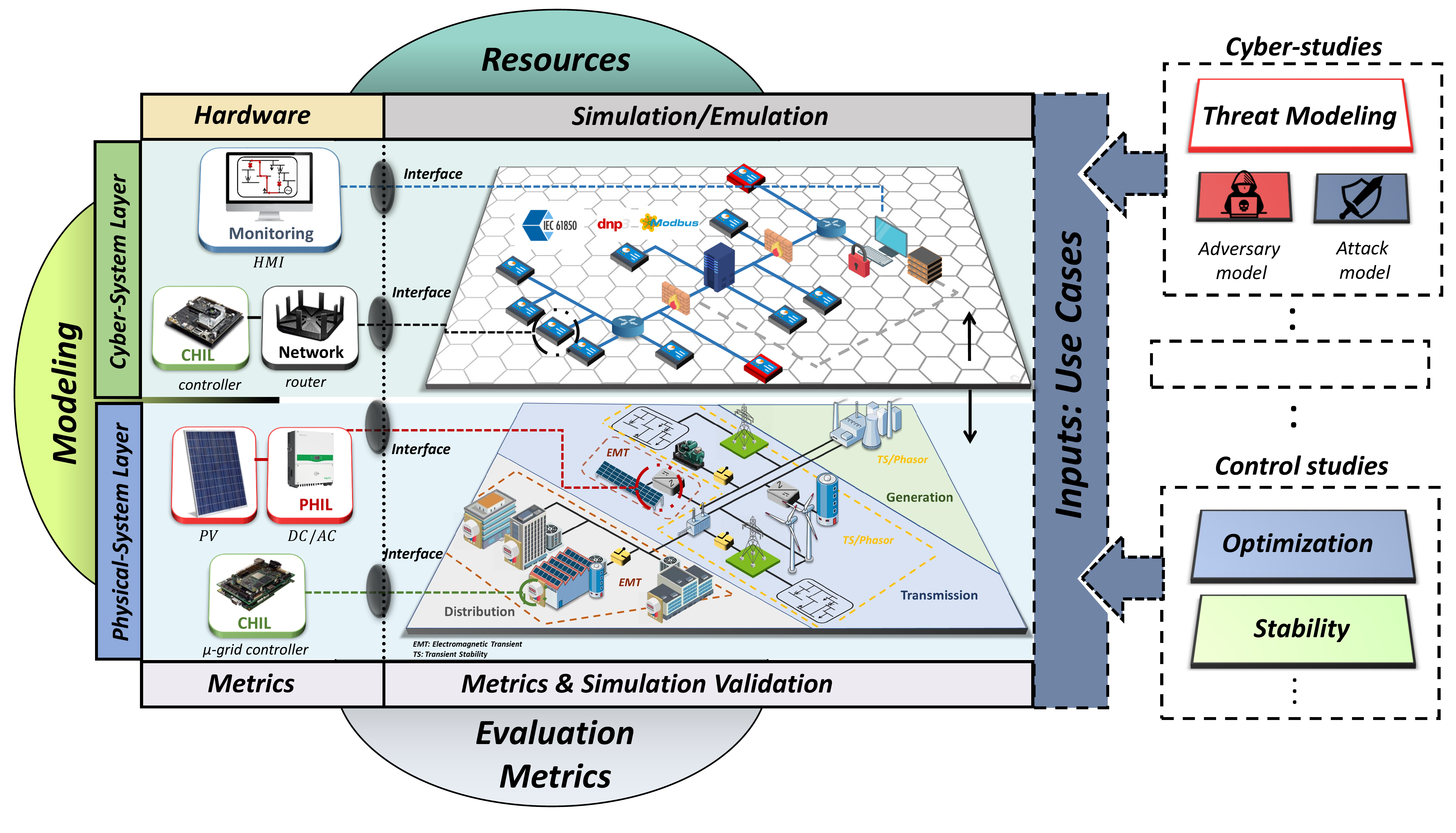}}
  \caption{Cyber-physical system (CPS) framework: the cyber-system and physical-system layers are presented with their respective factors, i.e., modeling, resources, and evaluation metrics, needed for conducting cyber-physical studies. Different use cases requirements can be adjusted to perform CPS related investigations.}
  \label{fig:cps_framework_overall}
\end{figure*}

The framework depicted in Fig. \ref{fig:cps_framework_overall} shows the different domains in which the proposed CPS framework is divided. The main objective of the framework is to provide a clear understanding of all the underlying concepts and components being considered in CPS investigations. Specifically, the presented conceptual framework is intended to assist researchers in identifying the models, resources, and metrics required to perform reliable CPES studies. Based on the study objectives, the framework can be treated as a `\textit{how-to guide}' towards the implementation of use cases and the development of CPES testbeds. This section, first, describes the \textit{cyber-system}  and \textit{physical-system} layers that need to be considered for the CPES representation. Then, we describe the different factors that need to be taken into account when performing CPES studies, i.e., the modeling techniques, resources, and metrics.

\textit{Physical-System Layer}: NIST's definition for CPS establishes that the physical-system layer of a CPS is composed of hardware and software components embedded into the system environment. These components have the capability of interacting with other physical-layer units through physical means, i.e., via sensors and actuators, or through the cyber-system layer using standard communication protocols. Some sectors where CPS can be extensively found are smart manufacturing \cite{smartmanufact_wan}, healthcare \cite{healthcare_zhang}, robotics \cite{robotics_chang}, transportation \cite{vehicles_naufal}, and EPS \cite{johnson2019power}. In this paper, the developed framework focuses on the EPS sector, i.e., the models, resources, and metrics used in the physical-system layer are based on elements encountered in the generation, transmission, and distribution systems that comprise CPES. Example components within the physical-system layer of CPES are PV panels, Li-ion BESS, wind energy systems, power converters, generators, voltage regulators, transformers, and T\&D lines.  
    
\textit{Cyber-System Layer}: The cyber-system layer of a CPS is composed of the ICT structures deployed in the system. It encompasses communication and networking components such as hubs, modems, routers, switches, cables, connectors, databases, and wired and/or wireless network interface cards (NIC) \cite{Vellaithurai2015, osti_1526728}. These components allow the interconnection of multiple computing devices using common communication protocols over digital links with the purpose of sharing, storing, and processing resources and data located across networking nodes. In this paper, our developed framework focuses on elements that make up communication networks in CPES, i.e., the models, resources, and metrics used in the cyber-system layer are related to components such as smart-meters, PMUs, EPS-related communication protocols (e.g., DNP3, IEC61850, IEEE 37.118, etc.), and other networking devices that support communication in EPS operations.

\subsection{Modeling}
Models able to represent systems by describing and explaining phenomena that cannot be experienced directly \cite{sciencemodel}. Such models are built from mathematical equations and/or data that are used to explain and predict the behavior and response of complex systems. Specifically for CPES, researchers focus on creating models capable of replicating the behavior of the components that comprise the cyber-system and physical-system layers of EPS, e.g., models for components such as PV systems, wind energy systems, ESS, transformers, transmission lines, distribution lines, smart meters, PMUs, routers, switches, etc. In this part, we describe the different modeling techniques used to model both the cyber and physical layers of CPES.  

\subsubsection{Physical-System Layer}
The design and modeling of the physical-system involve areas such as hardware design, hardware/component sizing, connection routing, and overall system testing. All components in this layer must be categorized based on their respective temporal and spatial requirements along with their intrinsic physical characteristics. In EPS, some of these characteristics and requirements are related to rated voltage, current, and power values, location of the generation and load resources, and physical characteristics of the lines (i.e., resistance, reactance, capacitance, and length). 
These features are utilized in developing models that represent the physical devices in the system. The objective is to capture and simulate system behavior so that a \textit{digital twin} of the real system can be implemented. This `virtualization' capability provides a significant advantage by allowing the analysis and study of different types of scenarios that can arise during the operation of the CPS. We can analyze and track physical processes, replicate potential harmful operating conditions or scenarios, and accelerate the testing of software and hardware components. More specifically, for EPS modeling, the current state-of-the-art simulation technology is based on electromagnetic transient (EMT) and transient stability (TS) simulation techniques \cite{marandi, new_emtandts1, new_emtandts2}.

\begin{enumerate}[label=\alph*), leftmargin=*, labelwidth=0.1mm, labelindent=0pt, wide=0pt]
    \item \textit{Electromagnetic transient (EMT)}: EMT simulation is a technique used to precisely reproduce the system response to fast dynamic events and system perturbations, that occur in the range of tens of microseconds or lower,  caused by fast switching electromagnetic fields or loading events. Due to requirements, such as the unsymmetrical and instantaneous modeling of the signals and values that characterize the behavior of the system, nonlinear ordinary differential equations (ODE) are used to represent the system behavior in the EMT simulation environment. This detailed modeling provides improved accuracy, compared to TS-type simulations, when capturing the system behavior and response to fast transient events. However, it requires high computational resources for the simulation of systems with a large number of components. Typical applications where EMT studies are used include the simulation of power electronic devices, unbalanced distribution systems, and the impact evaluation of DER integration into modern power networks.
    
    \item \textit{Transient Stability (TS) Simulation}: TS simulation is a technique used to capture the slow dynamic events, i.e., events in the range of tens of milliseconds and higher, that occur in power systems. These events are related to the voltage stability, rotor angle stability, and frequency stability phenomena. In TS, the EPS is represented by nonlinear differential algebraic equations (DAE). These equations are used to solve the system states assuming that the fundamental power frequency (e.g., 50 or 60 Hz) is maintained throughout the system. Commonly, TS-type simulations are used for studies related to the analysis, planning, operation, and control of EPS elements with large time-steps, i.e., in the milliseconds range. Given that large time-steps and positive-sequence phasor-domain simulations are used in TS-type simulations, they allow users to simulate large-scale T\&D networks while requiring significantly less computational resources when compared to EMT-type simulations \cite{marandi}. 

     \item \textit{Hybrid-Simulation (TS+EMT)}: Hybrid-simulation models make use of both EMT and TS simulation tools to leverage the benefits of two or more simulation environments, hence allowing even more comprehensive and accurate simulation studies. Some examples of these types of simulations are found in recent literature \cite{huang2017open, tdcosim2, tdcosim3}. Integrated T\&D co-simulations are a major field of study enabling the use of hybrid-simulation environments. Such environments can provide ways of simulating in detail, for example, power electronic converters interfaced with large-scale power networks. T\&D co-simulation also provides an effective way of studying the diverse impacts that anomalous events (e.g., unintentional faults or intentional malicious attacks) may have locally and globally in the overall physical-system layer of the CPES. 
\end{enumerate}

\subsubsection{Cyber-System Layer}
The design and modeling of the cyber-system layer involve communication network modeling, communication protocol implementation, design of information systems, and data storage processing. To model this layer, researchers must have a deep understanding of the communication infrastructure that needs to be replicated using the respective cyber-system layer models. Some of the characteristics that need to be taken into consideration for modeling the communication infrastructure are: \textit{i)} the topology of the communication network, \textit{ii)} physical characteristics (cable lengths, physical components, delays, etc.), and \textit{iii)} Quality-of-Service (QoS), among others \cite{osti_1526728}. In a real-world CPS (e.g., cellular networks, military zones, or SCADA systems), multiple and diverse networking and computing components comprise the cyber layer. This hinders the implementation of tests and studies designed to evaluate the operation and performance of the actual network or to simply conduct any other CPS-related investigation.

As discussed in Section \ref{s:related}, carrying out evaluation type of studies in real systems can be dangerous for human safety, excessively costly, and may cause interruption or degradation of the network performance and the QoS (as perceived by the users). To address these issues, models can be used to simulate or emulate the behavior and performance of the cyber-system layer under different scenarios. In essence, simulation allows replicating the behavior of cyber-system layer components, while emulation duplicates the behavior of these components and allows them to be used alongside real devices. The simulation and emulation of the cyber-system layer are fundamental tools for understanding and studying topics related to complex network deployment, networking architectures, communication protocol features, and deployment of new services. 

\begin{figure}[t]
\centering
\includegraphics[width = 0.48\textwidth]{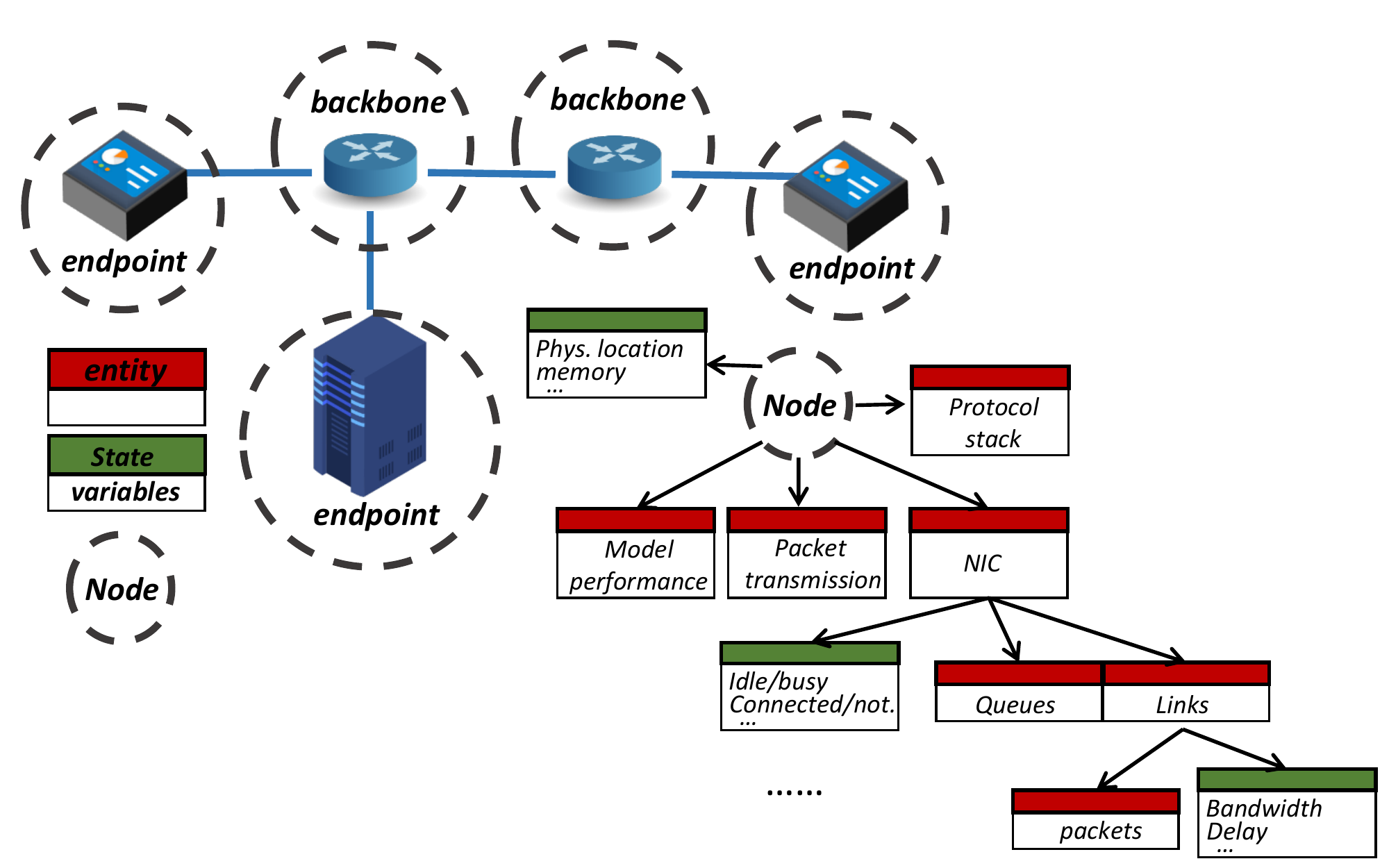}
\caption{\label{fig:network_simulation_modeling}Conceptual diagram of the modeling and simulation process of communication networks.}
\end{figure}

The simulation/emulation modeling process is often instantiated by identifying all the network components, commonly referred to as communication network entities. These entities, i.e., nodes and links configurations, constitute the network topology. Fig. \ref{fig:network_simulation_modeling} depicts a conceptual illustration of how the modeling process is performed in a communication network simulation. 
As seen in Fig. \ref{fig:network_simulation_modeling}, in a network simulator/emulator architecture, a node is a key entity that represents any computing device connected to the overarching network. This abstraction encapsulates all the possible representations of computing devices that may exist in a network setup. Some of these computing devices can refer to routers, switches, and hubs which embody the backbone of the network, while computers, RTUs, PLCs, meters, and servers constitute the endpoints of the network. A node is primarily characterized by its packet transmission entity attribute. In this packet transmission attribute, endpoints delineate the source or destination of the data packets while all backbone elements perform the forwarding tasks related to these packets. Other parameters, known as state variables, differentiate the behavior for each one of the modeled nodes. Some of these parameters are memory consumption, physical location, battery power, and CPU utilization. Additionally, other simulation entities, such as NIC, help to identify nodes in the network. These interfaces also have individual state variables that represent their state (i.e., idle or busy, and installed or not installed) while being in charge of transmitting, receiving, and processing the packets exchanged with other network nodes. 

Similar to the nodes, interfaces include other entities, such as queues and links, which represent realistic packet processing scenarios. Queues are modeled as buffers in the outgoing and incoming packet processes. Links are modeled as the connections between the two nodes communicating via the corresponding interfaces (i.e., communication medium). More specifically, links are modeled by defining communication parameters such as the available bandwidth, propagation delays, jitter, and pre-defined packet loss rates. Furthermore, packets are modeled as entities that contain the data exchanged between nodes in the network. For each node in the network, entities that represent the protocol stack must also be defined, while the packet sizes are determined by the corresponding communication protocol (e.g., TCP, UDP, etc.). 

A protocol entity is responsible for managing the outgoing and incoming packets by adding and removing packet headers. Protocol modeling is also a key process. It covers the specific steps required to accurately emulate the behavior of the protocol stack. In this process, models are developed to capture elements and properties from the network access layer, internet layer, transport layer, and application layer. Finally, models for performance evaluations, which do not represent real elements in the network, are also defined as additional entities that facilitate the implementation and evaluation of the network. Some representative examples of such entities are logging and helper utilities which can aid the network evaluation process \cite{obaidat2015modeling}.

\subsection{Resources}

The `\textit{resources}' represents the different hardware and software systems that form, and can be used to model and simulate, the cyber- and physical-system layers of the CPES being studied. In this part, we make a distinction between the hardware and simulation/emulation resources that need to be considered for modeling the cyber- and physical-system layers using tools and techniques such as offline simulation, emulation, real-time simulation, and HIL.

\subsubsection{Physical-System Layer}
The simulation and hardware resources for the modeling and implementation of the CPES physical-system layer are presented below.

\begin{figure}[t]
\centering
    \subfloat[]{
            \includegraphics[width=0.95\linewidth]{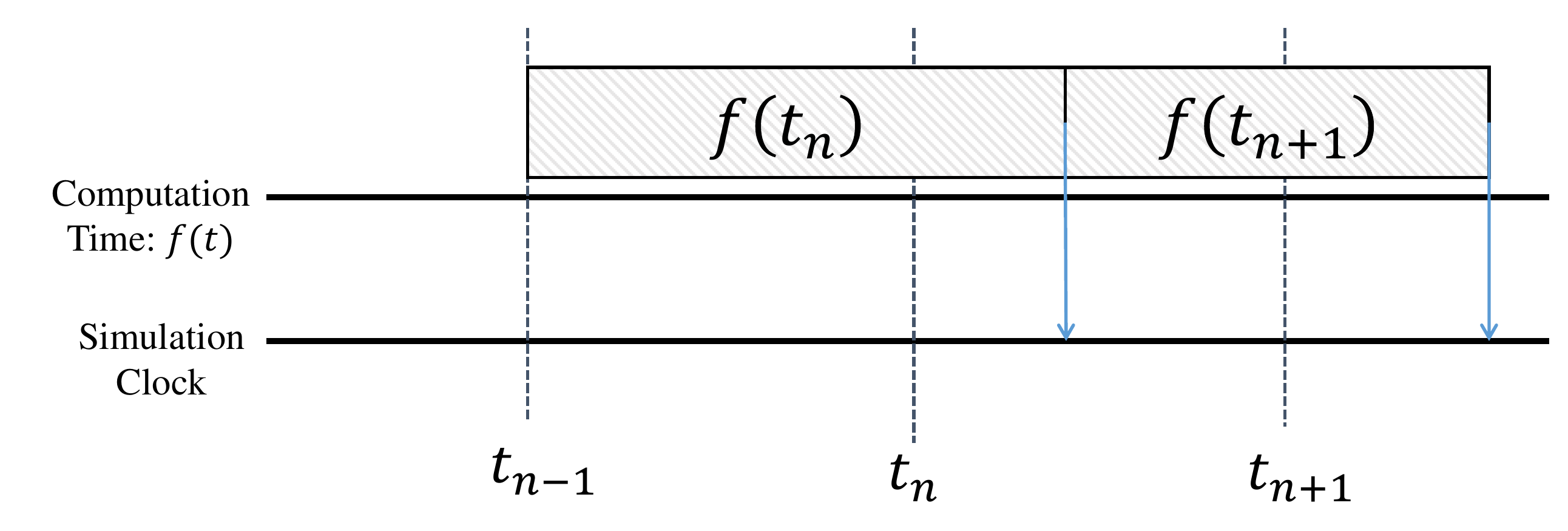}
            \label{subfig:slowerthan}
    } \\
    \subfloat[]{
            \includegraphics[width=0.95\linewidth]{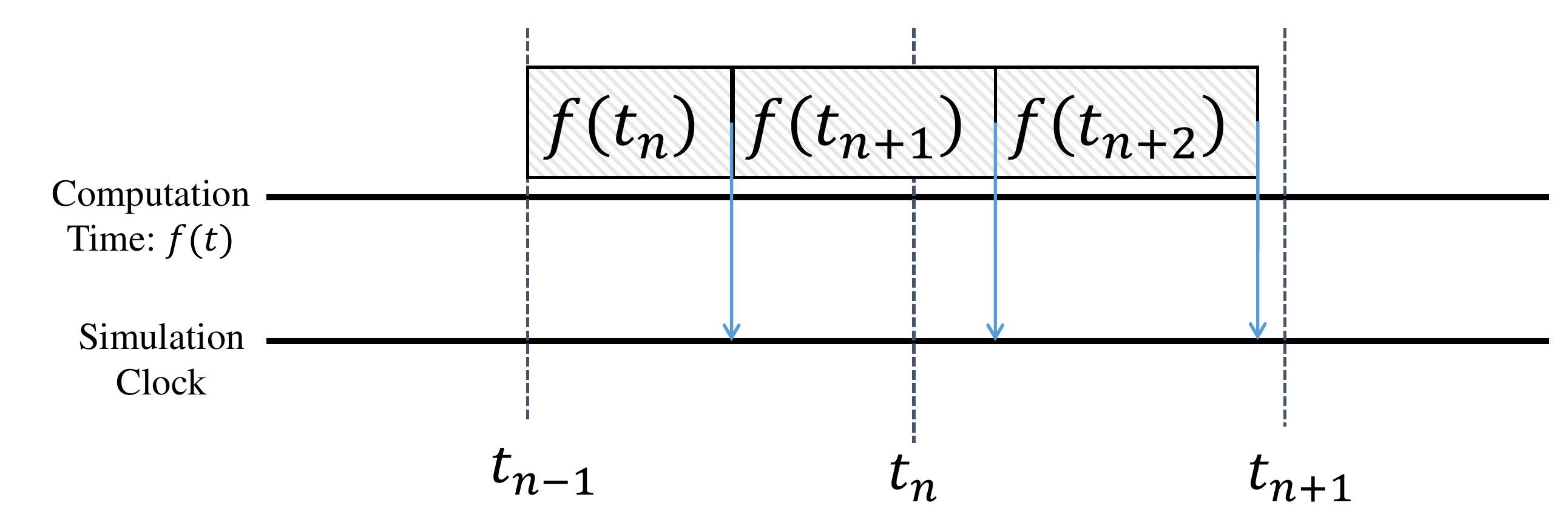}
            \label{subfig:fasterthan}
    } \\
   \subfloat[]{
            \includegraphics[width=0.95\linewidth]{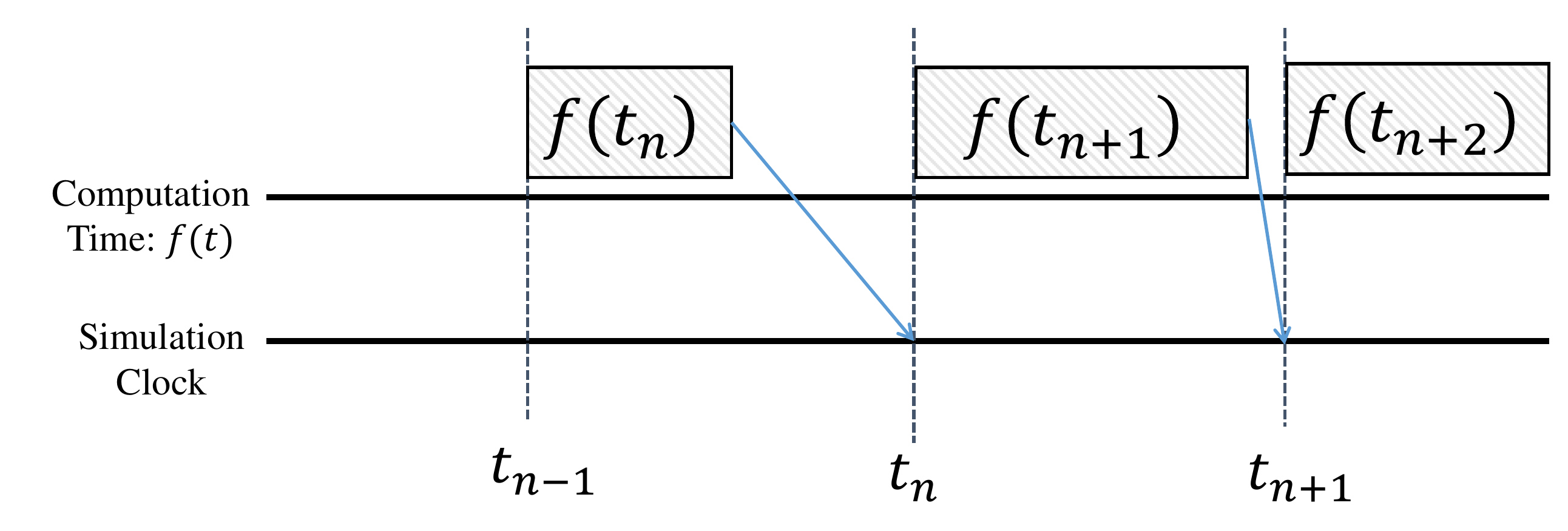}
            \label{subfig:realtime}
    }
\vspace{-1mm}
\caption[CR]{ Differences in the computation timing of offline simulation and real-time simulation: (\subref{subfig:slowerthan}) slower-than real-time, (\subref{subfig:fasterthan}) faster-than real-time, and (\subref{subfig:realtime}) real-time simulation.} 
\label{fig:realtime}
\end{figure}

    \begin{enumerate}[label=\alph*),leftmargin=*, labelwidth=0.1mm, labelindent=0pt, wide=0pt]
    
        \item \textit{Simulation}: A simulation provides a set of models or representations used to reproduce the behavior or operation of different processes of a particular system over time. 
        Particularly for EPS, EMT- and TS-type simulations are the most prominent tools used to investigate the behavior of different system components. These simulation classes can be further classified into two main categories: offline and real-time simulations  \cite{emt_dinavahi}.

        \begin{enumerate}[label=\roman*), leftmargin=*, labelwidth=0.1mm, labelindent=0pt, wide=0pt]
            \item \textit{Offline simulation}: Offline simulation tools provide a simple and cost-effective way of conducting simulations on any generic computing device. These tools can execute models at slower or faster-than-real-time speeds depending on the complexity of the model as well as the availability of computing resources. Figs. \ref{subfig:slowerthan} and \ref{subfig:fasterthan} show how the computation time of the system models, for both slower and faster-than-real-time offline simulations, is not synchronized with the simulation clock, i.e., the real-time clock. Offline simulations allow the simulation of complex systems without considering real-time constraints, which for instance, enable researchers to simulate large periods of time, e.g., months or years, in a few minutes or seconds. Some tools and software which are available for this type of simulations include: \textsl{MATLAB/Simscape Electrical} (EMT \& TS), \textsl{OpenDSS} (TS), \textsl{Gridlab-D} (TS), \textsl{eMegaSim} (EMT), \textsl{ePhasorSim} (TS), and \textsl{ETAP eMTP} (EMT).  
            
            \item \textit{Real-time simulation}: Real-time simulation tools provide the capability of generating results that are synchronized with a real-time clock. This allows physical devices to be interfaced with the simulated system via realistic data exchanges synchronized using a real-time clock. Fig. \ref{subfig:realtime} demonstrates how for real-time simulation the computation time for the system is synchronized with the simulation clock. The computation time needed to solve all the states of the simulated system needs to be lower or exactly the same as the simulation clock, i.e., the real-time clock. Real-time simulation setups allow researchers to connect real devices using HIL techniques such as CHIL and PHIL. Some tools and software which are available for this type of simulations include: \textsl{eMegaSim} (EMT), \textsl{ePhasorSim} (TS), \textsl{HyperSim} (EMT), \textsl{RTDS} (EMT), and \textsl{Typhoon HIL} (EMT).
        \end{enumerate}
        
        \item \textit{Hardware}: Real-time HIL implementations allow the interconnection of external hardware devices to a real-time simulation environment through the appropriate I/O or networking interfaces. Two of these HIL techniques are CHIL and PHIL. 
        
         \begin{enumerate}[label=\roman*),leftmargin=*, labelwidth=0.1mm, labelindent=0pt, wide=0pt]
               \item \textit{Controller Hardware-in-the-Loop (CHIL)}: In CHIL, physical devices are in constant communication and interaction with a simulation running in the real-time environment. This interconnection includes sending control signals and receiving feedback signals through I/O and/or networking ports \cite{rtsimulation, newaz2019controller}. As seen in the hardware section of Fig. \ref{fig:cps_framework_overall}, a physical device connected using a CHIL implementation can be interfaced directly: \textit{i)} with the physical-system layer simulation using the appropriate interface, or \textit{ii)} through the cyber-system layer using standard communication protocols and corresponding networking components.
        
                \item \textit{Power Hardware-in-the-Loop (PHIL)}: In PHIL, a power hardware system such as a PV panel, inverter, or battery system is physically connected to the RTS through analog and digital I/O ports. A PHIL implementation needs the use of a power amplification unit that is responsible for the amplification and conversion of the digital voltage and current data signals -- coming from the simulation environment -- into analog voltage and current signals required by the connected actual/physical device. Interfacing algorithms are also essential to facilitate the interconnection between the software models and the physical-system \cite{phil_rt}. 
         \end{enumerate}
    
    \end{enumerate}

\begin{table*}[t]
\setlength{\tabcolsep}{1.2pt}
\centering
\caption{Physical-system layer performance metrics. These metrics are divided according to the domain where they can be measured. }
\label{table:physical_metrics}
\renewcommand{\tabularxcolumn}[1]{m{#1}}

\resizebox{\textwidth}{!}{%
\begin{tabularx}{\textwidth} { 
      || >{\hsize=0.80\hsize\linewidth=\hsize\raggedright\arraybackslash}X 
      | >{\hsize=1.70\hsize\linewidth=\hsize\centering\arraybackslash}X
      | >{\hsize=.50\hsize\linewidth=\hsize\centering\arraybackslash}X || }

\hline \hline
\textbf{Name} & \multicolumn{1}{c|}{\textbf{Description}} & \textbf{Domain} \\ \hline
Rise time & Evaluates the time required for the output to rise from \%x to \%y of the steady-state response & Control \\ \hline
Percent overshoot & Evaluates the maximum peak value of the output minus the step value divided by the step value &Control \\ \hline
Settling time & Evaluates the time required for the output to reach and remain within a defined error band & Control \\ \hline
Steady-state error & Evaluates the difference between the input (command) and the output of the system & Control \\ \hline
Integrate absolute error (IAE) & Evaluates the absolute error of the system over time & Control \\ \hline
Voltage stability & Metrics that evaluate the voltage stability and regulation of the EPS according to defined limits & EPS \\ \hline
Frequency stability & Metrics that evaluate the frequency stability of the EPS according to defined limits & EPS \\ \hline
Optimization & Optimization metrics used to evaluate energy and power management functions, e.g., energy cost, efficiency & EPS \\ \hline
Power quality & Power quality metrics such as THD, transients, flickering, and voltage sags used to evaluate EPS operation \cite{key2005security} & EPS \\ \hline
Reliability & Reliability indices to evaluate EPS operation, e.g., SAIFI, SAIDI, ASAI, lost load \% \cite{key2005security} & EPS \\ \hline
Command vs. Measured \% error & Percentage error between signal command coming from controller and signal measured & Simulation (CHIL) \\ \hline
PHIL Accuracy & Metrics that evaluate the accuracy of the PHIL integration \cite{ren2007accuracy} & Simulation (PHIL) \\ \hline \hline

\end{tabularx}
}
\end{table*}

\subsubsection{Cyber-System Layer}
The simulation/emulation and hardware resources related to the modeling and development of the cyber-system layer for the communication network are presented below.

 \begin{enumerate}[label=\alph*), leftmargin=*, labelwidth=0.1mm, labelindent=0pt, wide=0pt]
    
        \item \textit{Simulation/Emulation}: As mentioned before, the main difference between simulation and emulation is that in a simulation, the models used are designed to replicate the behavior of the system while emulation is designed to duplicate the behavior of the system. A more detailed description of the difference between simulation and emulation is given below in the context of the resources required to effectively replicate the cyber-system layer. 

        \begin{enumerate}[label=\roman*), leftmargin=*, labelwidth=0.1mm, labelindent=0pt, wide=0pt]
        \item \textit{Simulation}: In network simulations, theoretical and mathematical models are developed to create entirely virtual models of the corresponding networking components. Network simulation tools use discrete-event simulation approaches that generate sequences of discrete events that characterize the discrete cyberspace. The two critical components of such discrete-event driven simulators include the simulation time variable and a list of pending future events. The simulation time variable represents the current time at which the state of the system is known (in the simulation), while the list of pending future events contains all the state changes that have been scheduled to occur in the future, which guide the flow of the simulation. In a network simulation, external devices cannot be interfaced with virtual simulated devices, contrary to a network emulation, hence, the entire communication network needs to be simulated. Some of the available software tools that support this type of simulations are: \textsl{ns-2} \cite{ns2_website}, \textsl{ns-3} \cite{ns3_website}, \textsl{SimPy} \cite{simpy_website}, and \textsl{EXata} \cite{exata_website}, \cite{obaidat2015modeling}. 
        
        \item \textit{Emulation}: In network emulation, hardware and software solutions are designed to accurately replicate the behavior of networking components,  exactly as if they were actual parts of an external network. Network emulation tools enable the configuration and manipulation of network parameters and constraints (e.g., packet loss, delays, jitter, etc.) to mimic the mirrored network. Some of the available software tools that support this type of network modeling are: the \textsl{Common Open Research Emulator (CORE)} \cite{corenetwork}, \textsl{NetEm} \cite{netem_website}, and \textsl{EXata} \cite{exata_website}. Notably, some tools are capable of adapting network simulation models for emulation purposes by adding real-time synchronization mechanisms between the virtualized simulated environment and the real networking components \cite{weingartner2011slicetime, weingartner2008synchronized}.
        
        \end{enumerate}

        \item \textit{Hardware (HIL)}: Similarly to the HIL implementations realized in the physical-system layer, HIL implementations of network components in the cyber-system layer can also be performed using the corresponding networking interfaces. Networking HIL provides emulation capabilities that allow the integration of real equipment into the emulated network through standard communication protocols. Commonly, a larger portion of the network or system is emulated and connected with external (real) devices. Such a method provides high-fidelity responses -- as expected from the actual device -- while maintaining the scale of the emulation. Some software tools that support HIL with communication network models are \textsl{EXataCPS} \cite{exata_website}, \textsl{ns-3}\cite{ns3_website}, and \textsl{CORE} through the \textit{RJ45} utility \cite{corenetwork}.
    \end{enumerate}

\subsection{Performance Metrics}

A multitude of metrics exists to evaluate the performance of the modeled cyber- and physical-system layers. The use of metrics allows the concise evaluation of the overall system alongside its corresponding subsystems. In essence, these metrics provide quantitative ways to measure and evaluate the performance of the system's operation at a particular time, both at the cyber- and the physical-system layers. 

\subsubsection{Physical-System Layer}
Some of the most commonly used metrics employed to evaluate the performance and operation of different functions that exist in the physical-system layer of CPES are presented in Table \ref{table:physical_metrics} and described below:

    \begin{enumerate}[label=\alph*), leftmargin=*, labelwidth=!, labelindent=0pt, wide=0pt]
        \item \textit{Control systems}: Metrics related to control systems can be used to examine the performance of different control routines present at the physical-system layer. The evaluation can include the steady-state response of the system or other system performance indicators such as rise time, percent overshoot, settling time, steady-state error, and integrate absolute error. 
        \item \textit{EPS resiliency, stability, and optimization}: Performance metrics can be defined in order to evaluate the performance of the system according to a predefined baseline behavior. For instance, in an EPS where the operation of a new MG controller is investigated, performance metrics related to voltage regulation, frequency regulation, energy cost, and power quality can be utilized. Similarly, especially for controllers, which are limited by their computing resources, different performance metrics can be utilized to determine execution times, CPU utilization, and memory utilization. 
    
        \item \textit{Simulation accuracy}: The simulation accuracy, either offline or real-time, can also be assessed based on different performance metrics dependent on the stability and accuracy of the system response, respectively. The main objective of these metrics is to validate the response of different physical systems (being simulated) when compared to the actual response expected from the system under examination.
        
    \end{enumerate}

\begin{table*}[t]
\setlength{\tabcolsep}{1.20pt}
\centering
\caption{Cyber-system layer performance metrics. These metrics are divided according to the OSI model layer and connection where they can be measured. }
\label{table:cyber_metrics}
\renewcommand{\tabularxcolumn}[1]{m{#1}}

\resizebox{\textwidth}{!}{%

\begin{tabularx}{\textwidth} { 
      || >{\hsize=0.8\hsize\linewidth=\hsize\raggedright\arraybackslash}X 
      | >{\hsize=1.40\hsize\linewidth=\hsize\centering\arraybackslash}X
      | >{\hsize=.4\hsize\linewidth=\hsize\centering\arraybackslash}X
      | >{\hsize=.4\hsize\linewidth=\hsize\centering\arraybackslash}X || }

\hline \hline
\textbf{Name} & \multicolumn{1}{c|}{\textbf{Description}} & \textbf{Layer} & \textbf{Connection} \\ \hline
{Bit rate (R)} & Number of bits conveyed per unit of time & \textit{L1/L2} & Wired/Wireless \\ \hline
{Bit-error rate (BER)} & Ratio of the number of received bits altered during transmission to the total number of bits sent & \textit{L1/L2} & Wired/Wireless \\ \hline
{Packet-error rate (PER)} & Ratio of the number of packets received incorrectly to the total number of packets received & \textit{L1/L2} & Wired/Wireless \\ \hline
{Nominal channel capacity (NCC)} & Maximum number of bits that can be transmitted per unit of time & \textit{L1/L2} & Wired/Wireless \\ \hline
{Channel utilization (CU)} & Ratio between NCC and the total number of bits received per transmission time & \textit{L1/L2} & Wired/Wireless \\ \hline
{Signal-to-noise ratio (SNR)} & Ratio of the signal power to the background noise & \textit{L1/L2} & Wired \\ \hline
{Signal-to-interference-plus-noise ratio (SINR)} & Similar to SNR but considers the interference power from other signals & \textit{L1/L2} & Wireless \\ \hline
{Spectral efficiency (SE)} & Number of received bits per unit of time per unit of bandwidth and per unit areas (i.e., $\frac{b/s}{Hz \cdot m^2}$) & \textit{L1/L2} & Wireless \\ \hline
{Received signal strength indication (RSSI)} & Signal strength measured at the receiver's antenna during packet reception & \textit{L1/L2} & Wireless \\ \hline
{Hop count} & Minimum hop-count from source node to destination node & \textit{L3} & Wired/Wireless \\ \hline
{Round trip time (RTT)} & Time that it takes for a signal to be sent and the acknowledgement received & \textit{L3} & Wired/Wireless \\ \hline
{Expected transmission time (ETT)} & Time needed for a data packet to be correctly transmitted over a link & \textit{L3} & Wireless \\ \hline
{Betweenness} & Number of shortest paths between any two nodes that pass through the evaluated node & \textit{L3} & Wired/Wireless \\ \hline
{Node degree} & Number of nodes that depend on the evaluated node. & \textit{L3} & Wired/Wireless \\ \hline
{End-to-end delay} & Time required to transmit a packet along the path between source and destination nodes & \textit{L4/L5/L6/L7} & Wired/Wireless \\ \hline
{Jitter} & Packet delay variation & \textit{L4/L5/L6/L7} & Wired/Wireless \\ \hline
{Bandwidth} & Overall bandwidth consumption in the network & \textit{L4/L5/L6/L7} & Wired/Wireless \\ \hline
{Link stress} & Number of packet replicas traversing the same physical link & \textit{L4/L5/L6/L7} & Wired/Wireless \\ \hline \hline
\end{tabularx}
}
\end{table*}

\subsubsection{Cyber-System Layer}
Different metrics can be utilized to evaluate the performance of the modeled cyber-system layer communication network. Here,  we demonstrate, as a practical example, some of the most widely used metrics designed to evaluate the network performance at different layers of the open systems interconnection (OSI) model \cite{obaidat2015modeling}. Table \ref{table:cyber_metrics} outlines some representative network performance metrics. 

    \begin{enumerate}[label=\alph*),leftmargin=*, labelwidth=0.1mm, labelindent=0pt, wide=0pt]
        \item \textit{Physical (L1) and Data Link Layers (L2)}: These layers describe how data should be generated and transmitted by network devices over the corresponding physical media. 
    
        \item \textit{Network Layer (L3)}: This layer describes how data packets are transferred between a source and a destination node inside the network. It represents layer 3 of the OSI model. The main performance metrics described below are designed to evaluate two main routing functions: path selection, and network topology management. Path selection aims to determine the best path from source to destination, while network topology management defines how network entities are interconnected for data forwarding purposes. 
    
        \item \textit{Transport (L4), Session (L5), Presentation (L6), and Application (L7) Layers}: These layers describe the shared communication protocols and interfacing methods used by the nodes in the network. In essence, these are the layers responsible for providing full end-user access to the communication network infrastructure.
    \end{enumerate}

It is important to note that many other network and physical performance metrics can be used to evaluate specific scenarios. The presented lists include a subset of the available metrics discussed in the literature. There are also application-specific metrics that can be defined according to each study's requirements. Overall, researchers should carefully model their systems as well as select the corresponding resources and metrics to accurately represent the cyber- and physical-layer of the CPES under test. This will allow the integration of any external physical device, either through CHIL and/or PHIL, and ensure the holistic validation of the system's operation.

%% file: sections/5-Case_studies.tex
\section{Experimental Setup \& Case Studies} \label{s:studies}

\begin{table*}[t]
\normalsize
\setlength{\tabcolsep}{1.2pt}
\centering
\caption{Threat model of the attack case studies. }
\label{table:Attack_threat_connection}

\renewcommand{\tabularxcolumn}[1]{m{#1}}

\begin{tabularx}{\textwidth \centering} { 
  || >{\hsize=.45\hsize\linewidth=\hsize\raggedright\arraybackslash}X 
  | >{\hsize=0.64\hsize\linewidth=\hsize\centering\arraybackslash}X
  | >{\hsize=0.64\hsize\linewidth=\hsize\centering\arraybackslash}X
  | >{\hsize=0.63\hsize\linewidth=\hsize\centering\arraybackslash}X
  | >{\hsize=0.64\hsize\linewidth=\hsize\centering\arraybackslash}X || }
  
    \hline
    \hline 
        \textbf{Threat Model} & {\textbf{Cross-layer Firmware Attacks}} & {\textbf{Load-changing Attacks}} & {\textbf{Time-delay Attacks}} & {\textbf{Propagating Attacks in Integrated T\&D CPES}} \\ 
    \hline
        {Knowledge} & {Oblivious}  & {Limited or Oblivious} & {Oblivious} & {Strong} \\
    \hline
        {Access} & {Possession} & {Non-possession} & {Non-possession} & {Possession} \\
    \hline
        {Specificity} & {Non-targeted} & {Targeted} & {Targeted} & {Targeted} \\
    \hline
        {Resources} & {Class I or II} & {Class II} & {Class I or II} & {Class II}  \\
    \hline
        {Frequency} & {Iterative} & {Iterative} & {Iterative} & {Non-Iterative}  \\
    \hline
        {Reproducibility} & {Multiple-time} & {Multiple-times}  & 
        {Multiple-time} & {One-time} \\
    \hline
        {Functional Level} & {L$1$} & {L$1$ or L$2$} & {L$1$} & {L$2$} \\
    \hline
        {Asset} & {PLC} & {PLC or HMI} & {Control Server} & {Engineering workstation}  \\
    \hline
        {Technique} & {Modify control logic} & {Modify control logic or wireless compromise} & {Wireless compromise, MitM, Spoofing, DoS} & {Engineering workstation compromise} \\
    \hline
        {Premise} & {Physical: Invasive, or non-invasive  \newline Cyber: Asset control commands} & {Cyber: Communications and protocols or Asset control commands} & 
        {Cyber: Communications and protocols } & {Cyber: Asset control commands} \\
    \hline
    \hline
\end{tabularx}
\end{table*}

The case studies discussed in this section demonstrate how the presented threat modeling approach, the CPS framework, and risk assessment methodology can be utilized to perform detailed CPES studies. Table \ref{table:Attack_threat_connection} describes how each study can be formalized using our proposed threat modeling method. Following, the corresponding modeling layers, resources, and evaluation metrics are identified for each case study according to the conceptual CPS framework. Additionally, for each attack scenario, the specific background, and mathematical formulation are described and the corresponding threat model is provided based on Section \ref{s:CPSSecurity}. The threat model describes the assumptions made for the adversary intentions and capabilities as well as the attack-specific details, demonstrating the practicality of our modeling approach for diverse attack scenarios. 
Furthermore, we demonstrate how the proposed risk assessment procedure can be applied to each case study and assist in prioritizing mitigation strategies. In our work, the objective priority for CPES is outlined in Table \ref{table:priority}. It should be noted that the order of objectives might change depending on the system's component being analyzed or the stakeholders' priorities. For instance, the impact of the ``uninterrupted operation and service provision'' objective could indicate less priority in the case of a compromised inverter serving as an ancillary power generation source in a residential deployment, in contrast to a T\&D system-wide attack.

\begin{table}[t]
\setlength{\tabcolsep}{1.2pt}
\centering
\caption{Objective Priority for CPES Risk Assessment. }
\label{table:priority}
\begin{tabularx}{\linewidth} { 
      || >{\hsize=1.6\hsize\linewidth=\hsize\raggedright\arraybackslash}X 
      | >{\hsize=.4\hsize\linewidth=\hsize\centering\arraybackslash}X || }
      
     \hline
     \hline
     \textbf{Objective} & {\textbf{Priority}} \\ 
     \hline
     {People health and personnel safety} & {4} \\
     \hline
     {Uninterrupted operation and service provision} & {3} \\
     \hline
     {Equipment damage and legal punishment} & {2} \\
     \hline
     {Organization financial profit} & {1} \\
     \hline
     \hline
\end{tabularx}
\end{table}

\begin{table}[ht!]

\setlength{\tabcolsep}{1.2pt}
\centering
\caption{Symbols and notation for case studies formulation.}
\label{tab:notations_sce}
\resizebox{\linewidth}{!}{
\begin{tabular}{||l|c|c||}
\hline \hline
\textbf{Domain} & \textbf{Symbol} & \multicolumn{1}{c||}{\textbf{Definition}} \\ \hline 
 \multirow{14}{*}{CPS} & $k$ & timestep \\ \cline{2-3}
 & $n$ & total number of states \\ \cline{2-3}
 & $m$ & total number of measurements \\ \cline{2-3}
 & $l$ & total number of controls \\ \cline{2-3}
 & $x$ & system's states \\ \cline{2-3}
 & $y$ & system's measurements \\ \cline{2-3}
 & $y_a$ & attacked system's measurements \\ \cline{2-3}
& $u$ & control variables \\ \cline{2-3}
 & $u_a$ & attacked control variables \\ \cline{2-3}
 & $e$ & measurement noise \\ \cline{2-3}
 & $G$ & system matrix \\ \cline{2-3}
 & $B$ & input matrix \\ \cline{2-3}
 & $C$ & output matrix \\ \cline{2-3}
 & $H$ & control matrix \\ \hline 
 
 \multirow{5}{*}{DIA} & $\beta$ & multiplicative attack factor \\ \cline{2-3}
 & $\mathrm{W}$ & additive White-noise attack factor \\ \cline{2-3}
 & $\mathrm{T_{attack}}$ & attack period (time) \\ \cline{2-3}
 & $\Delta y$ & manipulated measurements used to create $y_a$ \\ \cline{2-3}
 & $\Delta u$ & manipulated controls used to create $y_a$ \\ \hline 
 
 \multirow{3}{*}{DAA} & $f_D$ & time-delay function \\ \cline{2-3}
 & $d$ & discrete constant value or time function. \\ \cline{2-3}
 & $s_r$ & compromised signal (could be either $y$ or $u$) \\ \hline \hline
\end{tabular}}
\end{table}

The attack cases presented in this section can be characterized as either \textit{DIA} or \textit{data availability attacks (DAA)}. Table \ref{tab:notations_sce} provides the essential notation for the case studies. Each scenario follows a mathematical background as part of a CPS plant formulation: 

\begin{equation}
\label{eq:cpsx}
    x(k+1) = Gx(k) + Bu(k)
\end{equation}
\begin{equation}
\label{eq:cpsy}
    y(k) = Cx(k) + e(k)
\end{equation}
\noindent where $x(k) \in \mathbb{R}^{n}$ represents the states of the system, $u(k) \in \mathbb{R}^{l}$ represents the control variables, and $y(k) \in \mathbb{R}^{m}$ represents the system measurements. $G \in \mathbb{R}^{n \times n}$, $B \in \mathbb{R}^{n \times l}$, and $C \in \mathbb{R}^{m \times n}$ represent the system matrix, input matrix, and output matrix, respectively. The term $e \in \mathbb{R}^{m}$ represents measurement noise in the system's input measurements. As for the cyber part of the CPS, it  can be generally expressed as: 
\begin{equation}
\label{eq:cpscyber}
    u(k+1) = Hy(k)
\end{equation}
\noindent where $H \in \mathbb{R}^{l \times m}$ represents the control matrix \cite{zhang2018novel}. 

\begin{figure}[t]
\centering
\includegraphics[width = 0.49\textwidth]{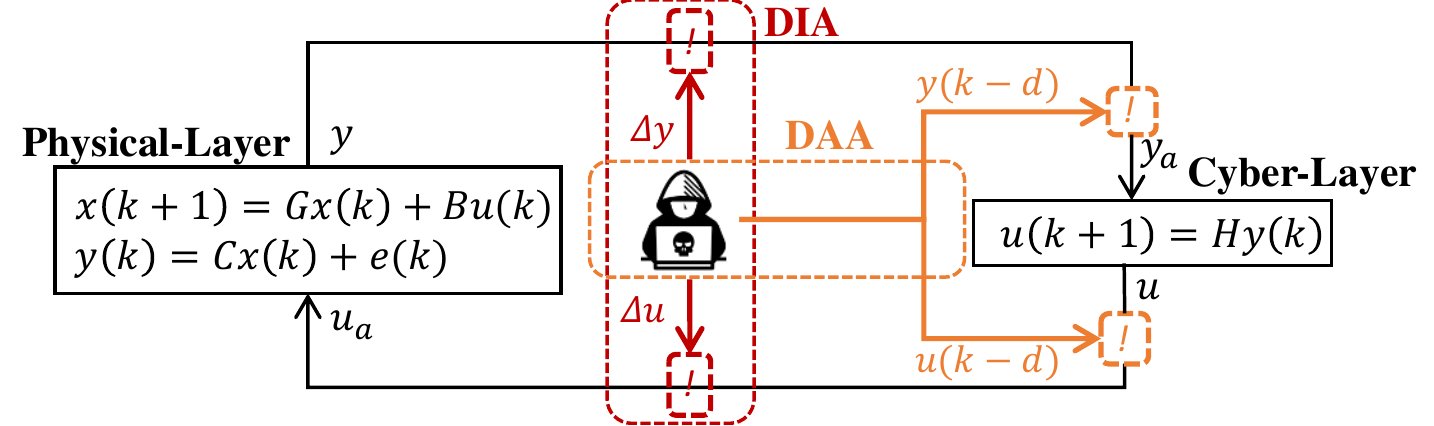}
\caption{\label{fig:cpsplant} Diagram of CPS plant under DIA and DAA scenarios.}
\end{figure}

Fig. \ref{fig:cpsplant} depicts a diagram of the CPS mathematical formulation and the respective variables compromised by attackers during DIA and DAA scenarios. In the DIA case, either the measurements ($y$) or the control variables ($u$) can be compromised by attackers via modification or fabrication. On the other hand, in a DAA scenario, either the measurements ($y$) or controls ($u$) can be compromised by attackers via interruption, i.e., delaying their acquisition or utilization by the system.

\subsection{Case Study 1: Cross-layer Firmware Attacks}

\noindent\underline{Background \& Formulation}:
Cross-layer firmware attacks refer to attacks targeting the firmware code of embedded devices (i.e., the device read-only resident code which includes microcode and macro-instruction level routines), aiming to generate and propagate impacts from the device layer to system and application layers, respectively. Typically, embedded devices in industrial CPS run on bare metal hardware without OS and directly boot monolithic single-purpose software. In such devices, tasks are executed on a single-threaded infinite loop. If the device firmware code execution is maliciously modified, adversaries could gain total control over the embedded device. The effects of such attacks can have a cross-layer impact affecting multiple components and processes of the CPS. For example, in a CPES, by modifying the firmware controlling grid-tied inverters connected to BESS or EV chargers, an adversary could compromise the system's measurements, thus causing frequency fluctuations, voltage sags, and system stability issues. Other scenarios could even cause wide-area outages, such as the Ukrainian power grid attack in 2015, in which attackers replaced the legitimate firmware of serial-to-Ethernet converters at substations causing them to become inoperable \cite{pbpo095e_ch15}. In general, cross-layer firmware attacks can be categorized as a DIA-type of attack since modifications at the firmware level could result in compromising the integrity of data at different CPS layers. 

In this type of DIA, the adversary (though firmware modifications) can tamper with the input/sensed measurements (e.g., modify, scale, etc.), $y(k)$, and thus directly affect the inverter control strategy and variables, $u(k)$, driving the system into instability. This type of attack can be characterized as a \textit{combined} DIA attack \cite{7820092,7171098,  8511631}. In more detail, the system's input measurements are modified using both an additive random/white noise component and an attack model in which nominal measurements are scaled (increased or decreased). These DIAs can be modeled as:
    
\begin{equation}
y_{a}(k)=\begin{cases}
    y(k) & , \text{when } k \notin \mathrm{T_{attack}}\\
    \beta y(k) + \mathrm{W} & , \text{when } k \in \mathrm{T_{attack}}\\
  \end{cases}
\end{equation}

\noindent where $\beta$ represents the multiplicative attack term, $\mathrm{W}$ represents the additive random/white noise attack, $ \mathrm{T_{attack}}$ represents the period of time when the DIA is performed, and $y_{a}$ represents the `altered'/attacked input measurements. $\beta > 1$ represents increasing-type of attacks, and $\beta < 1$ decreasing attacks.

Following this combined-type DIA mathematical formulation, we demonstrate how the inverter operation can be compromised by spoofing its energy conversion module. The results of this compromise affect not only the inverter behavior but also propagate and impact the MG operation as well.

\begin{figure}[t]
\centering
\includegraphics[width = 0.45\textwidth]{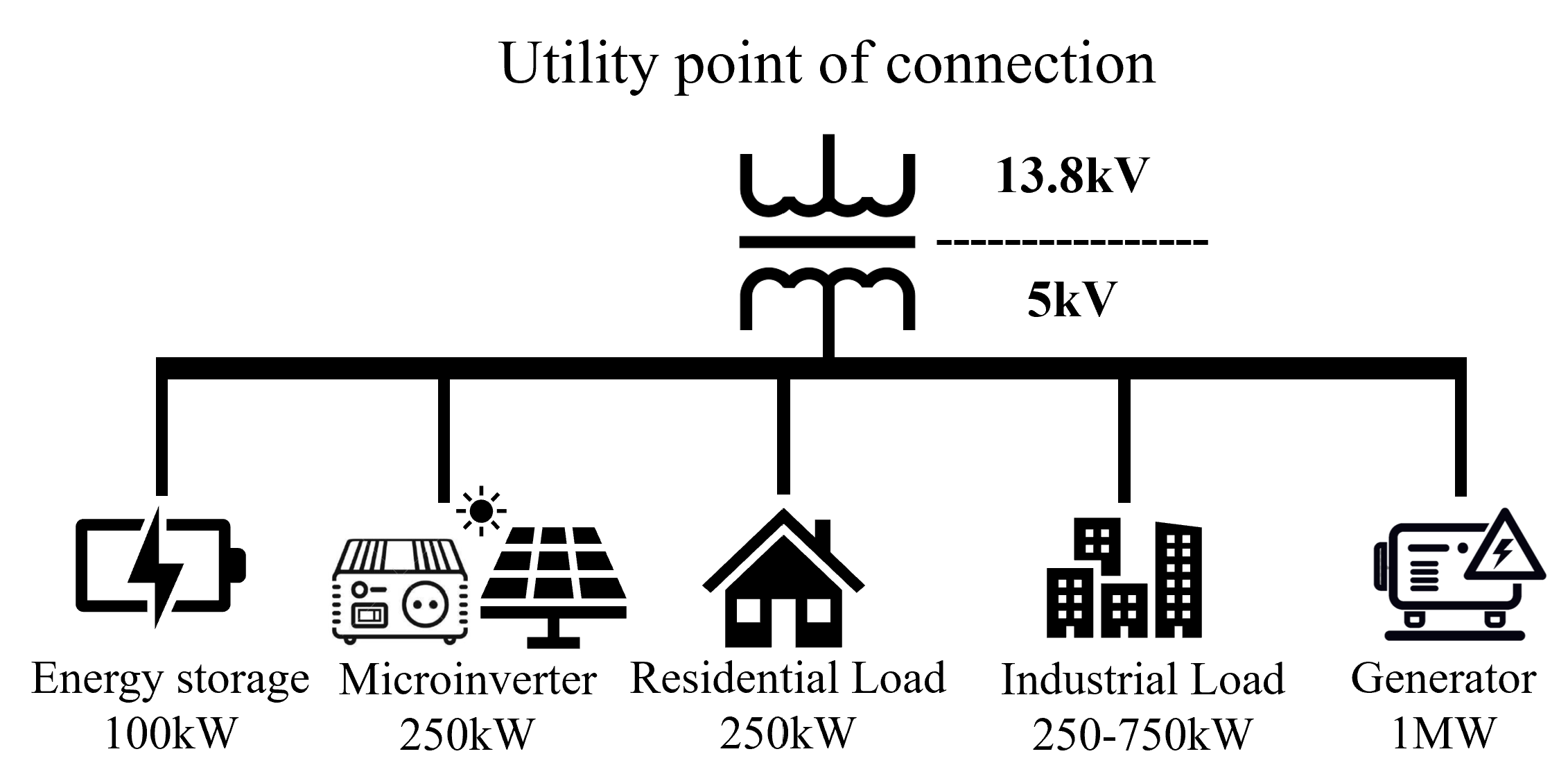}
\caption{\label{fig:crossmg} Conceptual illustration of the MG system used for the cross-layer firmware case study.}
\end{figure}

\noindent\underline{Threat Model}: As presented in Section \ref{s:CPSSecurity}, the threat modeling process for any attack can be characterized by the adversary model and the attack model formulations. Specifically, in this cross-layer firmware attack case, we assume an \textit{oblivious adversary} without full observability of the CPES, and who has direct physical access to the targeted hardware controller (i.e., \textit{adversary access: possession}). Regarding adversarial specificity, the attack is presumed to be a \textit{non-targeted attack}. The adversarial resources could range from the minimum, i.e., Class I, up to state-funded criminal organizations (Class II), in the worst-case scenario.

Furthermore, our case study assumes an attack that occurs \textit{iteratively} and can be reproduced \textit{multiple times}. The \textit{targeted asset is a solar inverter controller}, so the attack level is defined as  \textit{Level 1}. Finally, the technique employed to compromise the system involves \textit{control logic code modification}, and the attack premise can be categorized as either invasive or non-invasive (on the physical domain) or could target the inverter control (e.g., power conversion, power factor, active reactive injections, setpoints, etc.) using malicious commands (on the cyber domain).

\begin{figure}[t]
\centering
\includegraphics[width = \linewidth]{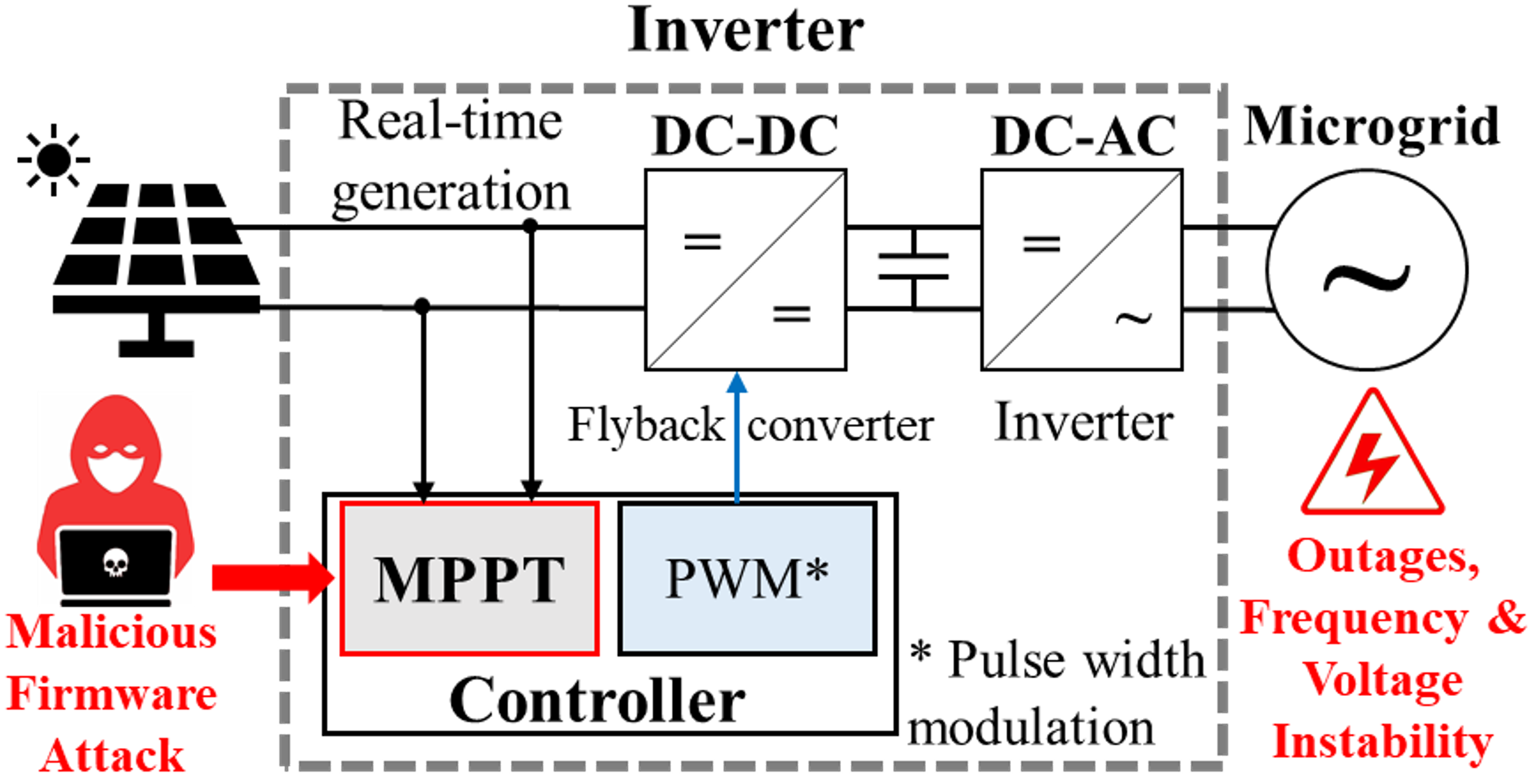}
\caption{\label{subfig:MPPT_attack} Cross-layer firmware attack targeting the inverter's maximum power point tracking (MPPT) controller.}
\vspace{-5mm}
\end{figure}

\noindent\underline{Attack Setup \& Evaluation}: In this case study, a cross-layer firmware attack is modeled as a DIA that compromises physical components, more specifically a PV inverter, at the \textit{physical-system layer} of the CPES. Both EMT and TS simulation modeling approaches are used to model a MG system comprised of a solar PV with its inverter, a Li-ion BESS, a diesel generator, and residential and industrial loads. The MG is connected to the main grid via a $13.8$ kV/$5$ kV distribution substation transformer with a capacity of $250$ MVA. The nameplate generation capacity for the diesel generator is set to $1$ MW. The maximum generation capacity that the PV inverter can reach is $250$ kW based on the provided solar irradiance profile. The BESS is capable of providing up to $100$ kW and storing $100$ kWh. The loads of the MG include aggregated residential loads with a constant power demand of $250$ kW and a variable lumped industrial load whose power demand ranges between $250$-$750$ kW. Fig. \ref{fig:crossmg} shows a conceptual illustration of the described MG. The main software resource used to conduct the EMT and TS offline-simulations of the physical-system layer for this case study is \textsl{MATLAB/Simscape} \textsl{Electrical}.

\begin{figure}[t]
    \centering
        \subfloat[]{
                \includegraphics[width=0.9\linewidth]{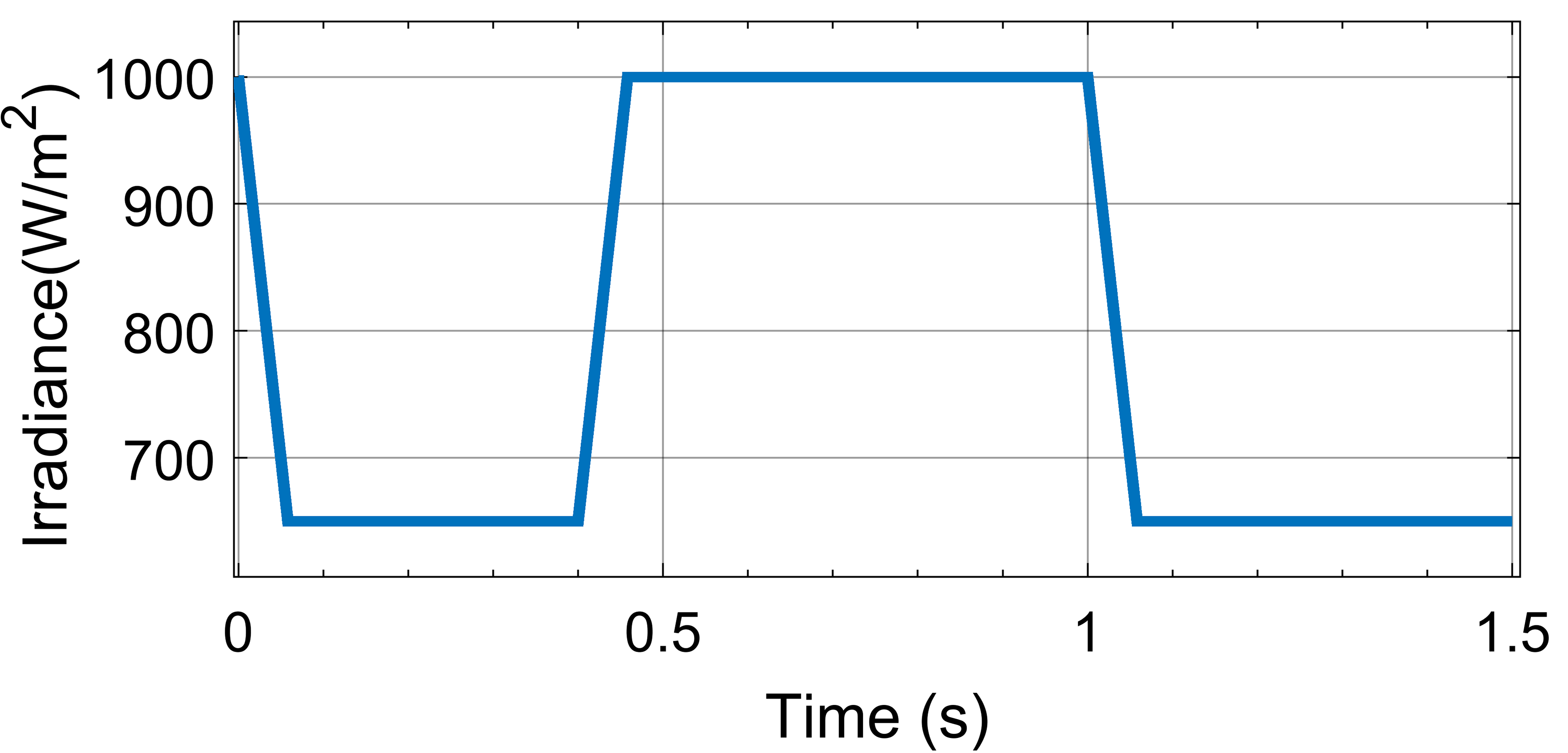}
                \label{subfig:irad}
        } \\
        \subfloat[]{
                \includegraphics[width=0.90\linewidth]{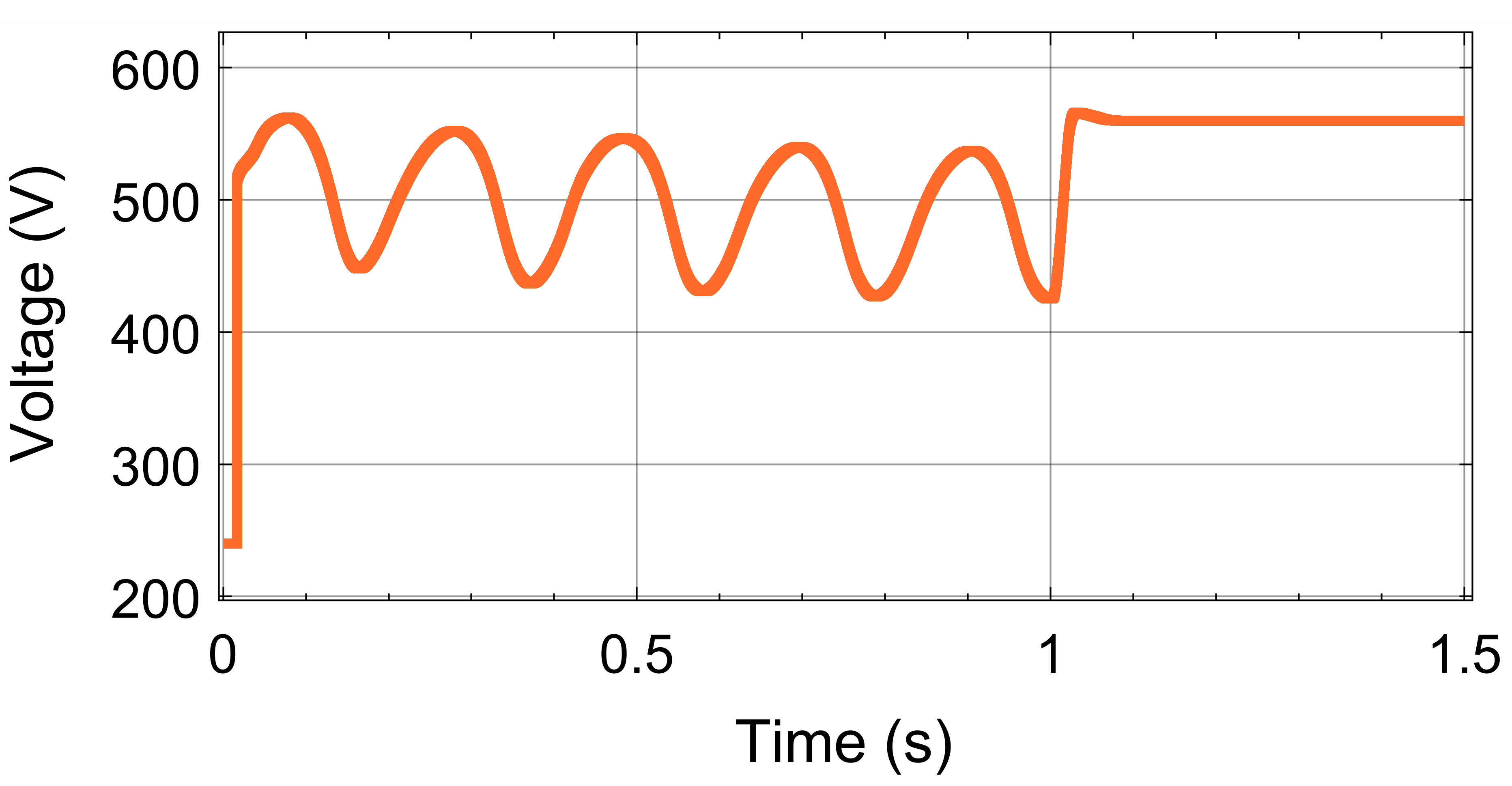}
                \label{subfig:DC_volt}
        } \\
       \subfloat[]{
                \includegraphics[width=0.9\linewidth]{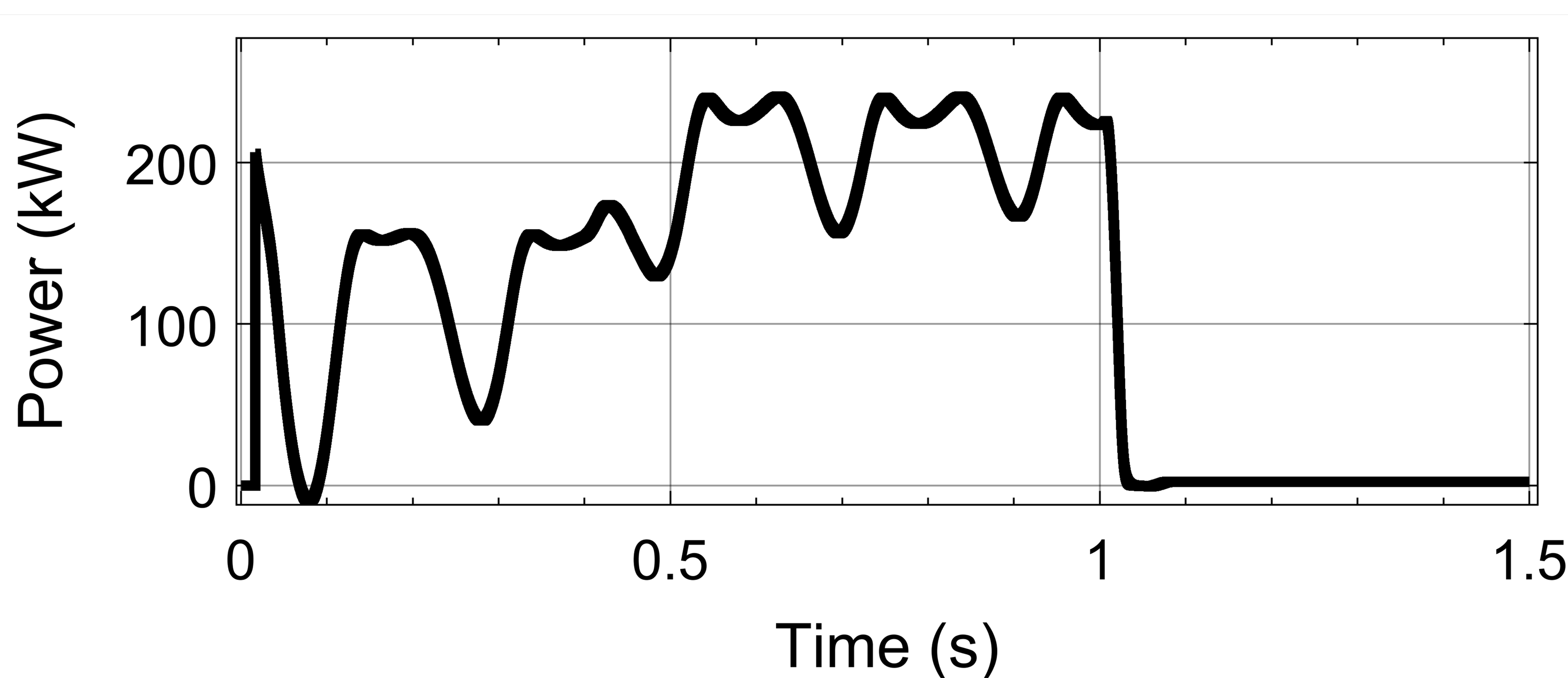}
                \label{subfig:DC_power}
        }
    \vspace{-1mm}
    \caption[CR]{Cross-layer firmware attack impact on the DC-DC converter stage: 
    (\subref{subfig:irad}) solar irradiance profile, 
    (\subref{subfig:DC_volt}) impact on the DC voltage output, and 
    (\subref{subfig:DC_power}) impact on the DC power output.} 
    
\vspace{-3mm}
\label{fig:MPPT_DC}
\end{figure}

\begin{figure}[t]
    \centering
        \subfloat[]{
                \includegraphics[width=\linewidth]{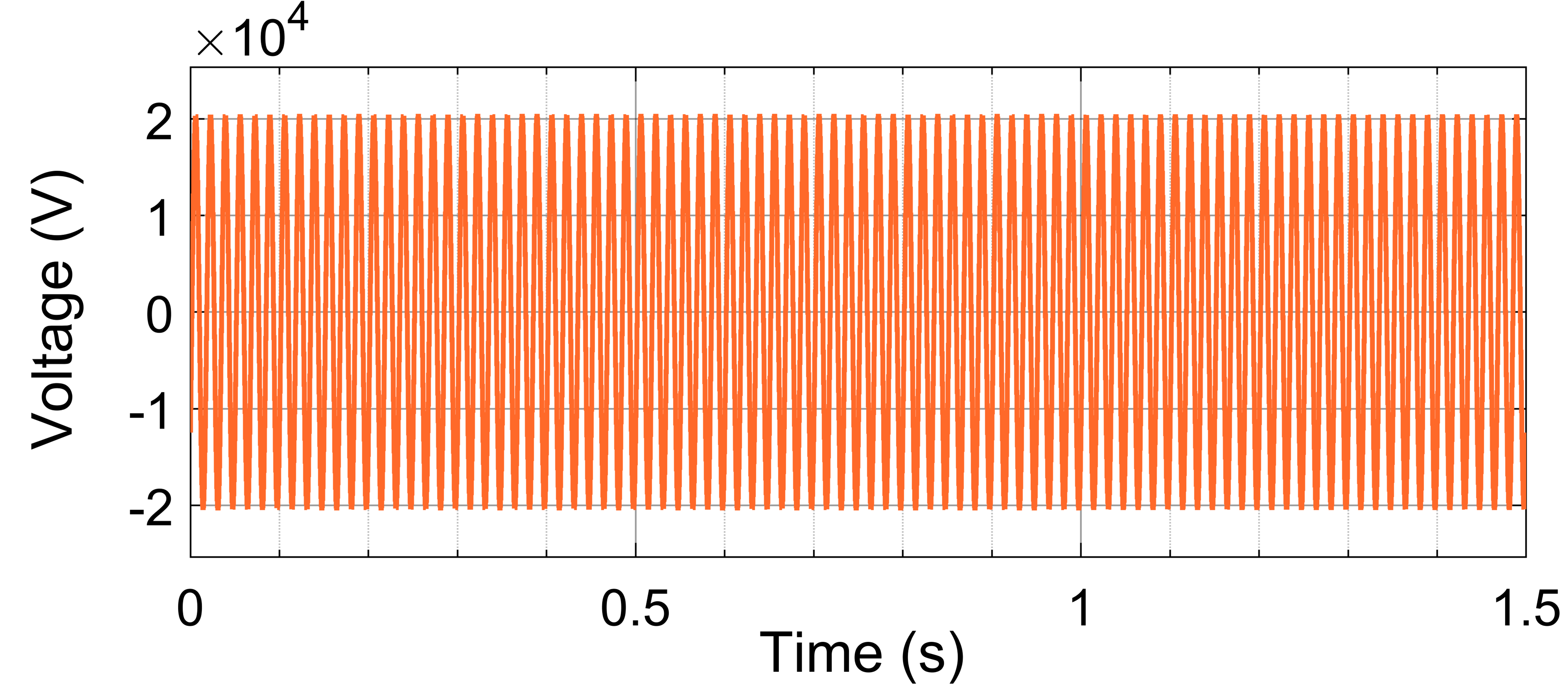}
                \label{subfig:AC_volt}
        } \\
        \subfloat[]{
                \includegraphics[width=\linewidth]{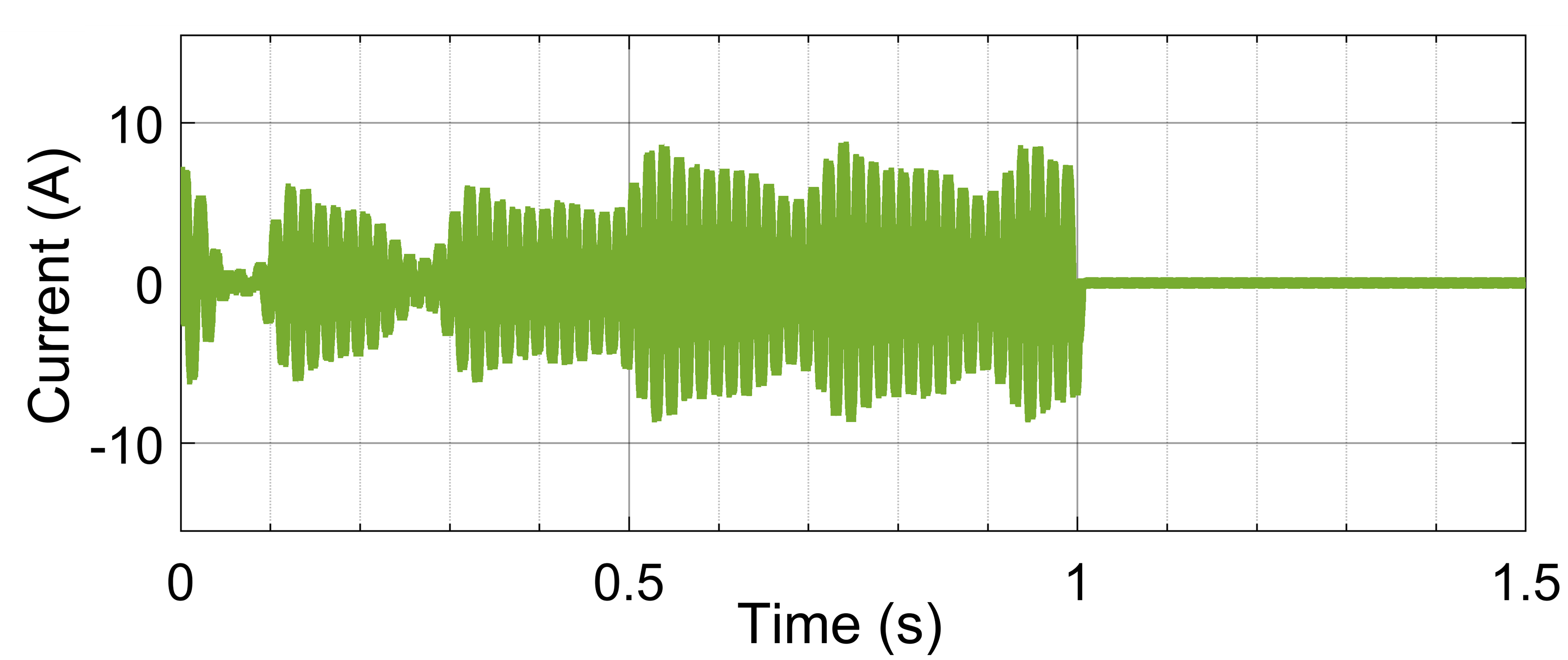}
                \label{subfig:AC_curr}
        } \\
       \subfloat[]{
                \includegraphics[width=\linewidth]{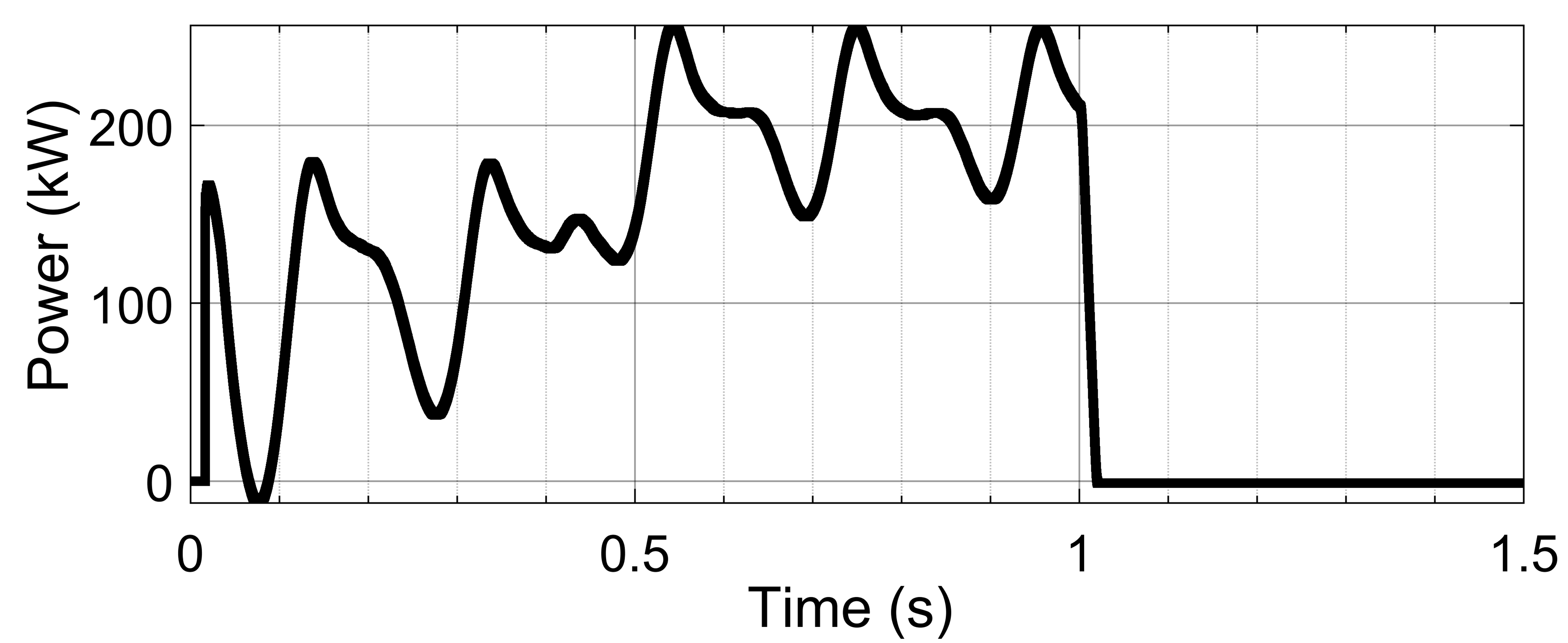}
                \label{subfig:AC_power}
        }
    \vspace{-1mm}
    \caption[CR]{Cross-layer firmware attack impact on the DC-AC converter stage (grid-tied inverter end):
    (\subref{subfig:AC_volt}) impact on the AC voltage output, 
    (\subref{subfig:AC_curr}) impact on the AC current output, and 
    (\subref{subfig:AC_power}) impact on the AC power output.} 
    
\vspace{-3mm}
\label{fig:MPPT_AC}
\end{figure}

In Fig. \ref{subfig:MPPT_attack} we illustrate, the top-level architectural overview of an inverter, the core components comprising it, and the maximum power point tracking (MPPT) controller block that is the main target of this attack use case. Attackers can disrupt the nominal inverter operation by tampering with the firmware subroutines which control both the DC-DC boost and the DC-AC conversion stages. In particular,   Figs. \ref{fig:MPPT_DC}, \ref{fig:MPPT_AC} demonstrate the specifics of how the operation of an inverter can be affected by an adversary capable of compromising the firmware. For our use case, we employ a grid-tied solar inverter module provided by Texas Instruments \cite{TIinverter}. The inverter leverages an \textit{F2803x} series control card which is responsible for managing the inverter's peripheral devices (e.g., sensing modules, analog-to-digital converters, transistor gate driver circuits, etc.) as well as the power conversion process (i.e., solar energy to electricity). By modifying the operation of the MPPT algorithm -- within the firmware code of the control card that the inverter utilizes to optimize the output power generated by the solar panels -- the attacker is able to destabilize the operation of the converter.

MPPT algorithms enable inverters to obtain high power conversion efficiencies. By constantly monitoring the solar PV outputs (i.e., PV generated voltage and current), MPPT algorithms regulate the converter's operating point achieving maximal power transfer. Given that the PV real-time generation measurements are critical for the MPPT operation, any perturbations of the sensed values can potentially compromise the inverter's nominal operation. For our case study, the modification of the inverter's firmware tampers with the inverter's MPPT function and the controls of the DC-DC and DC-AC converters. In the context of DIA attacks, the sensed inputs to the MPPT function, i.e., PV voltage and current, are maliciously modified. By tampering with the MPPT input measurements, down-scaling, and introducing additive sinusoidal noise (combined-type DIA attack), we are able to generate the oscillatory behavior depicted in Fig. \ref{fig:MPPT_DC}. 
This unstable behavior propagates through the inverter's power conversion process leading to anomalous behavior on the grid-tied inverter end, as seen in Fig. \ref{fig:MPPT_AC}. The result of this compromise is the eventual disconnection of the inverter-enabled power resource ($t=1sec$) in order to protect the rest of the MG devices and avoid operational disruptions. 

\begin{figure}[t]
\centering
    \subfloat[]{
            \includegraphics[width=0.9\linewidth]{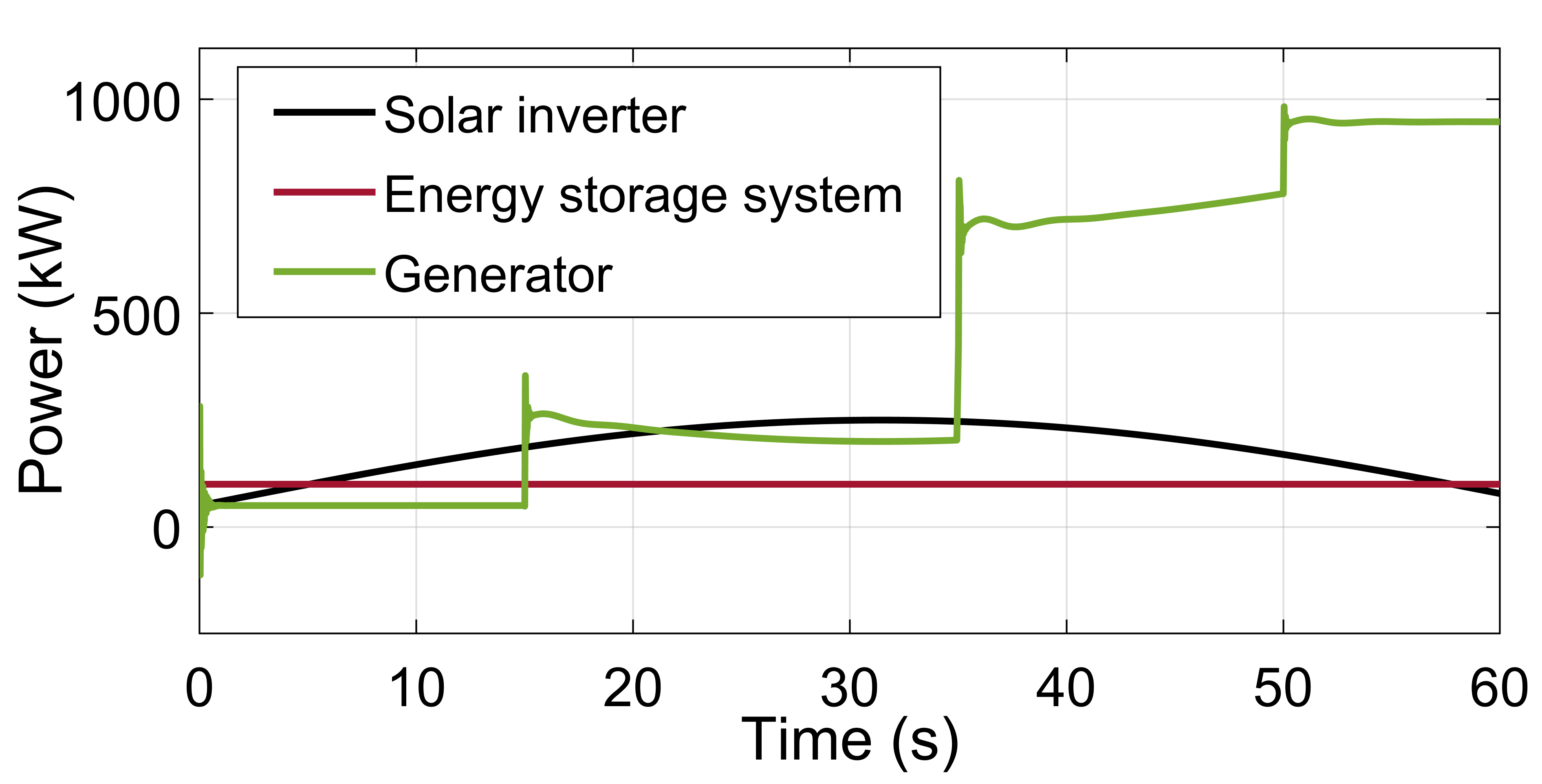}
            \label{subfig:generation}
    }\\
   \subfloat[]{
            \includegraphics[width=0.92\linewidth]{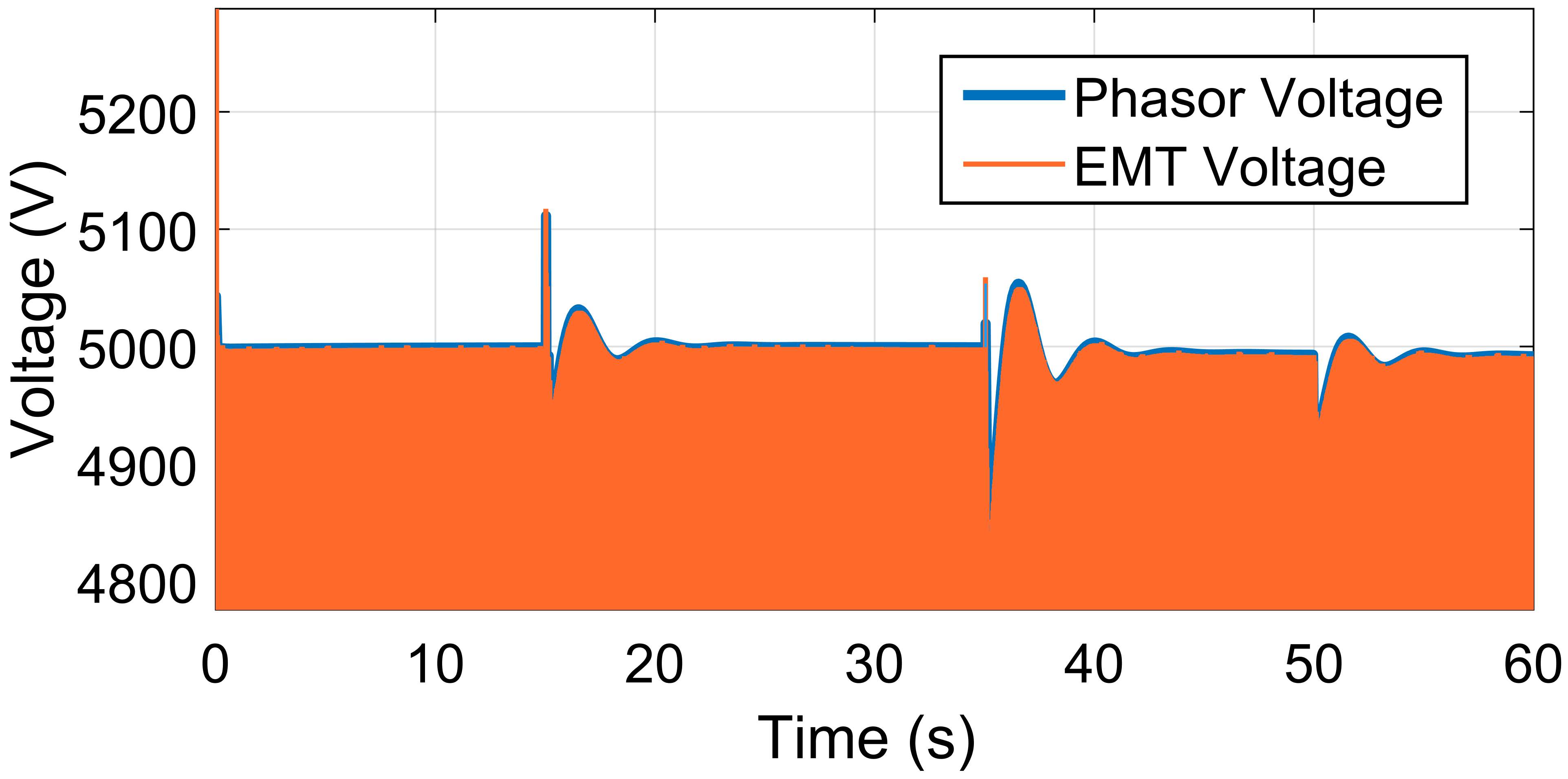}
            \label{subfig:voltage}
    }\\
    \subfloat[]{
            \includegraphics[width=0.9\linewidth]{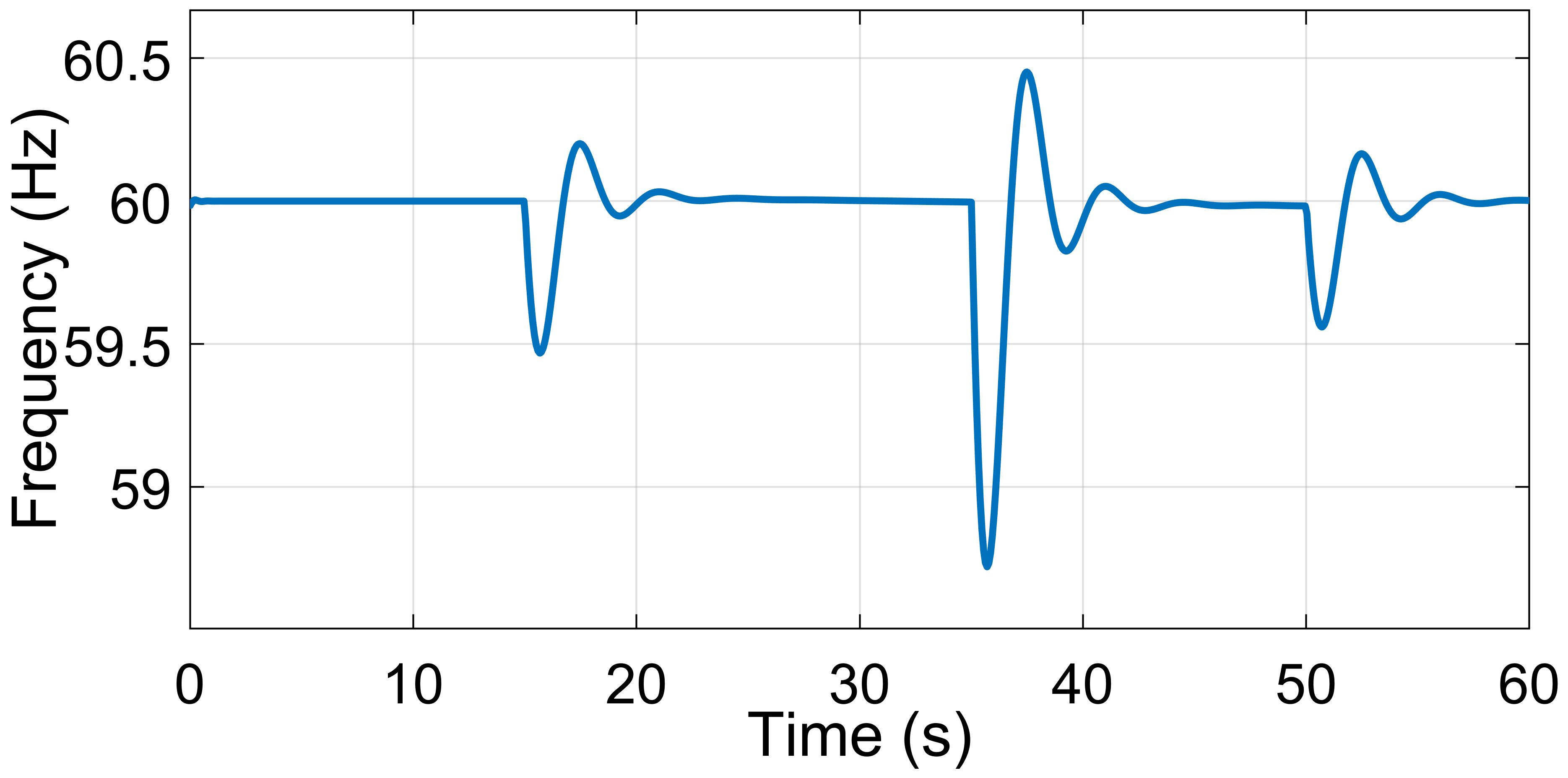}
            \label{subfig:freq}
    }
\vspace{-1mm}
 \caption[CR]{Cross-layer firmware attack impact on the power grid: 
(\subref{subfig:generation}) renewable source generated power, (\subref{subfig:voltage}) phasor and EMT magnitude of the MG voltage, and (\subref{subfig:freq}) MG operating frequency.} 
\vspace{-3mm}
\label{fig:CLFA}
\end{figure}

\begin{figure}[t]
\centering
\includegraphics[width = 0.48\textwidth]{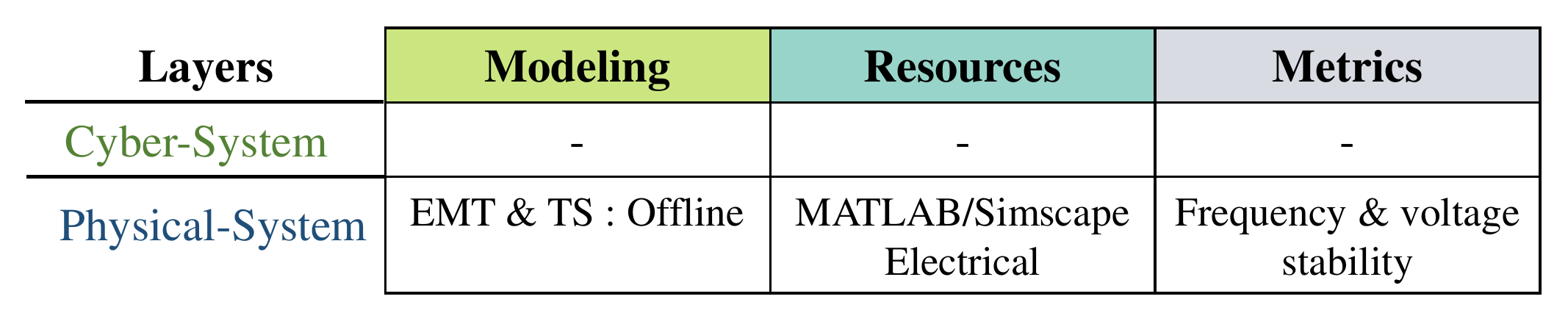}
\caption{\label{fig:crossmapping} Mapping of cross-layer firmware attack case study with CPS framework.}
\end{figure}

The metrics used to evaluate the performance and behavior of the MG operation, based on the presented CPS framework, are the physical-system layer performance metrics related to \textit{frequency stability} and \textit{voltage stability}. Fig. \ref{fig:CLFA} demonstrates the overall impact of malicious inverter operation on the MG, and how the grid's power, voltage, and frequency are affected. In more detail, we notice that at $t=35sec$ when a significant load increase in the MG occurs, the contribution of the anomalous inverter behavior significantly impacts the frequency causing potential stability issues. However, at $t=15sec$ and $t=50sec$, when the power generation of the inverter as well as its power contribution to the grid is much lower following the solar irradiance profile, the impact of the inverter's malicious behavior is reduced. Thus, from an adversarial perspective, targeting an inverter device during peak-hours when the solar generation is reaching its maximum can yield significant implications on the grid's operation. Fig. \ref{fig:MPPT_DC} shows the impact of the attack on the DC-side of the converter. It can be observed that both the DC voltage and current fluctuate, creating harmonic distortion at the output. Similarly, Fig. \ref{fig:MPPT_AC} demonstrates how the AC power generation is affected by the firmware modification attack. At $t=1sec$ the oscillatory behavior causes an islanding scenario that disconnects the PV system from the rest of the MG. Fig. \ref{fig:crossmapping} shows the mapping of the presented case study with the CPS framework.

It is important to note that in the presented case study, it is assumed that the cross-layer firmware attack is performed by an adversary with the capability of compromising the physical device, hence, modeling the cyber-system layer was not required. An extension of this study could involve the implementation of an over-the-air cross-layer firmware attack that compromises a device via the cyber-system layer. The implementation of such a scenario would also require the modeling of the cyber-system layer, i.e., the communication network that serves as the medium and entry point for the attack.

\vspace{1mm}
\noindent\underline{Risk Assessment}: Due to the inherent difficulty of getting simultaneous access to multiple devices in order to cause severe impacts on grid operation, the $ Threat \ Probability$ for this type of attack is set to \texttt{Medium (2)}. For the resulting damage part of the $Risk$ formula, we use the priorities indicated in Table \ref{table:priority}, and set the  ``People health and personnel safety'', ``Uninterrupted operation and service provision'', and ``Equipment damage and legal punishment'' attack impacts to \texttt{Low (1)}, while the ``Organization financial profit'' counterpart is set to \texttt{Medium (2)}. Thus, the comprehensive $Risk$ for the evaluated cross-layer firmware attack study is estimated to be $2* \sum(4+3+2+2) = 22$.

\subsection{Case Study 2: Load-changing attacks}

\noindent\underline{Background \& Formulation}: In load-changing attacks, an adversary triggers an unexpected or sudden demand increase or decrease of IoT connected high-wattage appliances and DERs, with the objective of causing grid instabilities \cite{ospina2020feasibility}. Although currently hypothetical, due to the low penetration rates of IoT-controllable high-wattage loads and DERs, load-changing attacks are projected to become a `real' threat in the near future as the number of controllable DERs and loads is anticipated to grow exponentially \cite{DER_outlook, antonakakis2017understanding, soltan2018blackiot}. Attackers able to install malware that could control DERs and load consumption, can therefore maliciously manipulate system operating conditions and affect the CPES. One example of such an attack can entail an adversary capable of synchronously switching on and off high-wattage devices at unexpected times, causing power, voltage, and frequency instabilities, i.e., an Aurora-type attack at the load side \cite{konstantinou2015impact}. This event could also potentially damage utility equipment or initiate cascading failures in distribution systems.

In terms of mathematical formulation, load-changing attacks can be framed as a DIA-type that maliciously modifies the control variables of loads in CPES, causing significant unexpected power variations that could, in turn, lead to circuit overflows or instabilities at certain vulnerable locations of the electric grid. This type of attack involves the malicious manipulation of high-wattage appliances and/or DERs that can significantly disturb the balance between power supply and demand. In order to perform this type of attack, we assume that the adversary accesses and controls multiple compromised elements through the cyber layer of the system, i.e., its communication network infrastructure, and then manipulates their control variables causing rapid fluctuations in the system's response. A load-changing attack is different from a `measurements-altering' DIA in the sense that, instead of measurements being affected, the control variables are the ones being directly manipulated by the adversary. Using the same CPS system described by Eqs. (\ref{eq:cpsx}) -- (\ref{eq:cpscyber}), the generalized DIA for the load-changing attack scenario is described by:
\begin{equation}
\label{eq:dialoadx}
    x_{a}(k+1) = Gx(k)+B\big(u(k)+\Delta u(k)\big)
\end{equation}
\begin{equation}
\label{eq:dialoady}
    y_{a} = C\big(x(k+1) + B\Delta u(k)\big) + e(k+1)
\end{equation}
\noindent where $x_{a}$ and $y_{a}$ represent the states and measurements, respectively, `altered' by the manipulation of the system's control variables $\Delta u$. 

In order to map the above formulation to the load-changing attack case within CPES, the term $u$ in Eq. (\ref{eq:dialoadx}) can be adapted to represent the controllable `altered' load demand in the system as:
\begin{equation}
\label{eq:loadchange}
    d_a(k) = d_{i}(k) + \Delta d(k)
\end{equation}
\noindent where $d$ represents the controllable load demand, $d_{i}$ is the initial `un-altered' load demand, $\Delta d$ is the portion of the total load demand affected by the attack, and $d_{a}$ represents the total load demand `altered' by the load-changing attack. If the attackers simultaneously compromise more than one load in the system, Eq. (\ref{eq:loadchange}) can be extended as:
\begin{equation}
\label{eq:loadchangegeneral}
    D_{T}(k) = \sum_{l=1}^{m} d_{i,l}(k) + \sum_{j=1}^{n} d_{a,n}(k) + P_{loss}
\end{equation}
\noindent where $D_T$ represents the total demand in the system, $m$ is the number of total `unaltered' loads, $n$ is the total number of loads compromised by adversaries, and $P_{loss}$ is the total loss in the distribution network. 

Based on the CPES requirement to balance load and generation in real‐time in order to maintain frequency stability in the system \cite{kundur1994power}, the summation of all generation output and all load demands and losses must be approximately equal:
\begin{equation}
\label{eq:balance}
    D_{T}(k) \approx \sum_{g=1}^{N_{g}} P_{g}(k)
\end{equation}
\noindent where $N_g$ represents the number of $g$ generators in the system. To understand the effect of sudden load changes in the frequency stability at each generator bus, we use the swing equations. The swing equations in Eq. (\ref{eq:swing}) -- (\ref{eq:Fequation}) describe the relationship between the input mechanical power ($P_m$), output electrical power ($P_e$), and the rotational speed of the generator ($\omega$) \cite{glover2012power}. The term $P_e$ is directly related to $P_g$, since it represents the generator power output plus electrical losses of the generating unit.
\begin{equation}
\frac{2 H}{\omega_{s}} \frac{d^{2} \delta}{d t^{2}}=P_{m}-P_{e}
\label{eq:swing}
\end{equation}
\begin{equation}
\frac{d \delta(t)}{dt}=\omega(t)-\omega_{\mathrm{s}}
\label{eq:FandPA}
\end{equation}
\begin{equation}
\frac{2 H}{\omega_{s}} \frac{d \omega(t)}{dt}=P_{m}-P_{e}
\label{eq:Fequation}
\end{equation}
\begin{equation}
P_e = \frac{V_s V_r}{X} sin(\delta)
\label{eq:Pe_V}
\end{equation}

\noindent In these equations, $H$ represents the constant normalized inertia, $\omega_{s}$ is the synchronous speed (i.e., 50 or 60 $Hz$), and $\delta$ is the power angle;  the angle between the generator's internal voltage, i.e., the voltage at the generator bus $V_s$, and its terminal voltage, i.e., the voltage at receiving bus $V_r$. $X$ is the reactance based on the classical model of a generator \cite{boldea2015synchronous}. The relationship between the electrical frequency $\omega(t)$ with the power angle $\delta$ is shown in Eq. (\ref{eq:FandPA}). Based on these relationships, any sudden change in load demand, caused by high-wattage loads turning on/off in the system, will affect $P_e$, and thus cause subsequent frequency fluctuations, as seen in Eq. (\ref{eq:Fequation}).  

\vspace{1mm}

\noindent\underline{Threat Model}: In the load-changing attack case study, the adversary is assumed to be either oblivious, i.e., having no knowledge of the system topology, or with limited knowledge. Such limited information regarding the CPES could assist in optimally coordinating the attack and could be acquired, for example, via open-source intelligence techniques. The adversary can perform the attack remotely, thus, \textit{non-possession} is presumed of the IoT devices controlling the high-wattage loads. Load-changing attacks are \textit{targeted} attacks aiming to destabilize grid operation by causing blackouts, voltage sags, and/or frequency fluctuations. As a consequence, determined adversaries with significant resources at their disposal (\textit{Class II} attackers) are required to successfully materialize such attacks.

As for the attack model of the load-changing scenario, the attack frequency component is considered \textit{iterative} due to the fact that in order to cause a significant effect on the system, a single attack incident may not be sufficient. The reproducibility of such stealthy and indirect attacks is set to \textit{multiple-times}. Furthermore, the attack functional level is at \textit{Level 1 or 2}, per the assets that are vulnerable and enable this load-changing scenario (e.g., PLCs, controllers, HMIs, etc.). Last, the attack techniques that the adversaries use can either include \textit{control logic modifications} if PLCs are targeted or \textit{wireless compromise} if a wireless controller is affected. In both cases, the attacks target the \textit{cyber domain}, and specifically, the integrity of the in-transit data issued from HMIs or SCADA MTUs (i.e., communications and protocols), or the control commands to PLCs.

\vspace{1mm}

\begin{figure}[t]
\centering
\includegraphics[width = 0.49\textwidth]{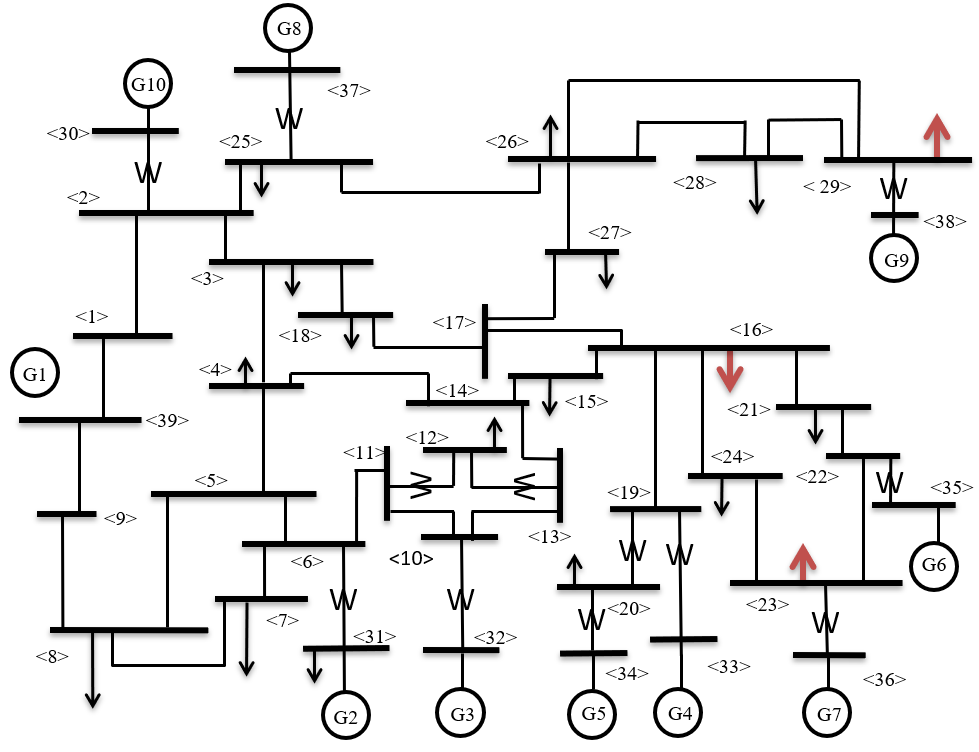}
\caption{\label{fig:D39system} Load-changing attack on IEEE-39 bus system. Load-changing attack targets are shown as red arrows in the system topology.}
\end{figure}

\noindent\underline{Attack Setup \& Evaluation}: In order to demonstrate the effects of load-changing attacks on CPES, we simulate such attacks targeting multiple load buses in the IEEE-39 bus system. Three vulnerable load buses (bus 16, 23, and 29) are selected as the targets for the load-changing attacks \cite{amini2016dynamic}, as shown in Fig. \ref{fig:D39system}.

\begin{figure}[t]
\centering
\includegraphics[width = 0.48\textwidth]{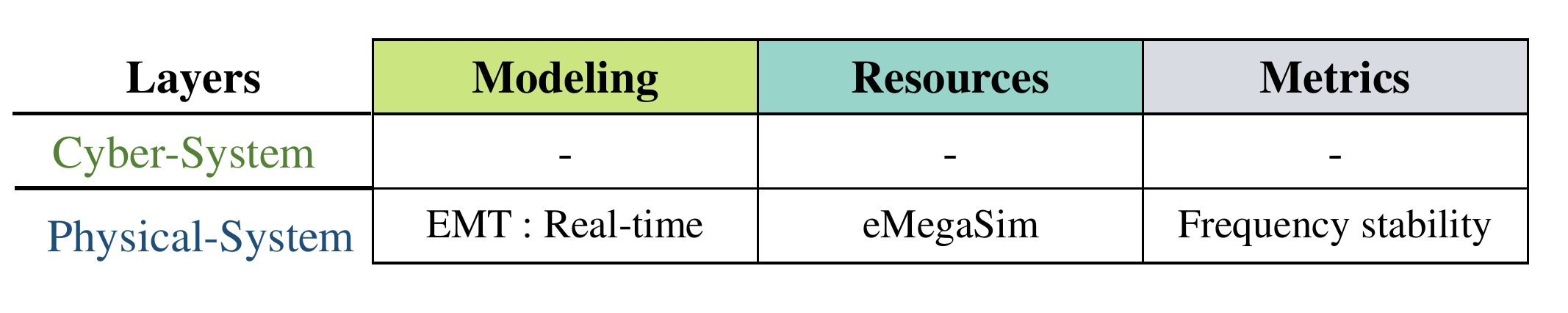}
\caption{\label{fig:demandmapping} Mapping of load-changing attack case study leveraging the CPS framework.}
\end{figure}

In this case study, it is important to examine the dynamic impact of frequency instabilities caused by load-changing attacks. Hence, to study these frequency instabilities, we model the physical-system layer using an EMT-approach with support from real-time simulation. At this layer, the generators are modeled as synchronous machines taking into consideration the dynamics of the stator, the field, and the damper windings. An excitation system is used for the system's control and protection functions designed to handle any disturbances measured in the power system \cite{singh2014development}. Loads are modeled as constant impedance, current, and power (ZIP) models \cite{bokhari2013experimental}. The mapping of this load changing attack case study to the CPS framework is presented in Fig. \ref{fig:demandmapping}. The main software resource used to conduct the EMT real-time simulations of the physical-system layer is \textsl{eMegaSim} (from \textsl{Opal-RT}). The metrics used to evaluate the performance and behavior of the IEEE-39 bus test system, based on the presented CPS framework, are the physical-system layer performance metrics related to \textit{frequency stability} (Table \ref{table:physical_metrics}).

\begin{figure}[t]
\centering
    \subfloat[]{
            \includegraphics[width=0.95\linewidth]{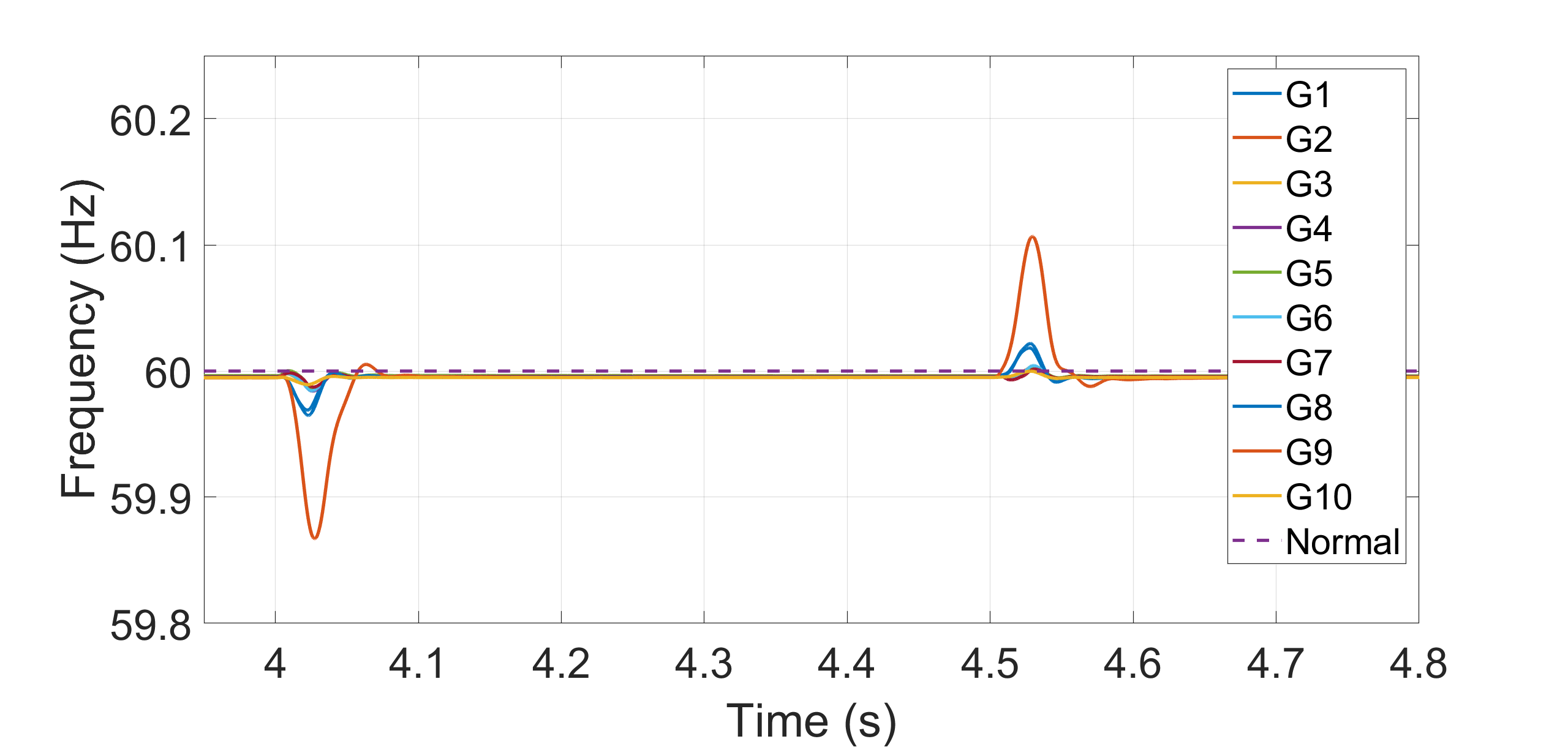}
            \label{subfig:1in}
    }\\
    \subfloat[]{
            \includegraphics[width=0.95\linewidth]{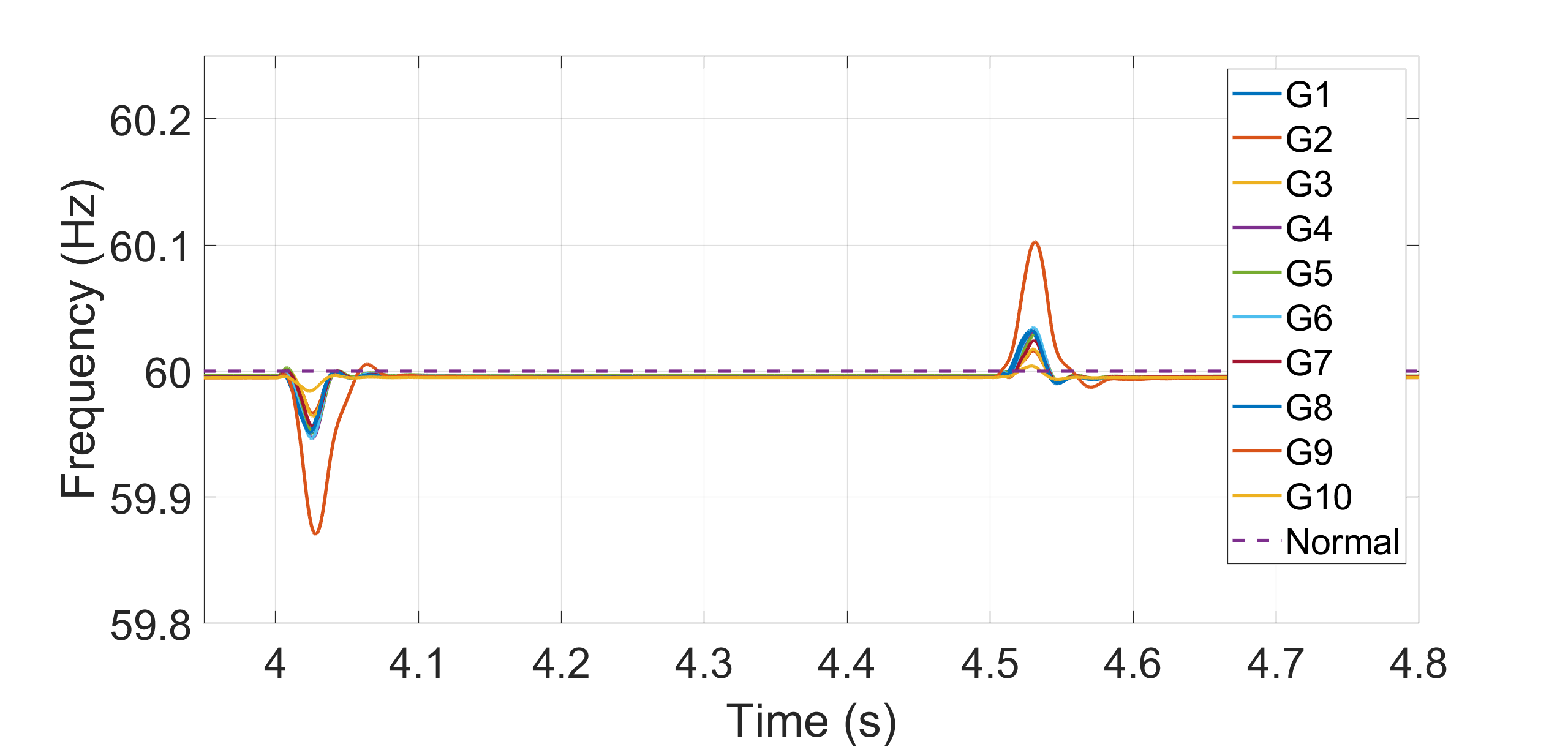}
            \label{subfig:2in}
    } \\
        \subfloat[]{
            \includegraphics[width=0.95\linewidth]{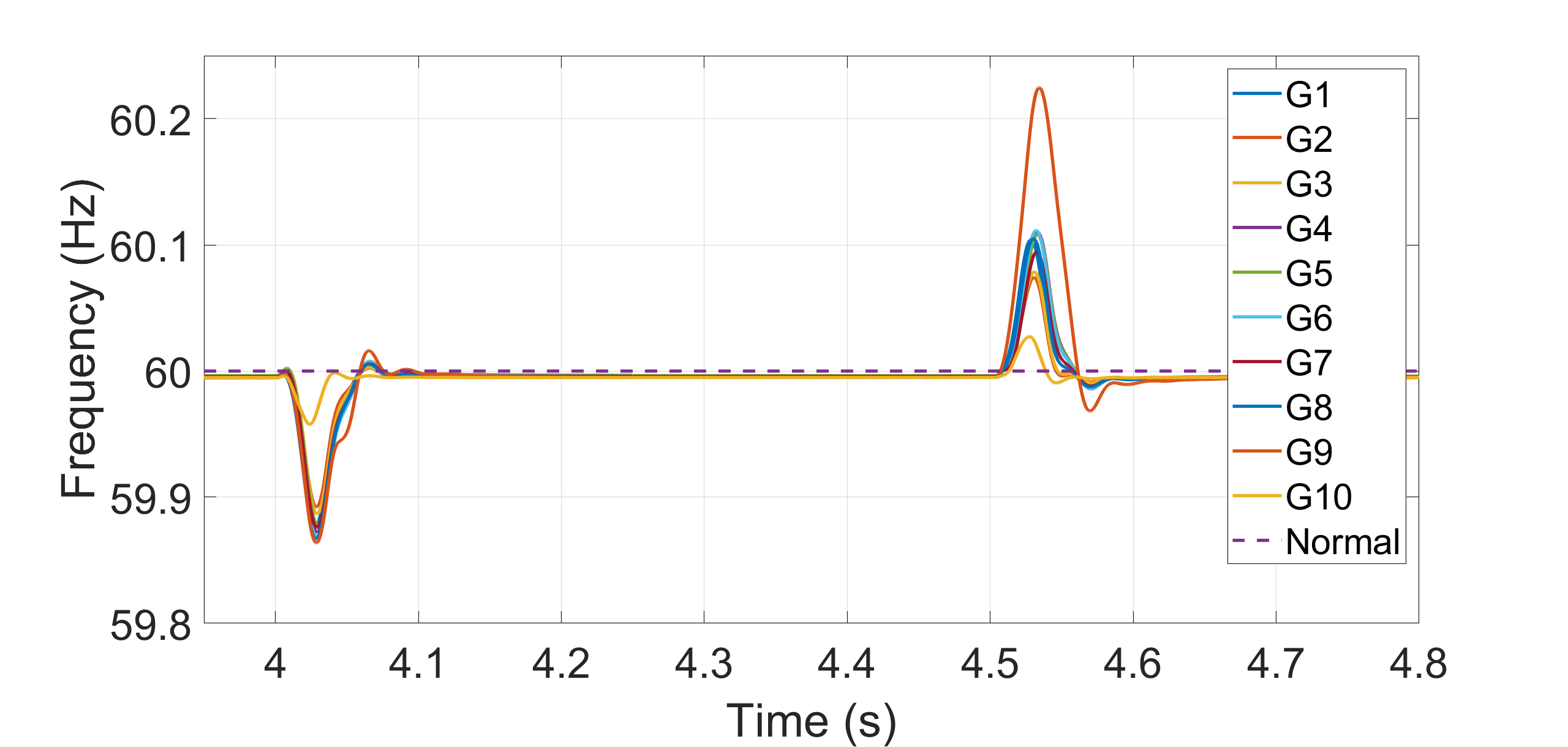}
            \label{subfig:3in}
    }\\
    \subfloat[]{
            \includegraphics[width=0.95\linewidth]{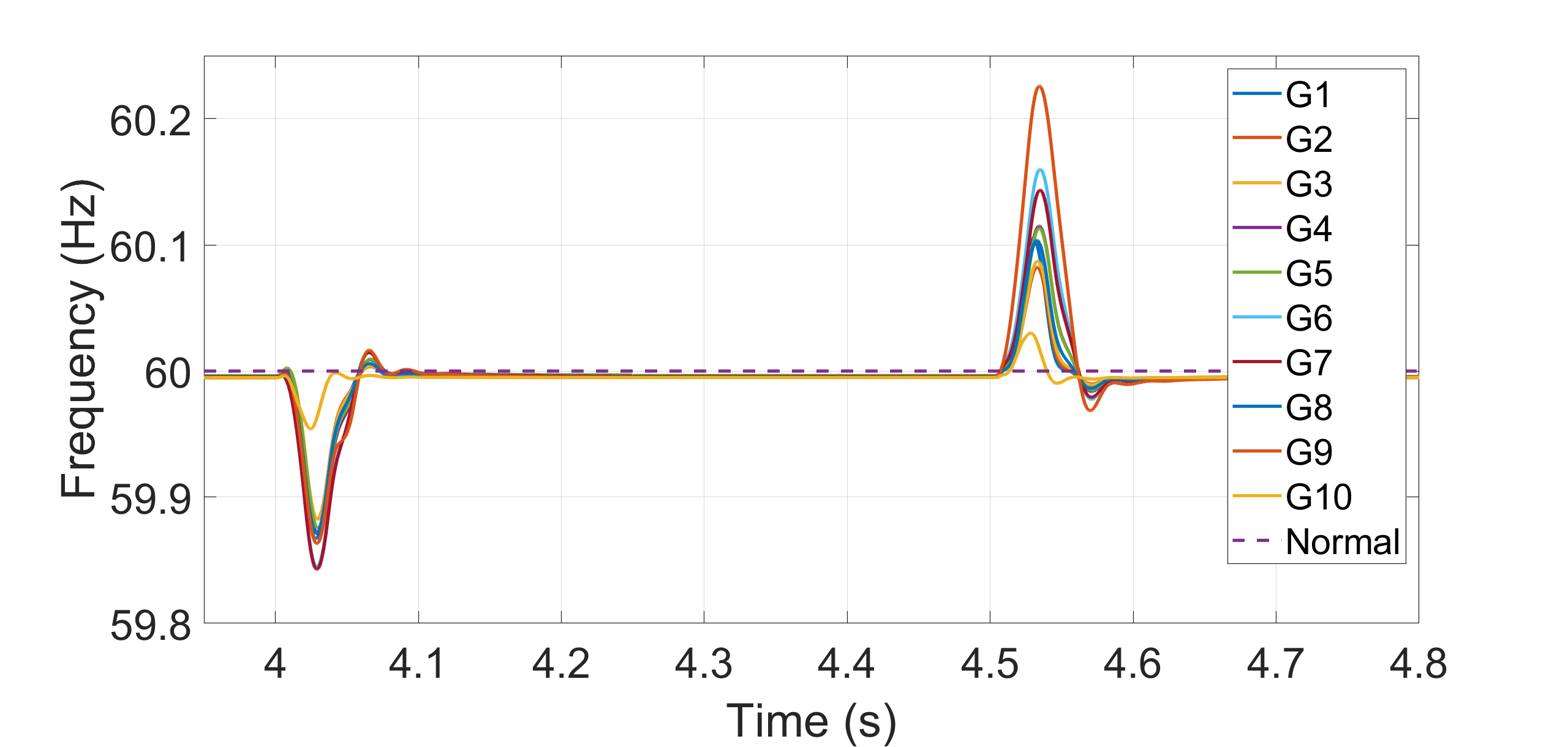}
            \label{subfig:4in}
    }  

\caption[CR]{Frequency variation impact on the power grid:  (\subref{subfig:1in}) with 20\% demand increased at bus 29, (\subref{subfig:2in}) with 20\% demand increased at buses 29 and 16, (\subref{subfig:3in}) with 50 \% load increase at buses 29 and 16,  and (\subref{subfig:4in}) with 50 \% load increase at buses 29, 16, and 23.} 

\label{fig:DS}
\end{figure}

In order to evaluate the impact of the load-changing attack on the power grid frequency, we observe the frequency variations measured at the generators' connections to the grid. We develop four different scenarios of load-changing attacks in which the system is initialized with original load values from literature  \cite{pai2012energy}, and the system frequency is kept at a nominal value of $60$ Hz. All load-changing attacks are triggered at $t = 4sec$ with a duration of $0.5sec$. Fig. \ref{subfig:1in} shows the effect of a 20\% load demand increase at bus 29. Such sudden load demand increase causes the measured frequency to decrease to around $59.87$ Hz on the nearby generator 9 while having a smaller impact on other generators. At $t = 4.5sec$, when the load demand increase is terminated, the frequency fluctuates and increases to around $60.11$ Hz at bus 29. Fig. \ref{subfig:2in} shows the results of a simultaneous load-changing attack that causes a 20\% load demand increase at buses 29 and 16. The main difference between this case compared to the first scenario is the higher number of generators that are affected by the attack.

A load-changing attack with greater system impact is depicted in Fig.\ref{subfig:3in}. In this scenario, an attack is simulated as a 50\% load demand increase that affects simultaneously buses 29 and 16. Here, we observe that the frequency measured at multiple generators approximately reaches $59.85$ Hz when the load-changing attacks are triggered at $t = 4sec$, and 60.23 $Hz$ when the load demand assumes nominal values ($t = 4.5sec$). The final scenario is shown in \ref{subfig:4in}, where we implement an attack that suddenly increases the load demand by 50\% at buses 29, 16, and 23. In this scenario, we observe how every generator in the system is heavily affected by the attack. The frequency measured at multiple generators reaches minimum and maximum values of $59.85$ Hz and $60.23$ Hz at the respective trigger and termination events of the attack. The most affected generators in this case study are generators 9 and 6.

\begin{figure}[t]
\centering
\includegraphics[width =\linewidth]{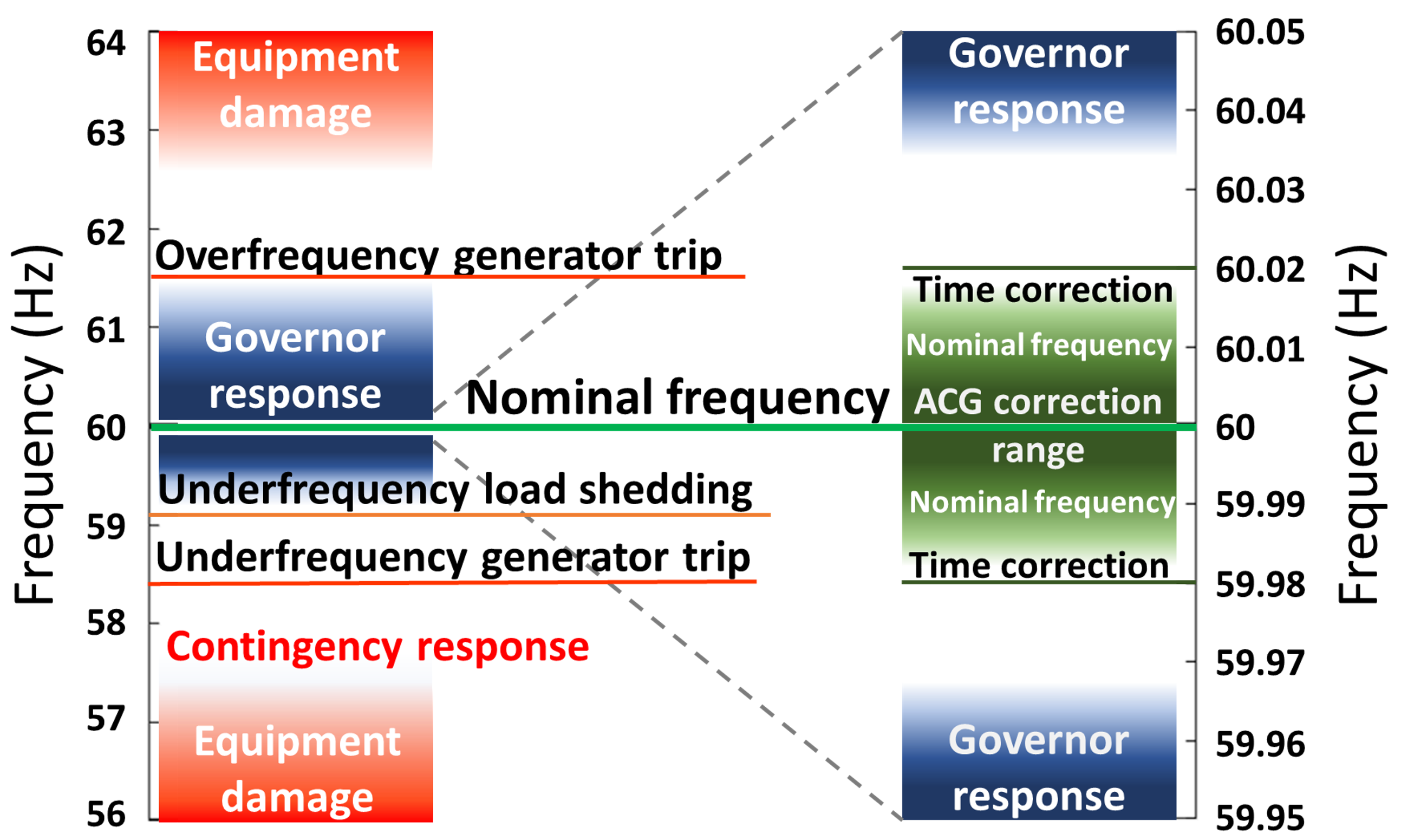}
\caption{\label{fig:Fre}Power system corrective mechanisms to maintain stability under different frequency deviations.}
\end{figure}

In the aforementioned scenarios, the attacker is assumed to be able to alter the power consumption profiles of IoT-connected controllable loads, and therefore cause sudden load demand increase. The presented results demonstrate the feasibility and impact of load-changing attacks. The frequency fluctuations from such adverse events can lead to exceeding the nominal EPS frequency limits \cite{nercfreq,NYISOfreq}, thus causing potential load-shedding incidents or equipment failures \cite{shekari2016adaptive}. As demonstrated in Fig. \ref{fig:Fre}, EPS have in-built control and protection mechanisms to maintain the power system frequency within its nominal range. For example, the AGC mechanisms can adjust minute frequency deviations from their nominal value. However, if the EPS frequency deviates more than  $0.036$ Hz from the predefined grid frequency (i.e., $60$ Hz), the generator governor systems are employed to account for such frequency discrepancies and stabilize the system. On the other hand, during more severe incidents, such as overfrequency (at or above at $62.2$ Hz) or underfrequency (at or below $57.8$ Hz) events, switching equipment and relays will automatically trip to protect generators from such instantaneous and potentially catastrophic frequency fluctuations \cite{NERCgenfreq}. Furthermore, during underfrequency incidents load shedding is typically employed to bring the system frequency within acceptable operational limits (between $58.4$ Hz-- $59.5$ Hz)  \cite{nercfreq}. An ancillary mechanism like the generator governors and AGC can then be utilized to bring the system back to its nominal frequency state. On the other hand, during severe events, where the frequency keeps decreasing even further, generators' CBs are tripped to protect the equipment from permanent damage.

\vspace{1mm}
\noindent\underline{Risk assessment}: Similar to case study 1, load-changing attacks require access to multiple devices to properly coordinate a successful attack. Thus, the $Threat \ Probability$ for this type of attack is set to \texttt{Medium (2)}. Following the same objective priorities depicted in Table \ref{table:priority}, we set the ``People health and personnel safety'', and  ``Equipment damage and legal punishment'' attack impact to \texttt{Low(1)}. On the other hand, since potential protection mechanisms could be triggered in the event of a load-changing attack causing potential brownouts, in order to avoid cascading system effects \cite{seyedi2009new}, the ``Uninterrupted operation, and service provision" as well as the ``Organization financial profit'' are set to \texttt{Medium (2)}. Consequently, the $Risk$ of the presented load-changing demand attacks is  estimated to be $2* \sum(4+2+6+2) = 28$.

\subsection{Case Study 3: Time-Delay Attacks}

\noindent\underline{Background \& Formulation}: Time-delay attacks (TDA) are a type of DAAs where attackers aim to destabilize the operation of a compromised control system by delaying measurements and/or control commands of sensors and actuators. This type of attack does not require a massive amount of attacker resources. For example, it can be implemented via network congestion, caused by flooding the network with a huge amount of data, thus disrupting the nominal operation of the attacked system.

The mathematical formulation of TDAs is formulated as follows. Consider the CPS system described by Eqs. (\ref{eq:cpsx}) -- (\ref{eq:cpscyber}). If $ \mathrm{T_{attack}}$ is defined as the period of time when the TDA is performed, then the TDA can be structured as: 
\begin{equation}
  f_D\big(s_r(k)\big) =
    \begin{cases}
      s_r(k-d) & \text{, if $k \in \mathrm{T_{attack}}$}\\
      s_r(k) & \text{, otherwise}
    \end{cases}       
\end{equation}
\noindent where $s_r$ represents the compromised signal (which can be either $u$, i.e., the control variable, or $y$, i.e., the measurements, in the CPS), $f_D$ represents a time-delay function, and $d$ represents either a discrete constant delay value or a time-varying delay function.

TDAs are considered a major threat to CPES due to their potential capability of disturbing the stability of islanded MGs, or even the overall power grid, by simply delaying measurements or control commands transmitted and received from sensing and control devices (e.g., smart meters, PMUs, etc.). Due to the importance of TDAs, existing literature aims to understand the complications such attacks could cause to CPES operations \cite{sargolzaei2013time, 10.1145/3411498.3422926, wang2017analysis}. For instance, in \cite{wang2017analysis}, the authors present an analysis of different TDA concepts (e.g., TDA margins, boundaries, surfaces, etc.) regarding effective conditions for TDA disruptions against grid stability. 

\vspace{1mm}
\noindent\underline{Threat model}: In the TDA case study, we assume an \textit{oblivious} adversary having essentially no knowledge of the system topology; such detailed information is not necessary to perform TDA events \cite{7352356, TEIXEIRA2015135}. Additionally, since this type of attack is performed by introducing substantial delays, mainly on the network level, \textit{possession} of the targeted device is \textit{not} required. Due to the objective of TDAs aiming to destabilize power grids by obstructing controls, crucial for the system's assets operation, TDA can be seen as a \textit{targeted} attack. Depending on the size and complexity of the compromised CPES, the adversaries might require fewer or an extensive array of skills and resources. Thus, adversaries' resources for performing TDAs can be classified in either Class I or Class II type of attackers. 

In order for TDAs to compromise CPES and severely impact their operation, TDAs should be performed \textit{iteratively} and \textit{multiple-times}. In addition, \textit{Level 2} assets are commonly the ones being targeted by TDAs. As mentioned before, typically, TDAs occur on the \textit{cyber domain}, i.e., \textit{communications and protocols}, and target asset availability by tampering with control commands issued by \textit{control server} devices. Consequently, \textit{wireless compromise, MitM, spoofing, and DoS attacks} are the most prominent techniques adopted by adversaries to cause anomalous incidents and cascading failures based on TDAs.  

\vspace{1mm}

\noindent\underline{Attack Setup \& Evaluation}: In this case study, we develop and simulate a TDA scenario in order to demonstrate its effect on a MG CPES. Specifically, in our study, a MG disconnects from the main grid by an intentional islanding command relayed from the MG controller at time ${t=10sec}$. Due to the insufficient generation capacity in the system, the MG controller sends a load shedding command to a breaker that controls a controllable load. At this point, the adversary performs a TDA that will delay this load shedding command sent from the MG controller to one of the controllable loads, thus causing major disturbances at the \textit{physical-system layer} of the CPES. The TDA occurs at the \textit{cyber-system layer} of the CPES, so for this particular case study, models for the cyber-system layer and the physical-system layer are required to perform a \textit{real-time co-simulation} of the respective layers.   

The physical-system layer is modeled using an EMT-simulation approach with support from real-time simulation. At this layer, the MG is modeled as a test system composed of a conventional generator operated using a frequency control mechanism rated at 1 MW, a Li-ion BESS rated at 100 kW/100 kWh, two controllable loads rated at 300 kW (load \#1) and 700 kW (load \#2), and a critical (non-sheddable) load rated at 200 kW. The main software resource used to conduct the EMT real-time simulations of the physical-system layer for this case study is \textsl{eMegaSim} (from \textsl{Opal-RT}). 
The cyber-system layer is modeled using a communication network emulation platform that supports co-simulation capabilities. Specifically, the software resource used to model the communication network that represents the cyber-system layer is \textsl{EXataCPS}.

\begin{figure}[t]
\centering
\includegraphics[width = 0.45\textwidth]{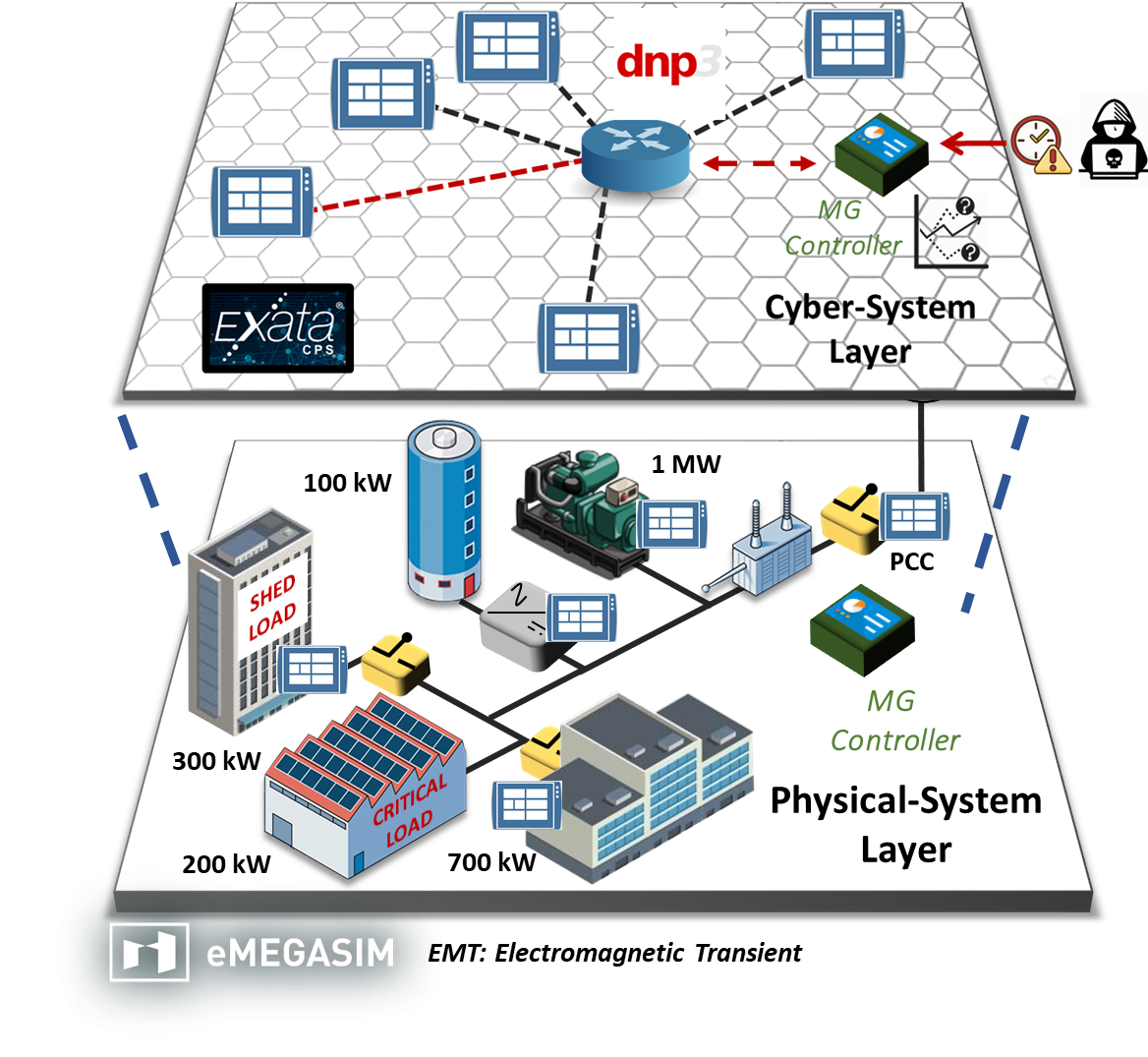}
\caption{\label{fig:micro_system}Conceptual illustration of the real-time co-simulation MG system testbed used in the TDA case study.}
\end{figure}

Every MG component from the physical layer is mapped with a virtual communication node inside the network emulation platform. The backbone of the communication network is represented by a network router. The network router is responsible for sending control commands and receiving measurements from the MG components, i.e., BESS, loads, and generator, to the MG controller, respectively. The communication protocol used is the IEEE Std 1815, commonly known as DNP3. IEDs in the network are modeled as DNP3 outstations and communicate with the MG controller which is modeled as a DNP3 master. The DNP3 master and outstation devices exchange data and control commands including, power generation, load consumption, breaker status, etc. The connections between the communication network nodes are modeled as wired 802.3 Ethernet connections with $100$ Mbps bandwidth. Fig. \ref{fig:micro_system} shows a conceptual illustration of the real-time co-simulation scenario designed to perform the described case study. The metrics used to evaluate the performance and behavior of the MG operation, based on the proposed CPS framework, are the physical-system layer performance metrics related to \textit{frequency stability}, and the cyber-system layer performance metrics related to \textit{average end-to-end delay} and \textit{total number of packets delayed} by the  TDA.

\begin{figure}[ht]
    \centering
        \subfloat[]{
            \includegraphics[width=\linewidth]{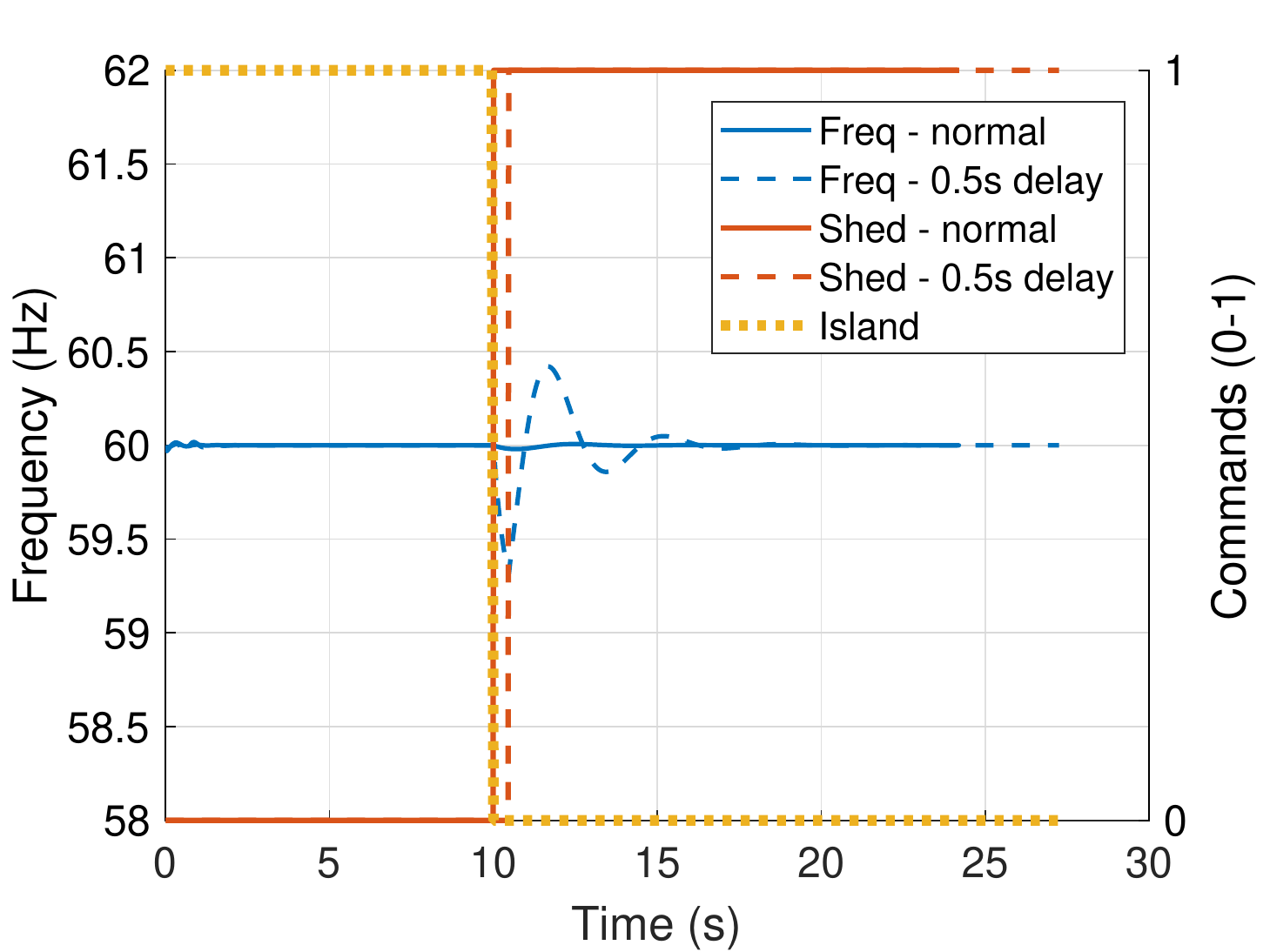}
            \label{fig:05sec_test_freq}
        } \\
       \subfloat[]{
                \includegraphics[width=\linewidth]{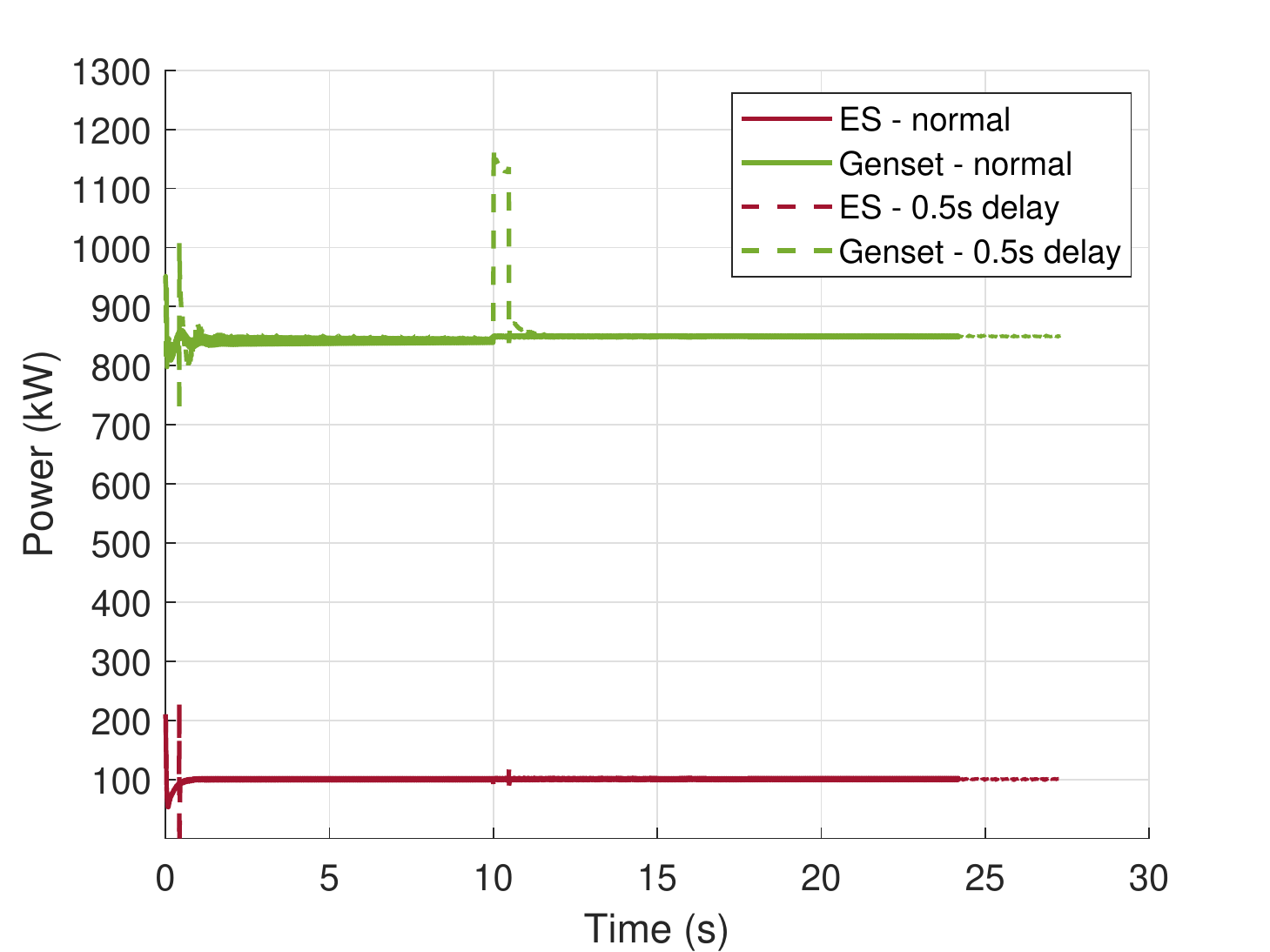}
                \label{fig:05sec_test_power}
        }
    \vspace{-1mm}
    \caption[CR]{Normal operation Vs. $0.5sec$ time-delay attack (TDA) scenario: 
    (\subref{fig:05sec_test_freq}) EPS frequency response during a $0.5sec$ time-delay attack (TDA) on the islanding command, and (\subref{fig:05sec_test_power}) generator power fluctuation during the $0.5sec$ time-delay attack (TDA).} 
\vspace{-3mm}
\label{fig:05sec_test}
\end{figure}

Based on the described setup, the impact of a malicious TDA in an islanded MG system is evaluated. An attacker compromises the communication link between the MG controller and the IED controlling the disconnection of the breaker at the controllable (sheddable) load \#1 (300 kW). Three different attack test cases are evaluated by varying the time-delay duration of the TDA. These delays are $0.5sec$, $5sec$, and $15sec$ approximately. In the communication network, the attacks are modeled by modifying the exchanged packets while introducing a timing delay between the DNP3 master and the corresponding outstation.

\begin{figure}[ht]
    \centering
        \subfloat[]{
                \includegraphics[width=\linewidth]{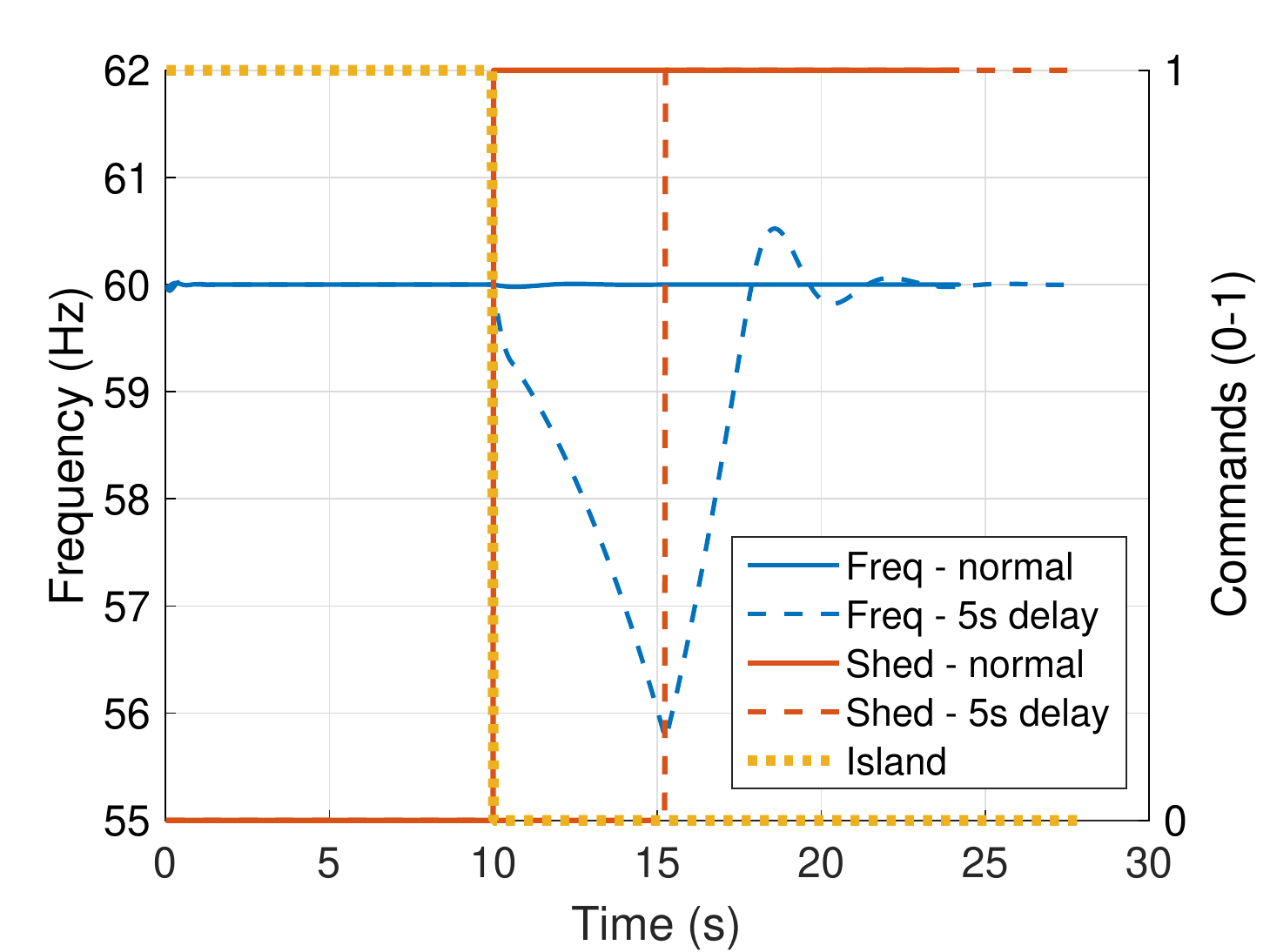}
                \label{fig:5sec_test_freq}
        } \\
       \subfloat[]{
                \includegraphics[width=\linewidth]{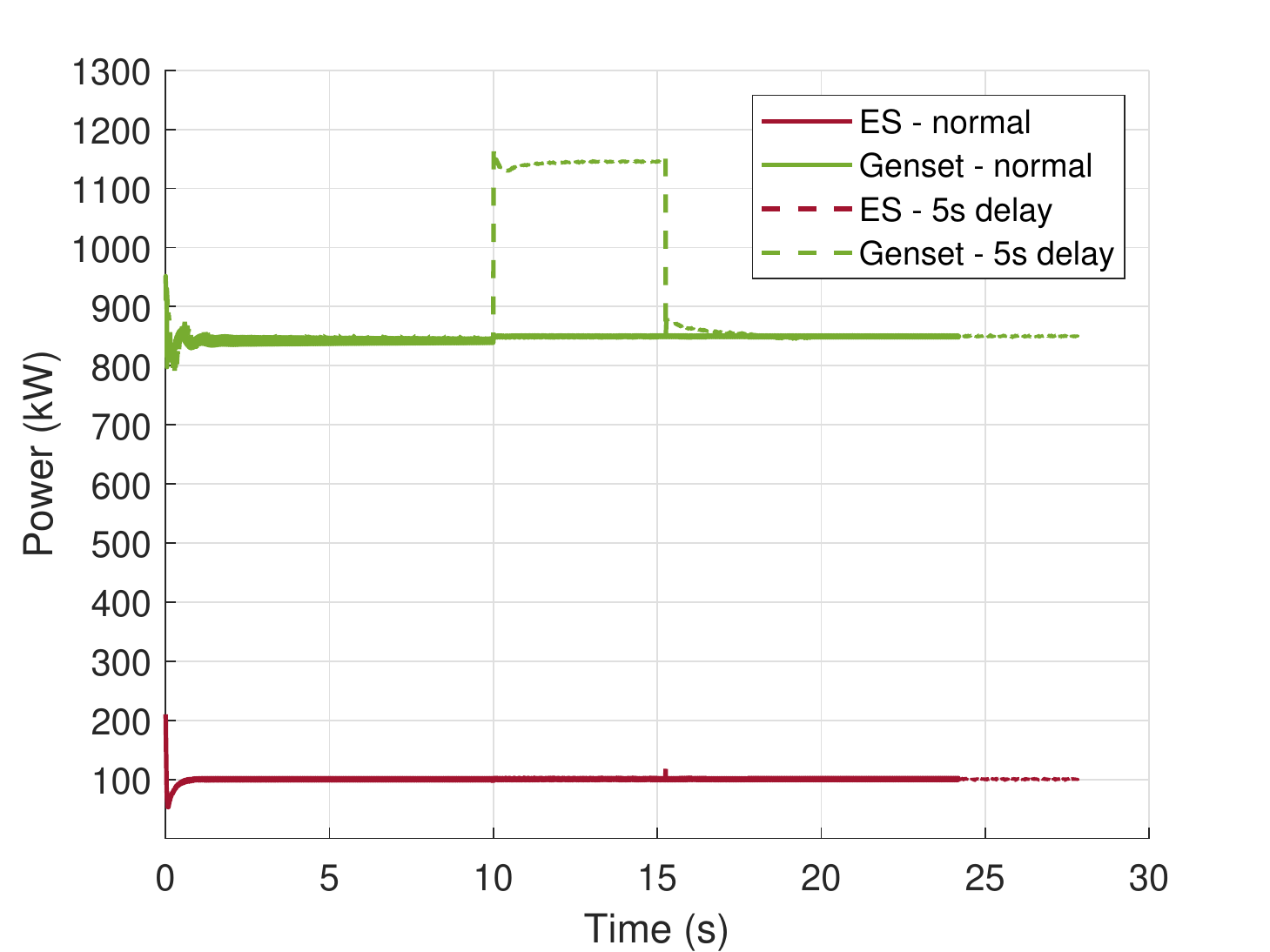}
                \label{fig:5sec_test_power}
        }
    \vspace{-1mm}
    \caption[CR]{Normal operation Vs. $5sec$ time-delay attack (TDA) scenario: 
    (\subref{fig:5sec_test_freq}) EPS frequency response during a $5sec$ time-delay attack (TDA) on the islanding command, and (\subref{fig:5sec_test_power}) generator power fluctuation during the $5sec$ time-delay attack (TDA).} 
\vspace{-3mm}
\label{fig:5sec_test}
\end{figure}

The first attack scenario shows a $0.5sec$ TDA that blocks the load shedding command performed by the MG controller. Fig. \ref{fig:05sec_test_freq} showcases the impact of the $0.5sec$ seconds TDA when compared to the normal operation of the MG system. In the graph, we observe how at ${t=10sec}$ the breaker at the point of common coupling (PCC) is disconnected, i.e., breaker command goes from 1 to 0,  in order to perform intentional islanding of the MG. Then, due to the insufficient generation capacity, the MG controller sheds controllable load \#1 (shed command goes from 0 to 1). In the normal operation case, the shedding procedure is performed as soon as the MG islands, while in the TDA scenario the shedding procedure gets delayed by the amount of the time-delay attack. 
Notably, the maximum and minimum values of the MG frequency during the normal operation scenario are $60.02$ Hz and $59.71$ Hz, respectively. On the other hand, the maximum and minimum values of the MG frequency during the $0.5sec$ TDA scenario are $60.42$ Hz and $59.32$ Hz, indicating (see Fig. \ref{fig:Fre}) that system operators would have to employ emergency corrective measures to maintain system stability. Fig. \ref{fig:05sec_test_power} depicts the output power of the generator set and the ESS during both scenarios.

\begin{figure}[ht]
    \centering
        \subfloat[]{
                \includegraphics[width=\linewidth]{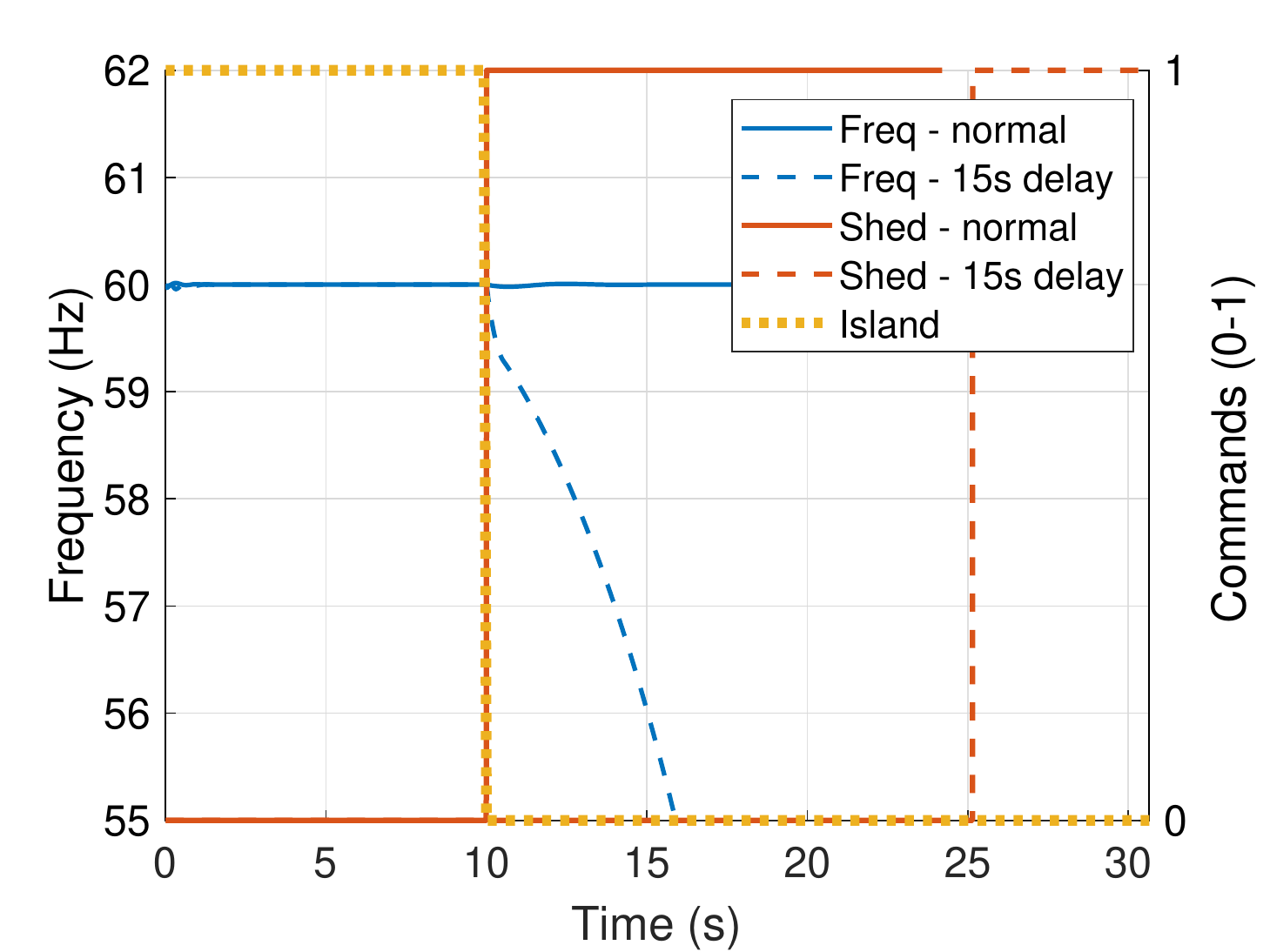}
                \label{fig:15sec_test_freq}
        } \\
       \subfloat[]{
                \includegraphics[width=\linewidth]{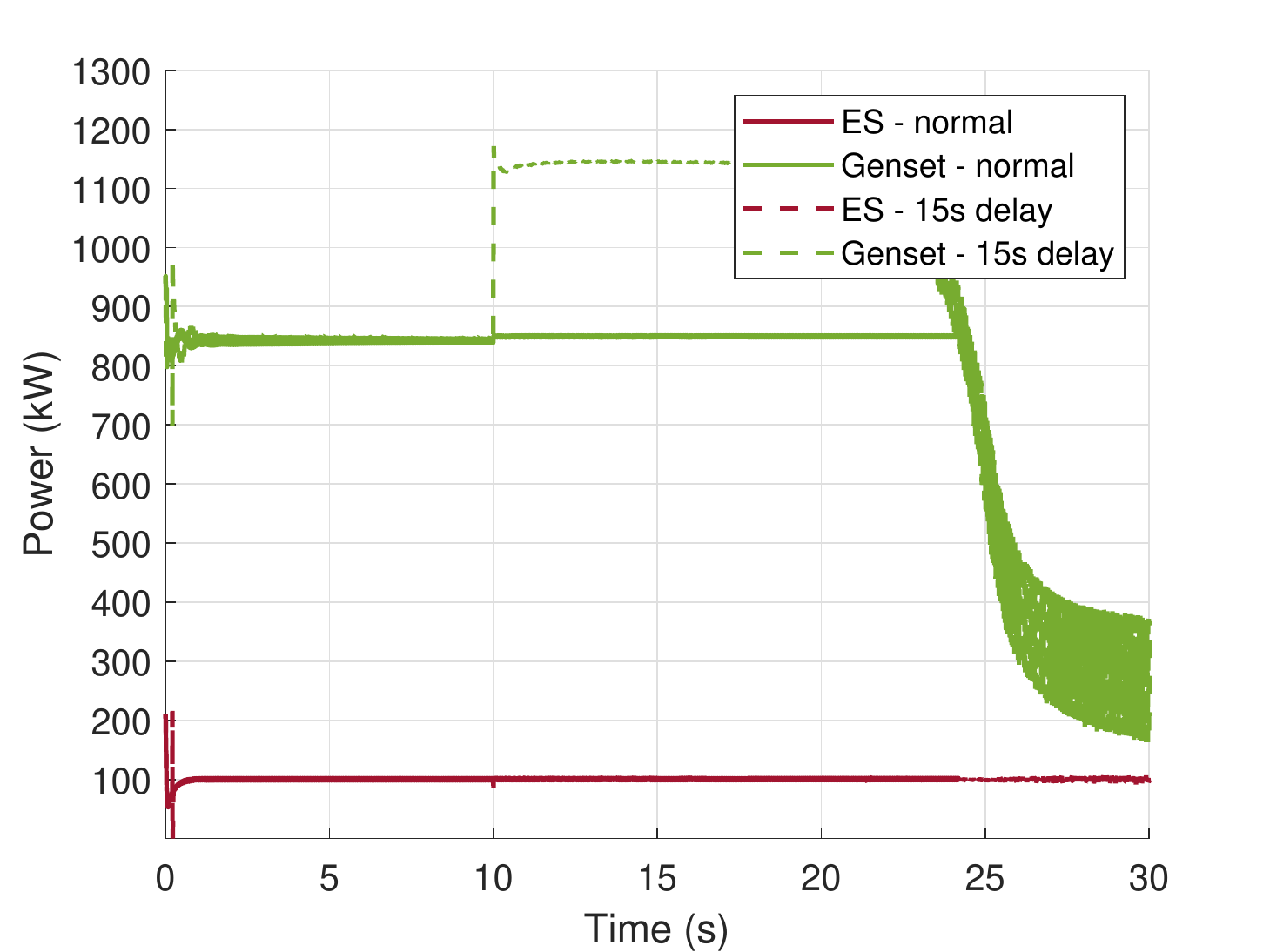}
                \label{fig:15sec_test_power}
        }
    \vspace{-1mm}
    \caption[CR]{Normal operation Vs. $15sec$ time-delay attack (TDA) scenario: 
    (\subref{fig:15sec_test_freq}) EPS frequency response during a $15sec$ time-delay attack (TDA) on the islanding command, and (\subref{fig:15sec_test_power}) generator power fluctuation during the $15sec$ time-delay attack (TDA).} 
\vspace{-3mm}
\label{fig:15sec_test}
\end{figure}

Similarly, the second test scenario demonstrates a $5sec$ TDA that blocks the load shedding command performed by the MG controller. Fig. \ref{fig:5sec_test_freq} presents the impact of the $5sec$ TDA when compared to the normal operation of the MG system. As seen, the impact on the operating frequency of the MG is greater than the first test scenario due to the sustained timing attack. The $5sec$ TDA causes a maximum and minimum MG frequency of $60.52$ Hz and $55.75$ Hz, respectively. Granted the substantial under-frequency incident, i.e., $55.75$ Hz, load-curtailment along with generator tripping would have to be enforced to protect the EPS equipment and avoid the incident propagation leading to a generalized grid collapse (Fig. \ref{fig:Fre}). As a result, this attack case demonstrates the potential of TDAs to greatly disrupt the operation of the system causing major equipment damages.

In the third test scenario, we perform a $15sec$ TDA that blocks the load shedding command performed by the MG controller. This case is analogous to a DoS attack, due to the long period of the TDA, which can greatly disrupt the operation of the MG's load shedding mechanism. As seen in Fig. \ref{fig:15sec_test_freq}, this scenario demonstrates the worst-case scenario of a TDA to the CPES. The MG frequency decreases rapidly until it hits a minimum value of $15.31$ Hz. Additionally, as depicted in Fig. \ref{fig:15sec_test_power}, the frequency-mode generator set is not capable of maintaining the stability of the system for such a prolonged period causing large oscillations in its power output. Notably, in realistic systems frequency violations should be averted before reaching such extreme values (e.g., $15.31$ Hz). However, by leveraging the CPES framework we can perform worst-case scenario analyses, evaluate the system behavior under coordinated attacks (e.g., if an attacker disables automated grid safety mechanisms), and identify critical system components and contingencies without endangering the EPS operation.

\begin{figure}[t]
\centering
\includegraphics[width = 0.46\textwidth]{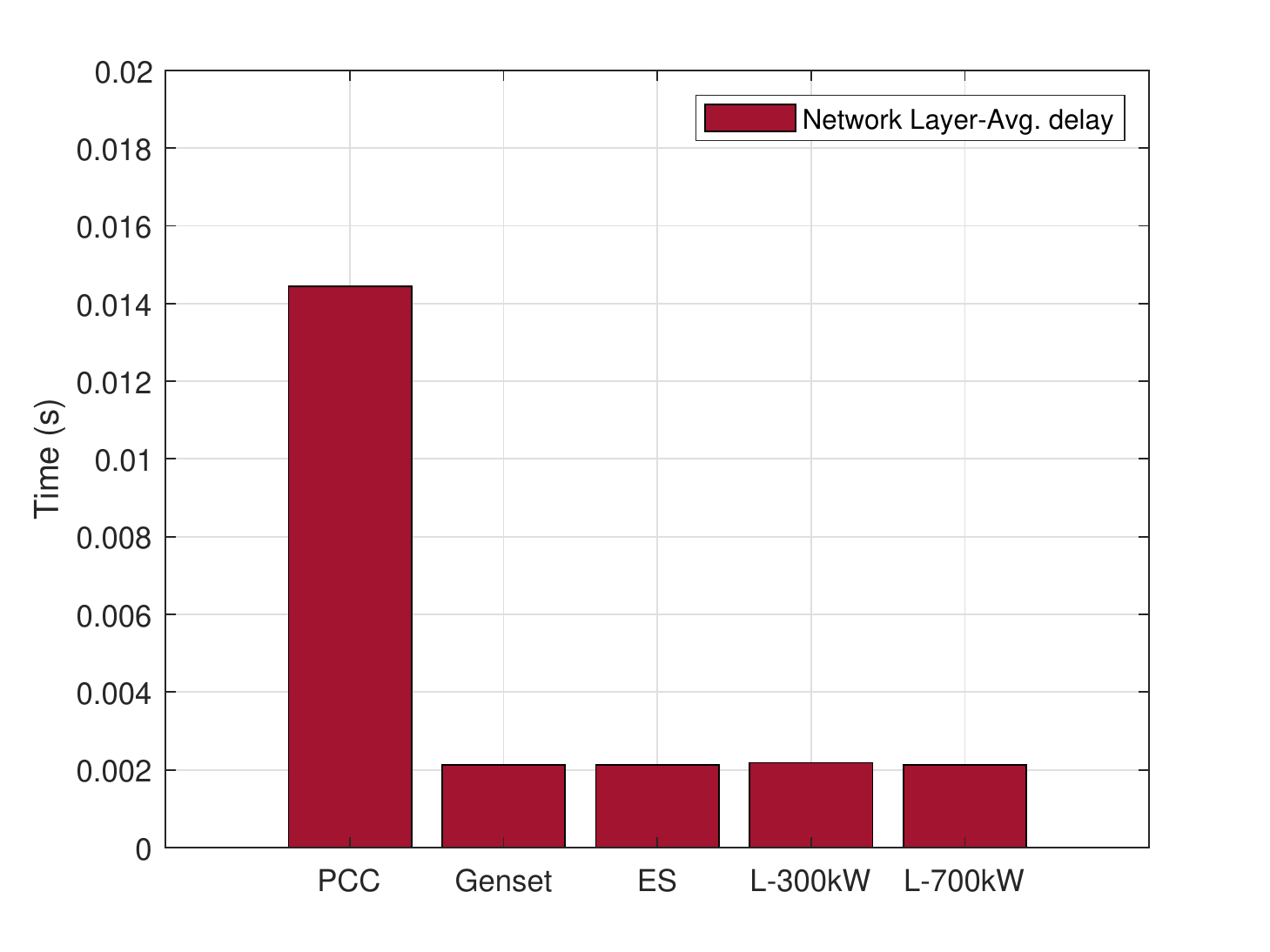}
\caption{\label{fig:comm_delay} Average end-to-end delay of all nodes in the communication network (cyber layer).}
\end{figure}

\begin{figure}[t]
\centering
\includegraphics[width = 0.46\textwidth]{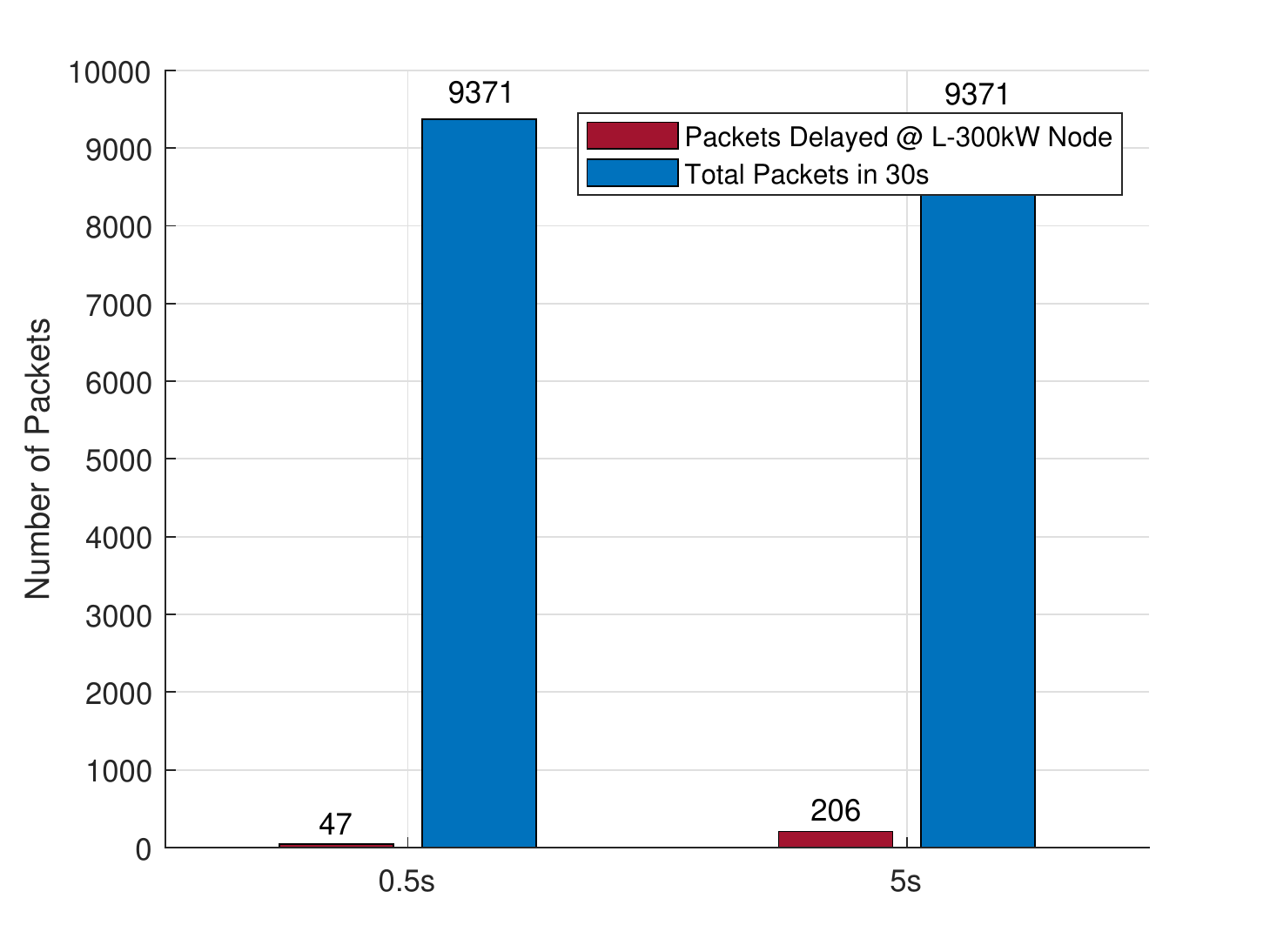}
\caption{\label{fig:comm_packets} Number of packets delayed by time-delay attack (TDA) at \textit{delay = $0.5sec$} and \textit{delay = $5sec$} Vs. \textit{total number of packets in $30sec$}.}
\end{figure}

\begin{figure}[t]
\centering
\includegraphics[width = 0.48\textwidth]{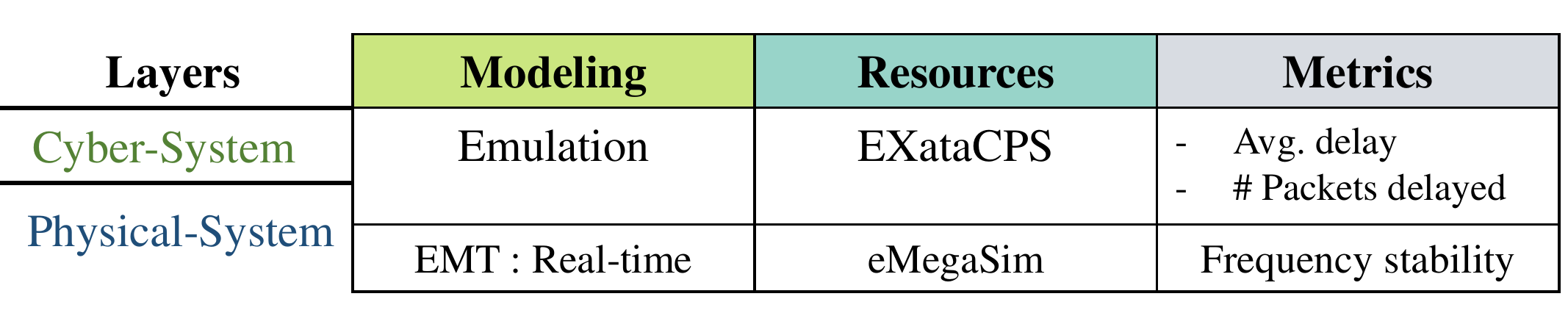}
\caption{\label{fig:tdamapping} Mapping of TDA case study with CPS framework.}
\end{figure}

In order to explore the behavior of the CPES at the cyber-system layer, we analyze two metrics that provide important information regarding the response of the communication devices to the TDA. These two metrics are the  \textit{average end-to-end delay} at the communication network, and the \textit{number of packets delayed} by the TDA. Fig. \ref{fig:comm_delay} shows the average end-to-end delay of all the network devices communicating using DNP3 at the cyber-system layer. Fig. \ref{fig:comm_packets} presents the total number of packets delayed due to the TDA that compromises the correct operation of the CPES according to two of the TDA scenarios ($0.5sec$ and $5sec$ TDA). As seen in Fig. \ref{fig:comm_delay}, the average end-to-end delay of the communication network, operating under normal conditions, has a maximum value of $0.0144sec$. This value is related to the \textit{master} DNP3 device located at the PCC that is communicating with all the DNP3 \textit{outstations}. This is the average time that the MG controller takes to communicate the load shedding signal to the respective sheddable load under normal operating conditions. In contrast, the TDA compromises the system's operation by delaying the load shedding signal based on the scenarios presented previously. In order to get more details regarding the attack study, the total number of packets delayed by the TDA are measured and plotted in Fig. \ref{fig:comm_packets}. Here, we observe a side-by-side comparison of the number of packets delayed in two of the presented test scenarios, i.e., $0.5sec$ and $5sec$ delay scenarios, and the total number of packets sent by the \textit{master} and \textit{outstation} devices in the $30sec$ real-time co-simulation.
Fig. \ref{fig:tdamapping} shows the mapping of the presented case study with the CPS framework.

\vspace{1mm}
\noindent\underline{Risk assessment}: In this type of attack, an adversary does not require significant resources or capabilities to compromise the CPES, as long as the system has not been fortified with state-of-the-art defense mechanisms. This ``low-bar'' requirement of resources increases the probability of successfully performing such an attack on a vulnerable CPES. As a result, the $ Threat \ Probability$ for the TDA case study is set to \texttt{High (3)}. The impact on ``People health and personnel safety'' as well as the ``Organization financial profit'' are set to \texttt{Low (1)}. However, TDAs can potentially cause severe impacts on the grid operation. For this reason, the ``Uninterrupted operation and service provision'' is set to \texttt{Medium (2)}, and the ``Equipment damage and legal punishment'' objective priority is set to \texttt{High (3)}. The resulting $Risk$ for the TDA is estimated to be equal to $3* \sum(4+1+6+6) = 51$.

\subsection{Case Study 4: Propagating Attacks in Integrated Transmission and Distribution (T\&D) CPES}

\noindent\underline{Background \& Formulation}: As mentioned in Section \ref{s:cps_framework}, T\&D integrated models for real-time simulation within CPES co-simulation testbeds can provide comprehensive and accurate simulation results able to capture the dynamic behavior of CPES. 
Specifically, integrated T\&D model co-simulations can be used to holistically evaluate the impact of disruptions (e.g., malicious attacks, faults, etc.) in EPS, and exhibit how maloperations on the transmission system extend to the distribution system and vice versa. Thus, in this case study, we present DIA-type attacks as propagating processes, similar to computer viruses, evaluated in real-time integrated T\&D simulation models. 

\begin{figure}[t]
\centerline{\includegraphics[width=\linewidth]{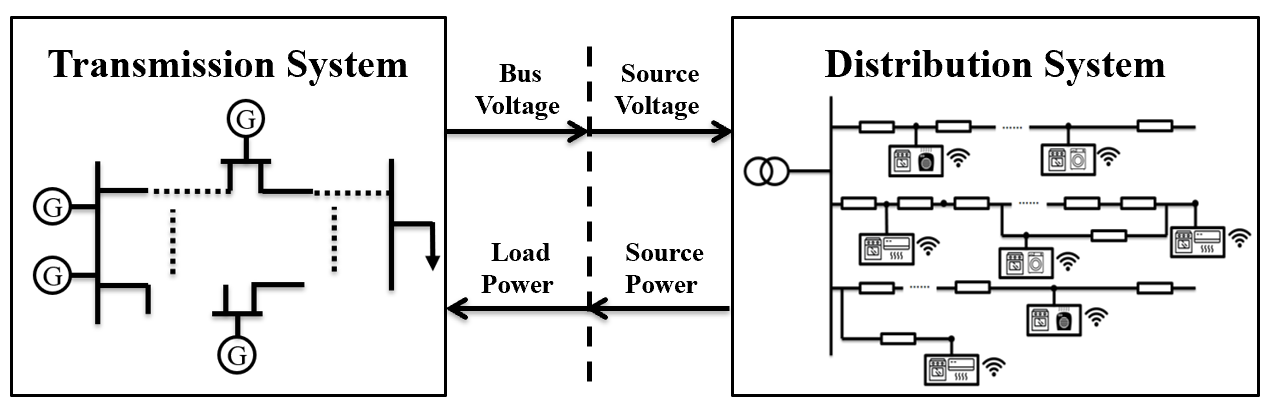}}
\caption{Transmission and distribution (T\&D) integrated simulation system setup.} 
\label{fig:TD}
\end{figure}

EMT and TS power system simulations often model only the transmission or the distribution system of power grids. This is mostly due to the high computational power required to have a real-time simulation model of an entire EPS \cite{huang2017open}. Aggregated distribution system sections are typically replaced by static or dynamic loads when simulating transmission system models \cite{kundur1994power}. Correspondingly, the transmission system's behavior is often abstracted using ideal voltage sources in distribution system modeling \cite{kersting2012distribution}. In addition, T\&D models are usually simplified to a single-phase representation \cite{emmanuel2017impact, pradhan2007fault}. Such modeling approaches lose key information related to the behavior of highly unbalanced distribution systems. In reality, T\&D systems are highly coupled \cite{singhal2017long}, and in order to perform  comprehensive and accurate security assessment and impact analysis studies in CPES, both T\&D domains need to be accurately modeled and simulated in a coordinated fashion. This coordination involves a clock-synchronized loop in which, even if the two models are executed on different cores of a machine, they communicate in parallel to match boundary conditions (i.e., voltages, power values, etc.) at every simulation step, as seen in Fig. \ref{fig:TD}.

There are different techniques that can be used to develop real-time integrated T\&D models. Different platforms provide different solutions and methods that allow the parallel execution of different systems in real-time EMT environments. In general, the overall T\&D system is separated into different groups (assigned to different cores of the machine) that are solved individually using a state-space approach.  State-space equations and matrices are used to describe the system group dynamics, while the interaction between the groups is solved using a nodal admittance method \cite{artemis_statenodal}. In the state-space approach the physical-system is modeled as:
\begin{equation}
    \label{statespace_s}
    s'=A_{q}s+D_{q}v 
\end{equation}
\begin{equation}
    \label{statespace_}
    o=E_{q}s+F_{q}v 
\end{equation}
\noindent where $s$ is the state vector, $v$ is the input vector, $o$ is the output vector, and $A$, $D$, $E$, and $F$ are the state-space matrices. The term $q$ represents the size of the matrices.

In a typical EMT state-space implementation, such as the one available in \textsl{Matlab Simscape Power Systems}, every time a switch changes status (on/off), the entire state-space solutions are re-computed. Using such an approach for real-time simulation ($<\approx$50$\mu$s simulation time-step) of large interconnected T\&D systems could be infeasible due to the required computational resources. With every single status change within the system model, the state-space outputs of the entire system would need to be re-computed. To address this computational issue, platforms such as \textsl{Opal-RT}, and its Advanced Real-Time Electro-Magnetic Solvers (\textsl{ARTEMiS}) package, use state-space nodal methods \cite{artemis_userguide}. \textsl{ARTEMiS} implementations discretize, pre-compute, and store into cache memory, the state-space matrices for all the combinations of switch topologies that can occur. Then, using a nodal method, the common voltages, admittances, and currents of the system (i.e., shared values between groups) are solved as:
\begin{equation}
    VY = I
\end{equation}
\noindent where $V$, $I$, and $Y$ are the respective common voltages, currents, and admittance matrices at the boundaries of the groups.  In essence, the use of this approach improves the accuracy and computational execution time of the entire system's solutions. As a result, this is a feasible way for simulating a real-time integrated T\&D system and evaluating the propagation impact of adverse disruptions, e.g., faults, attacks, etc.   

\vspace{1mm}

\noindent\underline{Threat model}: Integrated T\&D models can be seen as complex structures. Depending on the T\&D aspect targeted by an adversary and the type of the attack, the threat model may be adjusted to the specific details. For our use case, we assume an adversary with \textit{strong knowledge} of the system's topology and its components. Additionally, in our setup, the adversary aims to destabilize the integrated T\&D system by maliciously controlling switching devices, i.e., the CBs, thus \textit{possession} of the device is assumed. In the worst-case scenario analysis, the attackers could lead the CPES towards full system collapse, designating a \textit{targeted attack} by \textit{Class II} adversaries with abundant resources (e.g., nation-state funded groups).

In terms of the attack model formulation, the attack frequency is \textit{non-iterative}, since compromising a critical system asset (crown-jewel) could impact the overall system. The reproduction of such types of attacks can be seen as impractical due to their high system impact. Thus, we model them as \textit{one-time attacks}. The attack level is presumed to be \textit{Level 2} since critical system components need to be compromised. Such assets for our case include engineering workstations since the attacker targets -- in a DIA-type event -- the control and coordination between the T\&D systems. The attack technique is correspondingly an \textit{engineering workstation compromise}. Directly issuing malicious commands from an engineering workstation can also be a possible attack path, assuming a malicious insider scenario. However, in our case study, we assume a sophisticated and stealthy attack implemented on the \textit{cyber domain} targeting the data integrity of the issued control commands from the engineering workstations (DIA). For instance, disruptions on the T\&D can occur by falsifying the in-transit data exchanged between engineering workstations and CB control devices, triggering unexpected CB tripping and system sectionalization.

\vspace{1mm}

\begin{figure}[t]
\centerline{\includegraphics[width=2.8in]{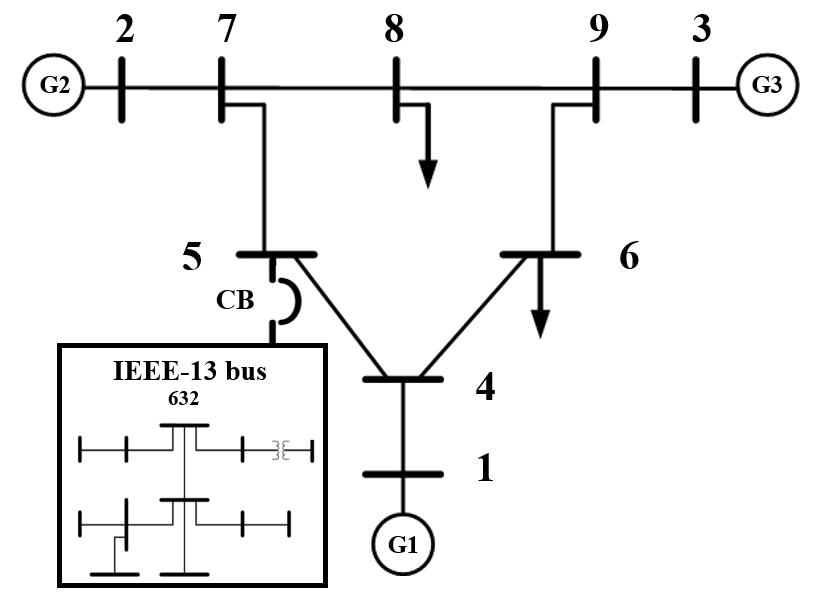}}
\caption{The integrated T\&D real-time simulation model: it includes the IEEE-9 bus system (transmission model) and IEEE-13 bus system (distribution model).} 
\label{fig:TDmodel}
\end{figure}

\noindent\underline{Attack Setup}: In this case study, an integrated real-time EMT T\&D system is modeled in order to investigate different interactions of propagating attacks and disturbances between a transmission and an unbalanced distribution system. 
Specifically, we integrate a transmission system, modeled as the IEEE-9 bus system, with a distribution system, modeled as the IEEE-13 bus test system. In order to match the power generation and load consumption between the power grid benchmarks, we scale some of the systems' parameters. For example, the active power and reactive power of the generators and the loads in the transmission system are reduced by an order of magnitude, while all the loads in the distribution system are increased by an order of magnitude. Additionally, as shown in Fig. \ref{fig:TDmodel}, the load at bus 5 of the IEEE-9 bus transmission model is `replaced' by the IEEE-13 bus distribution system. Generator 1 (G1) is used as the slack bus. The EMT modeling and real-time simulation of this case study's physical-system layer are performed using \textsl{eMegaSim} of \textsl{Opal-RT}. 

In order to evaluate the bi-directional impact of propagation attacks in integrated T\&D models of CPES, we develop two attack scenarios in this case study. The first scenario assumes that the adversary has the capability of altering the EPS topology. This can be achieved by decoupling the T\&D system at the PCC via a DIA attack on the EPS switch devices (i.e., the distribution feeder CBs). The second scenario demonstrates the impact on the distribution system when transmission system components are compromised. Following our CPS  framework, the metrics used to evaluate the performance and behavior of the T\&D system under the propagation attack scenarios, are the physical-system layer performance metrics related to \textit{frequency stability} and \textit{voltage stability}.

In the first propagation attack scenario, we assume that the adversary by tripping the CBs at the PCC between the T\&D system can disturb the EPS frequency impacting its operation and potentially causing damages to field equipment (e.g., transformers, commercial and residential loads, etc.). Different attack paths can be pursued to compromise and decouple T\&D systems. For instance, such adversarial objectives (i.e., T\&D decoupling) can be achieved by \textit{i)} intruding via the communication infrastructure and remotely manipulating the control tags issued by engineering workstations located at the system operation management facilities, \textit{ii)} implementing DoS attacks on the targeted PLCs, disabling the CBs, \textit{iii)} compromising the controller logic of IED-enabled switching equipment, or \textit{iv)} penetrating the utility SCADA network and maliciously manipulating control settings (e.g., over/under -voltage or current limits)  \cite{adepu2019attacks}.

\begin{figure}   
\centering
 \subfloat[]{
\includegraphics[width=3.4in]{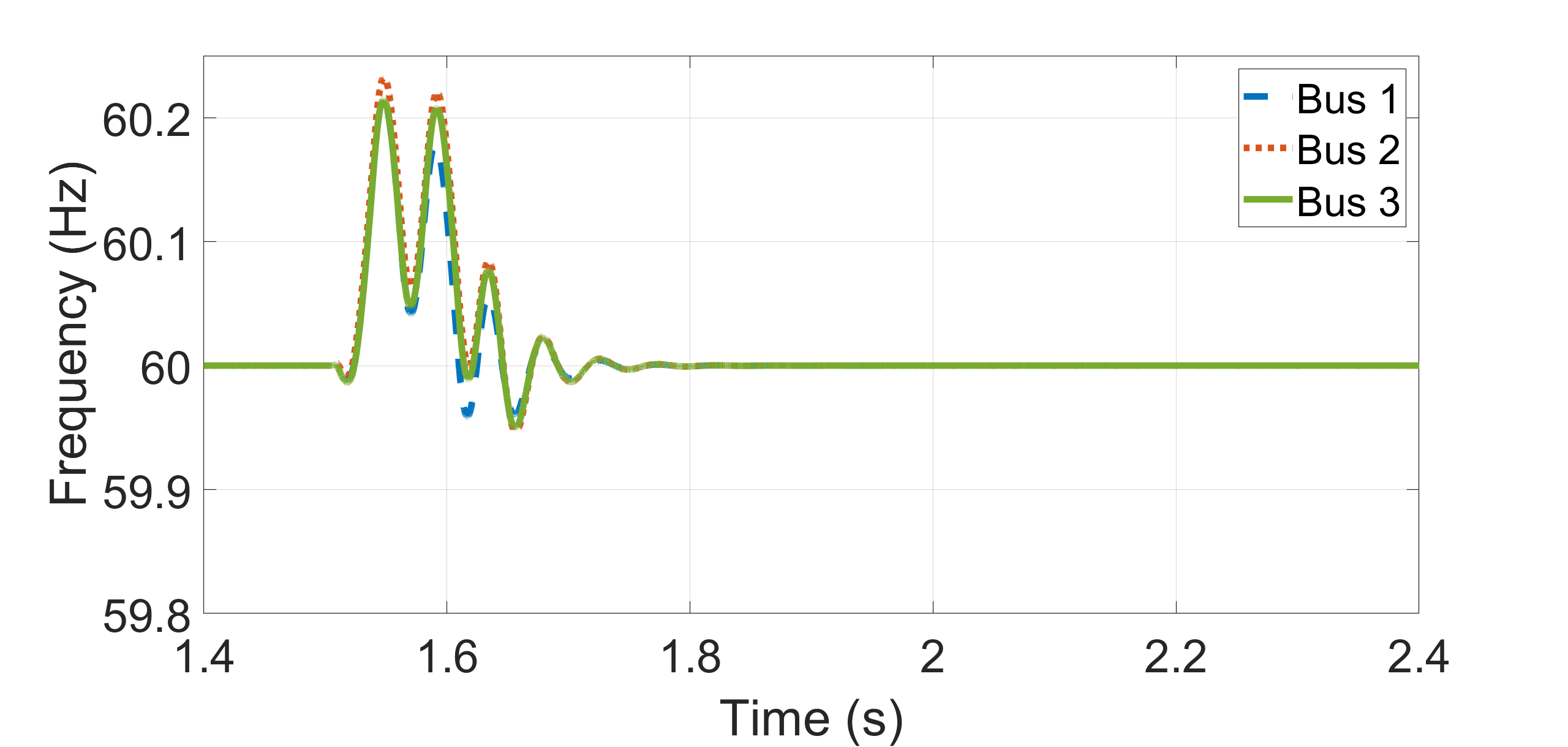}  
\label{subfig:Frequency1}}\\
\subfloat[]{
\includegraphics[width=3.4in]{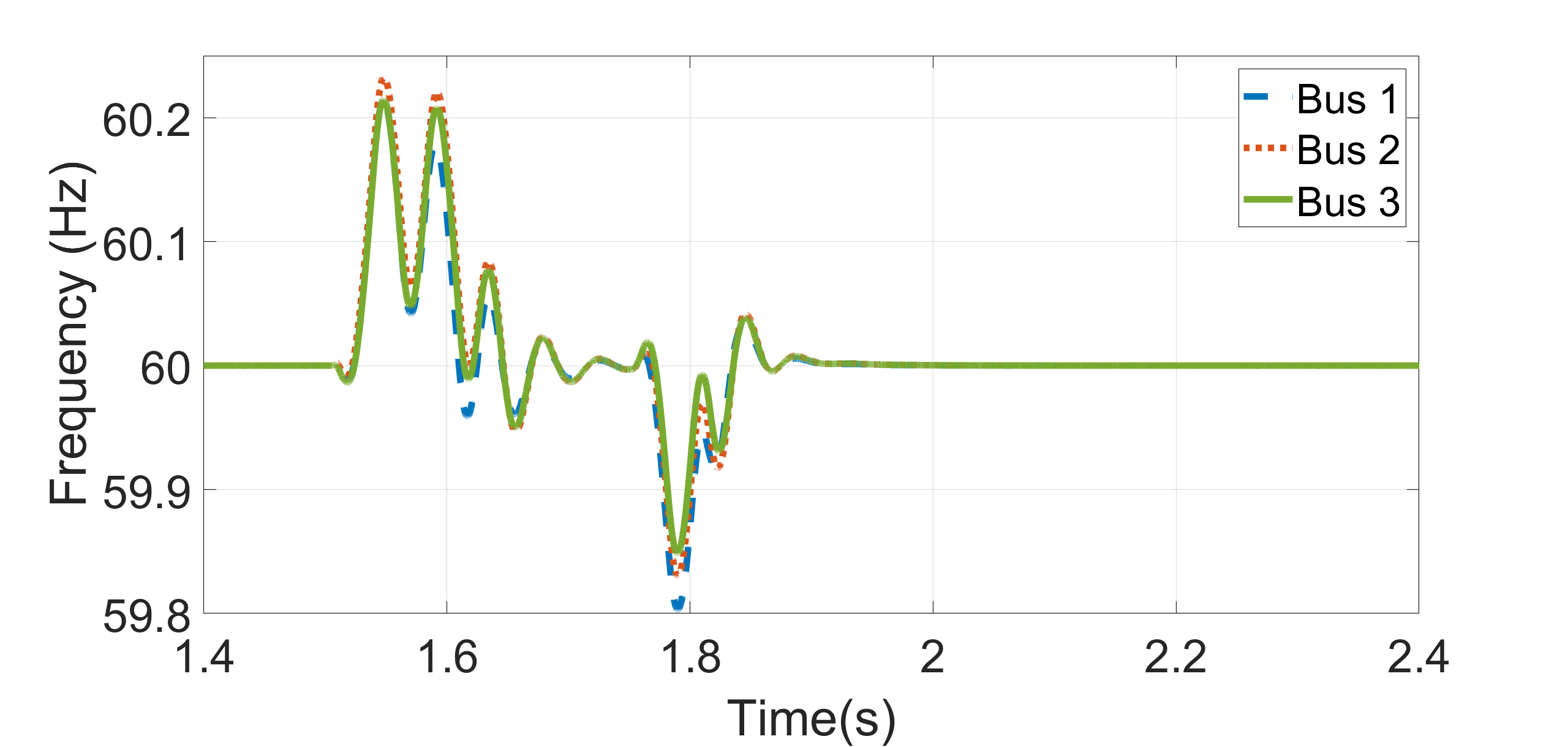}   
\label{subfig:Frequency2}}\\
\subfloat[]{
\includegraphics[width=3.4in]{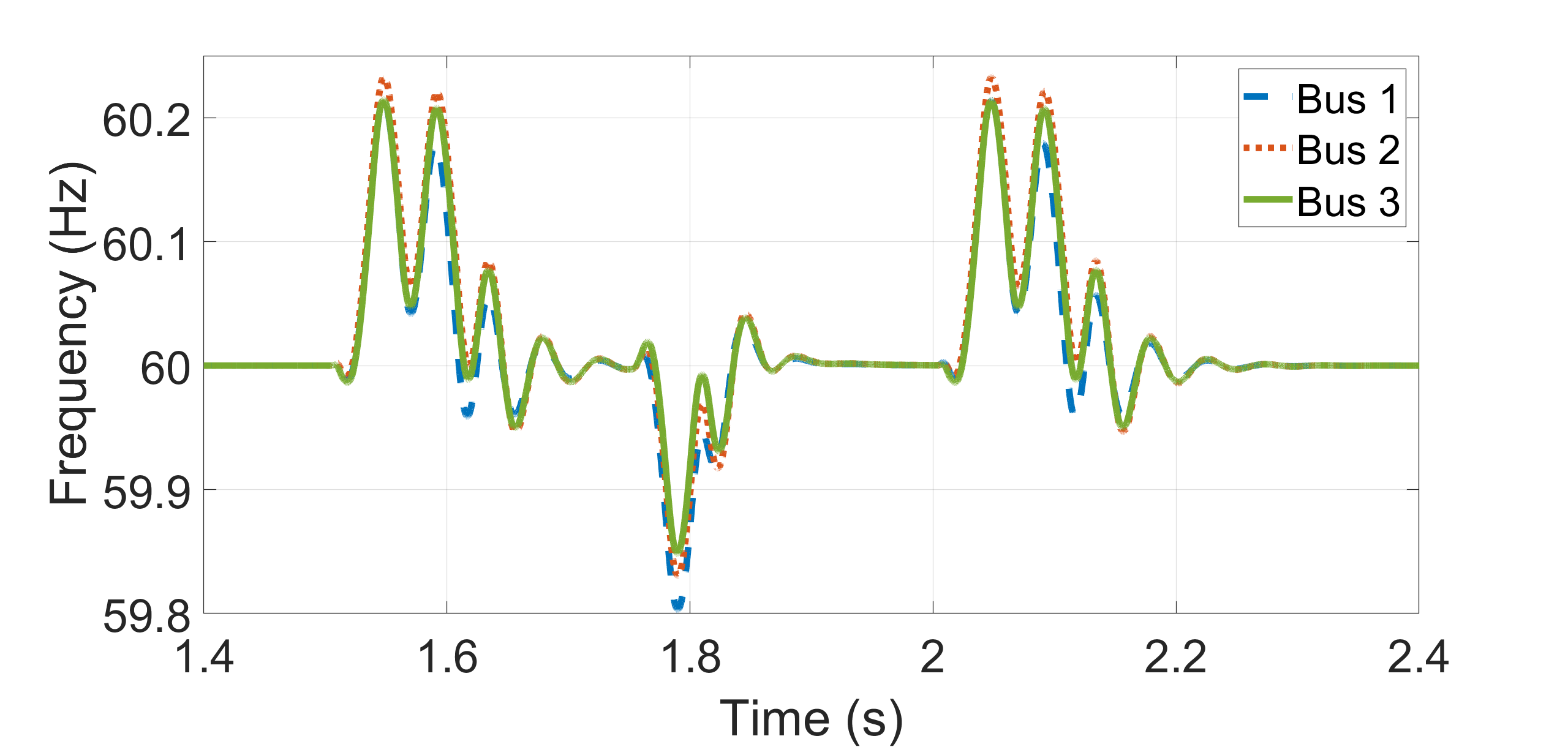}   
\label{subfig:Frequency3}}
\caption{Frequency response when the CB at bus 5 of the IEEE-9 bus system is: (\subref{subfig:Frequency1}) opened at $t=1.5sec$, (\subref{subfig:Frequency2}) opened at $t=1.5sec$ and then closed at $t=1.75sec$, and (\subref{subfig:Frequency3}) opened at $t=1.5sec$, closed at $t=1.75sec$, and opened again at $t=2sec$.} 
\end{figure}

The results presented in Fig. \ref{subfig:Frequency1} are measured at the generator buses, i.e., buses 1, 2, and 3 of Fig. \ref{fig:TDmodel}. The frequency at the transmission side of the network rapidly increases when the CB is tripped at $t=1.5sec$, and returns to its nominal values at around $t=1.8sec$. The peak frequency value is around $60.23$ Hz. Fig.\ref{subfig:Frequency2} shows the frequency response of the system when the attacker opens the CB between the T\&D system at $t=1.5sec$, and then closes it after $15$ cycles (approximately $0.25sec$ later) which would avoid triggering any protection countermeasures during this intermittent frequency transient \cite{konstantinou2015impact}. We observe how such attacks could stealthily destabilize the EPS just by tampering with the CB controls between different zones of the power grid. Here, two main fluctuations are observed following the CB tripping behavior, one between $t=1.5sec$ and $t=1.75sec$ when the CB is tripped open, and between $t=1.75sec$ and $t=2sec$ when the CB is closed. The last scenario assumes an attacker aiming to damage system components by asynchronously changing the status of the CB multiple times. Fig. \ref{subfig:Frequency3} demonstrates the frequency fluctuations on the generator buses when the CB between the T\&D systems is opened at $t=1.5sec$, closed at $t=1.75sec$, and then opened again at $t=2sec$. Notably, if safety mechanisms are not promptly enforced, the frequency instabilities occurring between $t=1.5sec$ and $t=2.4sec$ could affect frequency-sensitive grid components (i.e., consumer,  commercial, and industrial loads), and impact grid equipment and control functions (e.g., generators, transformers, automated voltage control, etc.).

In the second scenario, it is assumed that an adversary has the capability of compromising components at the transmission side of the power grid. For example, such types of attacks have been experimentally evaluated and indicate that if they last around three minutes, they can cause permanent damage on generators \cite{aurora}. In this use case, our aim is to evaluate how an attack on the transmission level can propagate on the distribution level and manifest as a voltage variation. Contingency analysis is employed in the integrated simulation environment to study the effects of transmission-side adverse events on the distribution system. 
In more detail, contingency analysis is a simulation-based system analysis tool used to assess the impact of various combinations of component failures occurring in transmission systems. The North American Electric Reliability Corporation (NERC) enforces a $N-1$ constraint for the U.S. power grid, which means that EPS transmission systems need to maintain nominal operation even if one component fails \cite{nerc-N1}. Such components include generators, transmission lines, transformers, etc. Depending on the security level of the EPS, a higher $N-k$ criterion may be required, where $k\geq 2$ represents two or more contingency events. For example, a nuclear plant may be required to satisfy a $N-2$ constraint, allowing the grid to withstand the simultaneous failure of two components.

For the purpose of our study, we assume that the attacker is able to compromise one or more components of the transmission system, causing under-voltage events at the distribution-side. When component failures occur, the system aims to maintain stability. However, the intermittent transmission system instability along with the potential inability to support power demand results in voltage deviations which are also propagated to the distribution level. Four main sub-cases are designed in this second scenario to illustrate the corresponding voltage impact at bus 632 of the distribution system. The first two sub-cases consider an attacker that compromises one generator (G2 or G3) once at a time ($N-1$), where the rest sub-cases consider an attack on two generators (G2 and G3) consecutively ($N-1-1$), or simultaneously ($N-2$). In all sub-cases, we evaluate the voltage variation (depicted in per unit -- p.u.) measured at bus 632. 

\begin{figure}[t]
\centering
    \subfloat[]{
            \includegraphics[width=\linewidth]{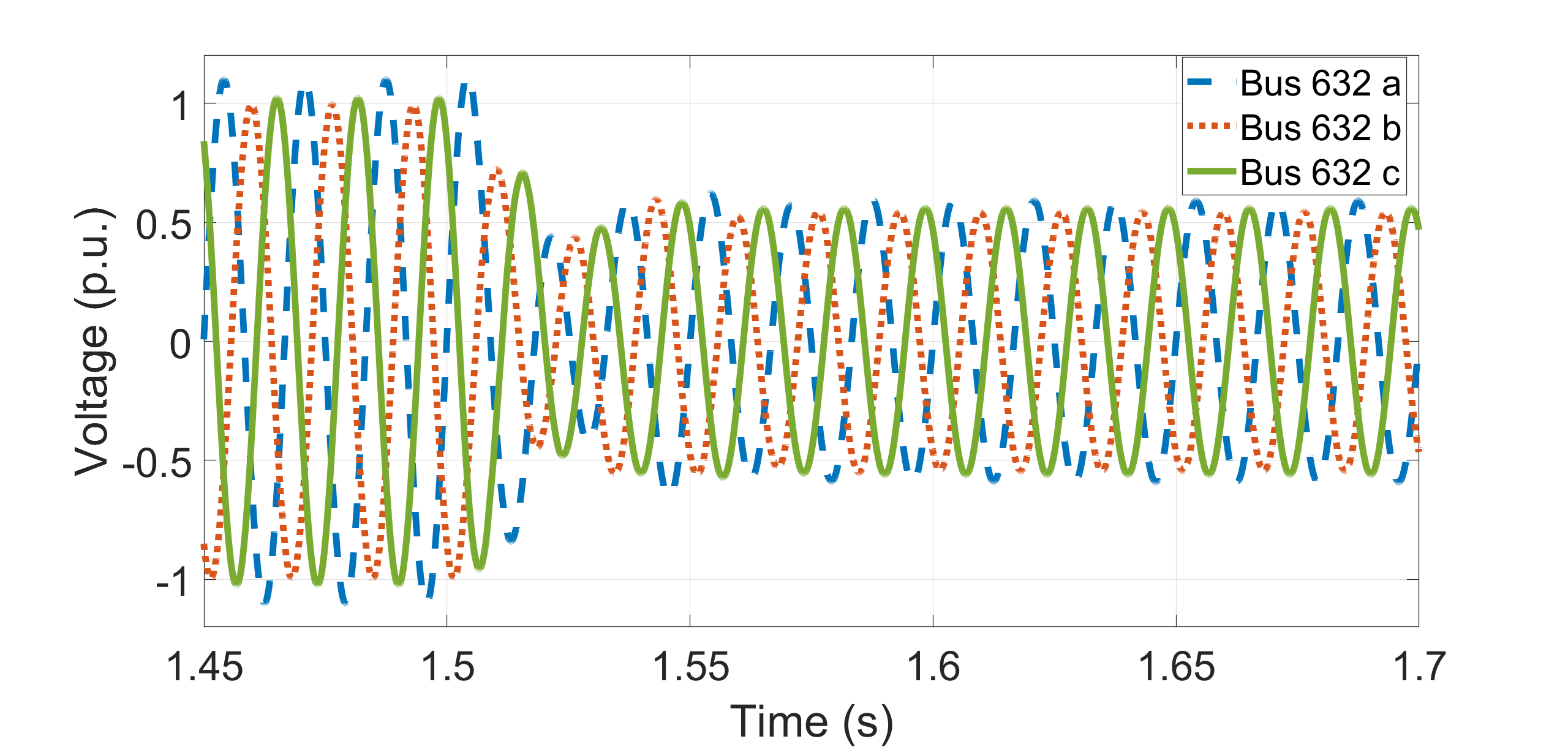}
            \label{subfig:N-1(1)}
    }\\
    \subfloat[]{
            \includegraphics[width=\linewidth]{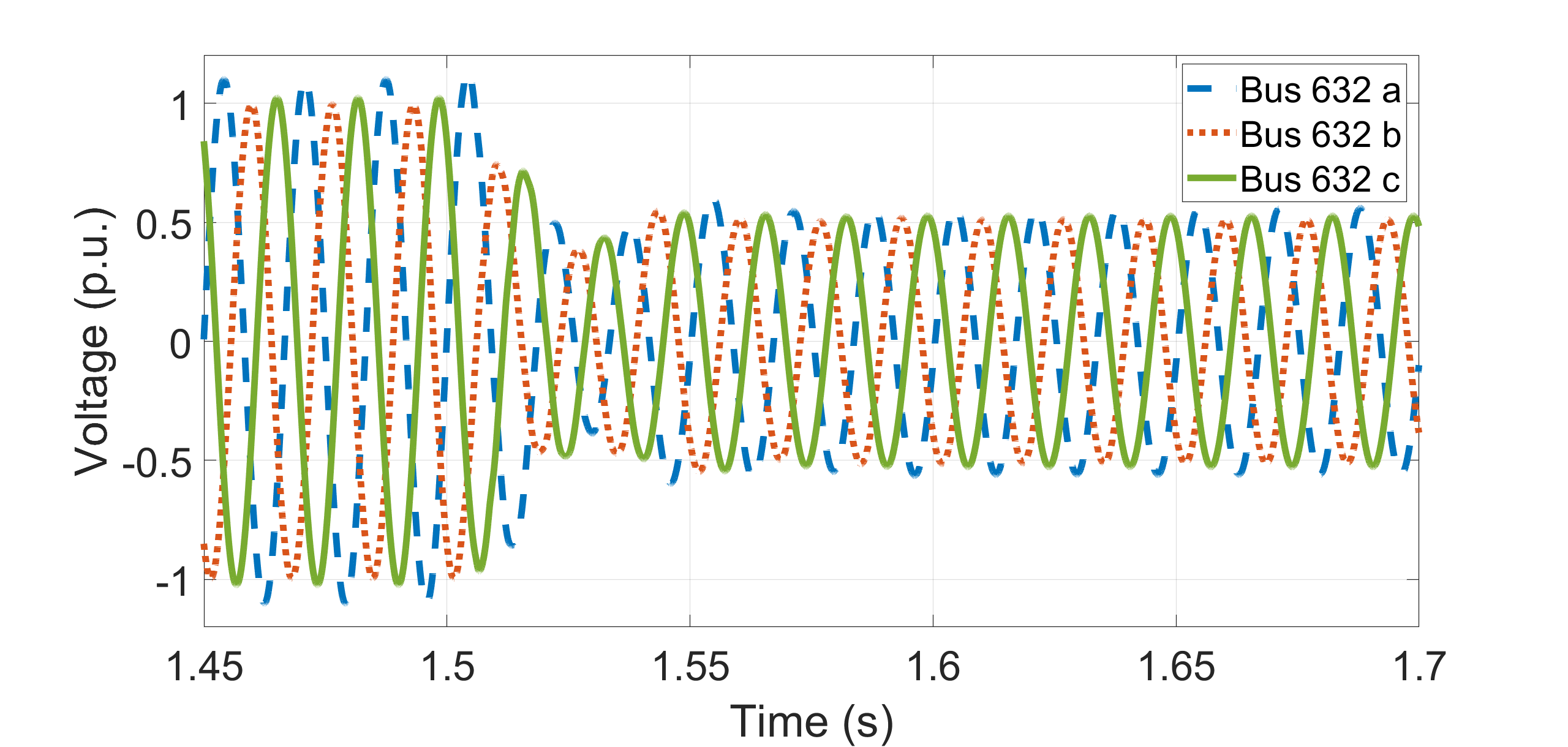}
            \label{subfig:N-1(2)}}
    
\caption{Voltage response at bus 632 (distribution system) during N-1 transmission system contingencies when:  \subref{subfig:N-1(1)}) generator G2 is disconnected at $t=1.5sec$, and
(\subref{subfig:N-1(2)}) generator G3 is disconnected at $t=1.5sec$.}
\label{fig:voltresults1}
\end{figure}

\begin{figure}[t]
\centering

   \subfloat[]{
            \includegraphics[width=\linewidth]{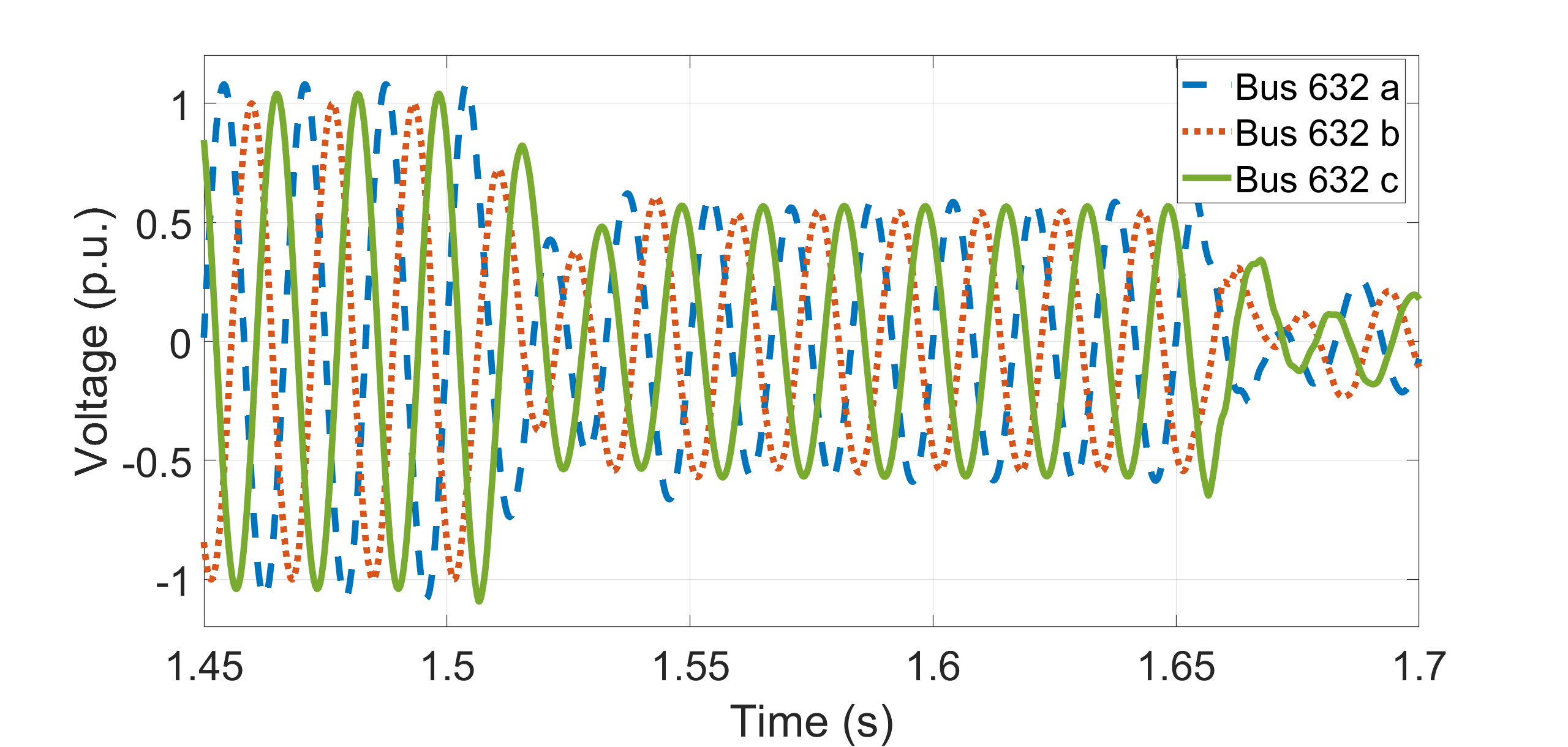}
            \label{subfig:N-1-1}
    } \\
    \subfloat[]{
            \includegraphics[width=\linewidth]{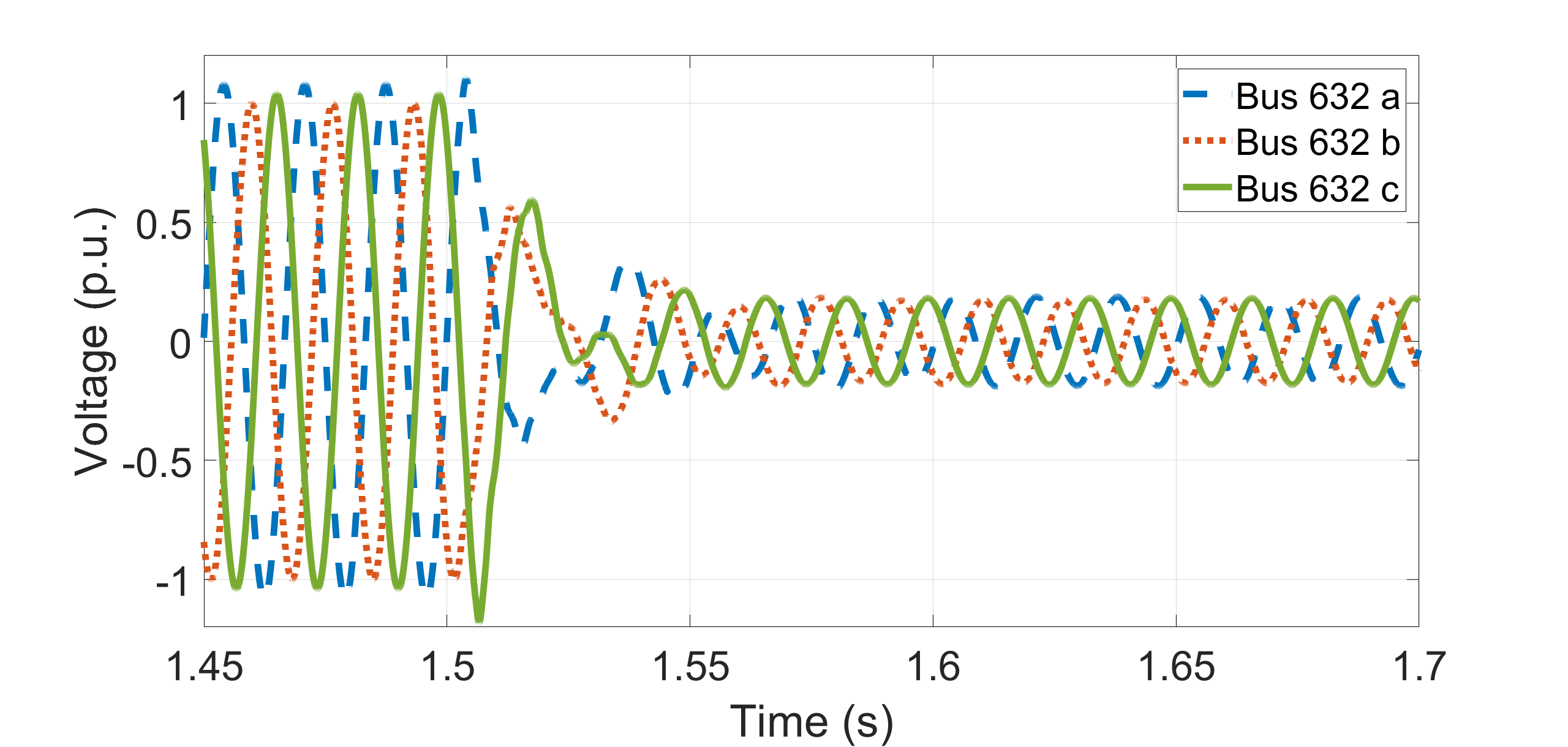}
            \label{subfig:N-2}}
    
\caption{Voltage response at bus 632 (distribution system) during transmission system contingency scenarios:\subref{subfig:N-1-1}) N-1-1 contingency where generators G2 is disconnected at $t=1.5sec$ and G3 at $t=1.6sec$ consecutively, and (\subref{subfig:N-2}) N-2 contingency where generators G2 and G3 are disconnected at $t=1.5sec$ simultaneously.}
\label{fig:voltresults2}
\end{figure}

\begin{figure}[t]
\centering
\includegraphics[width = 0.48\textwidth]{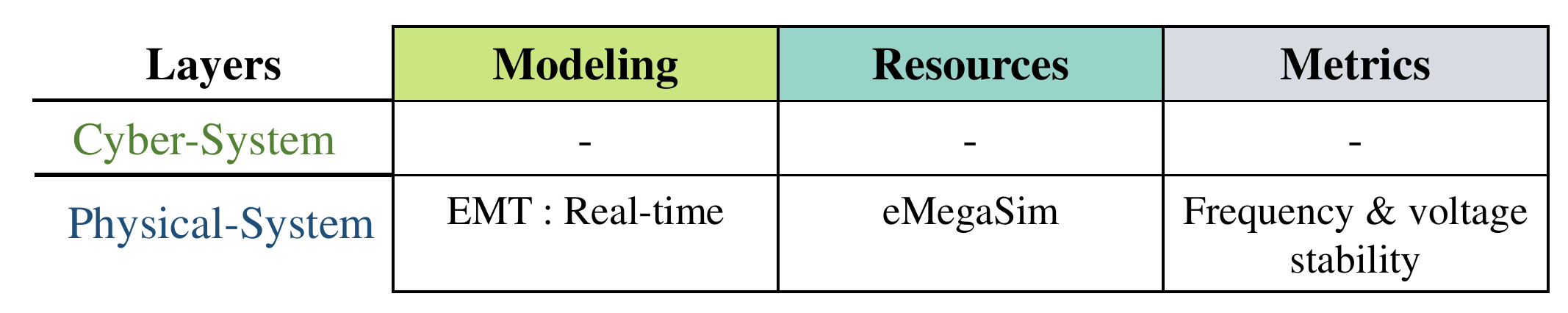}
\caption{\label{fig:tanddmapping} Mapping of T\&D case study with CPS framework.}
\end{figure}

As seen in Fig. \ref{subfig:N-1(1)} and Fig. \ref{subfig:N-1(2)}, the voltage measured at bus 632 of the  distribution system drops from $1$ p.u. to $0.5$ p.u. at $t=1.5sec$, i.e., when one of the generators (G2 or G3) is disconnected from the transmission system ($N-1$). Fig. \ref{subfig:N-1-1} demonstrates the voltage variations of the  $N-1-1$ contingency event in which G2 and G3 are disconnected at $t=1.5sec$ and \textit{t=1.6sec}, respectively. During this case, the bus voltage initially drops from $1$ p.u. to $0.5$ p.u. (G2 disconnection), and then to $0.2$ p.u. when G3 is also disconnected. In the $N-2$ case, presented in Fig. \ref{subfig:N-2}, the simultaneous disconnection of G2 and G3 from the system lowers the voltage significantly at $t=1.5sec$. The voltage measured at bus 632 of the distribution system decreases to under $0.2$ p.u. within $0.05sec$. Fig. \ref{fig:tanddmapping} illustrates the mapping of the propagating attack case study in T\&D systems with the CPS framework.

\vspace{1mm}
\noindent\underline{Risk Assessment}: Compromising T\&D systems requires determined adversaries possessing both strong knowledge of the system architecture as well as ample resources since these can enhance the probability of materializing successful attacks. Thus, we set the $ Threat \ Probability$ to \texttt{High (3)} \cite{NICHOLSON2012418, maglaras2019threats, 6016202, Wilson2014}. Attackers could perform stealthy and disastrous attacks by leveraging the knowledge of system topology, asset placement information, power demand profiles, etc. As a result, by targeting mission-critical system components during peak utilization periods (e.g., peak power demand time of the day), adversaries could maximize the corresponding attack impact \cite{PESGM2021}. A power system collapse (e.g., blackout), which could be the impact of the successful propagation of T\&D attacks, can significantly affect ``People health and personnel safety", ``Uninterrupted operation and service provision'', and ``Equipment damage and legal punishment''. Based on these assumptions, the $Resulting Damage$ is set to \texttt{High (3)}, while the ``Organization financial profit'' is set to \texttt{Low (1)}. The aggregated $Risk$ for this type of attack can be estimated to be $3* \sum(12+9+6+1) = 84$.

\subsection{Post-risk assessment discussion} 
The next step after the risk score calculation includes risk prioritization. The risk identification, assessment, and prioritization serve as preliminary steps and are critical for the decision making and formulation of mitigation plans. Specifically, in this work, we have considered four diverse attack use cases aimed at CPES while targeting different cyber or physical subsystems or components. In more detail, we discuss cross-layer firmware attacks with a calculated risk score equal to $22$, load changing attacks with risk score equal to $28$, TDAs with a risk score evaluated to $51$, and finally propagating attacks targeting integrated T\&D  CPES with an $84$ risk score. These risk scores provide a useful way to perform one-to-one comparisons between attacks even if their specifics are unknown. 

Attacks with higher risk scores (e.g., the T\&D propagation attacks) will induce a higher impact on the system when compared to other attacks such as the cross-layer firmware attack with a less pronounced risk score. In Section \ref{s:riskAssess}, we justify how the use of each case's risk score variations depend on the corresponding attack characteristics (e.g., threat probability, objective priorities, and potential impact on the CPES operation). As a result, attacks similar to the one targeting the integrated T\&D system, aim to affect almost every CPES operational objective. Furthermore, they are attractive from an adversarial perspective due to the maximization of the inflicted system disruption. Hence, such attacks will obtain high-risk scores. The same cannot be argued for attacks that can be sustained even post-compromise, targeting less critical CPES equipment.

The risk score-based ranking helps to categorize the attacks (and their corresponding risks) into pools \cite{Allegro}. For example, assuming that we have four pools, the most devastating attacks (i.e., with scores greater than a system-defined threshold) would be placed into pool $1$, while less critical attacks -- with smaller risk scores -- would be allocated to pools $2$ -$4$ in a descending risk score fashion. For each of the pools, predefined strategies are designed to mediate potential attacks. Typically, attacks belonging to pool $1$ should be mitigated at all costs since they can compromise the whole system (in our case the CPES). However, the mitigation of attacks belonging to lower-ranked pools might either be \textit{i)}deferred if they do not pose significant threats to system operation, \textit{ii)} transferred to other parties instead of allocating system resources to resolve them, or \textit{iii)} accepted if the cost of mitigating them outweighs the impact that could be inflicted on the system. Thus, the risk assessment does not only provide better awareness of system risks and an efficient way to perform risk comparisons, but it can also automate the process of handling risks and administering corrective measures.

%% file: sections/6-Conclusion.tex
\section{Conclusions} \label{s:conclusion}

In this work, we provide a comprehensive analysis of CPS security, with particular emphasis on CPES applications. The first step in this process encompasses an extensive threat modeling procedure, where adversary and attack models are constructed. The adversary and attack models provide an in-depth understanding of attackers' motives and capabilities, in addition to the attack's details including potential entry points, attack techniques, and end goals. The next step in the analysis includes the presentation of a CPS framework, where the resources, metrics, and modeling techniques needed to effectively evaluate CPS, and more specifically CPES, are discussed in detail. This framework is designed with the objective of assisting researchers and stakeholders identify the models and resources required to perform high-fidelity and reliable CPS studies. Furthermore, we present a risk assessment methodology that leverages both the treat modeling as well as the CPS framework to characterize system risks.

In order to illustrate the suitability of the overall methodology and description of the CPES security landscape, we investigate four attack case studies. For each scenario, we provide a fundamental background alongside its mathematical formulation and discuss the corresponding threat model and attack setups. The presented case studies are simulated under nominal and abnormal operating conditions to uncover their system-wide impacts. Risk assessment analysis is also performed as part of each case's security investigation. During the risk assessment stage, we calculate the relative risk scores indicating the severity of each compromise. The risk scores correspond to the discussed studies, the threat scenarios, and the targeted assets (e.g., microinverters, T\&D system, time-delay, etc.). These scores can be utilized for the ranking and prioritization of possible disruptions, and the determination of proper risk mitigation strategies to address malicious attacks implications. 

The holistic approach and studies presented in this paper provide guidelines for modeling CPS threats as well as designing, simulating, and evaluating detailed CPS models. The presented framework can promote rigorous security analysis of CPS. Our future work will extend this framework and advance its capabilities even further, allowing for:

\begin{itemize}[itemsep=0pt,parsep=0pt,leftmargin=*, wide=0pt]
    \item Secure and resilient CPES operation:
    In this work, we have stressed the importance of cyber-secure CPES as well as that the integration of contemporary cyber features and new physical components can increase the attack surface. The emphasis though, should not only be placed on detecting attacks, limiting and mitigating them but also in designing fault-tolerant and resilient CPES. Having identified potential vulnerabilities present in the CPES and leveraging our framework, we will define resiliency methodologies and metrics to assess CPES posture. In more detail, the resiliency methodologies will serve as CPES design best practices promoting the design of robust systems with in-built redundancy mechanisms if adverse scenarios occur. On the other hand, the resiliency metrics will be ported to our current framework and have a twofold objective, \textit{i)} they will indicate how effectively the system can handle adverse circumstances, and  \textit{ii)} they will serve as criteria for the categorizations of CPES based on their ability to withstand attacks.

    \item Autonomous CPES operation and simulation-aided risk assessments:
    CPES are becoming more sophisticated and support a plethora of automated processes (e.g., automated control mechanisms, PLCs, AGC, etc.). Such automated systems should be capable to make real-time decisions, especially for time-critical parts of CPES, and coordinate the dynamic system behavior. It is expected that CPES will become more complex and densely interconnected as they integrate more features (remote access and control, assets, communications protocols, etc.). During their autonomous operation, the system might encounter unexpected states (e.g., unintended faults during natural disasters, or malicious attacks) that might require specific handling. Thus, determining and evaluating their security should be facilitated in a dynamic, albeit abstract way. Following this approach,  guarantees that every unexpected scenario will be accounted for, and adverse situations will be timely prevented. Digital twin system configurations can achieve these objectives and enable the design and real-time evaluation of risk mitigation strategies. As a result, a CPES testbed will be designed to support the fully-automated operation, and incident-response structures, where attacks can be promptly detected and optimally mitigated, eliminating any adverse consequence on the actual system.
    
    \item Dynamic reconfiguration and self-healing capabilities:
    Securing CPES  should be viewed from two directions. The first direction includes the security measures and practices which should be employed to protect system operations and avert attackers. On the other hand, the second direction features the policies and strategies which should be pursued post-compromise or during dire circumstances. The first direction has been extensively discussed in this paper; we aim to account for the second direction in our future framework extensions. Specifically, utilizing our framework and system resources we will provide classes of crisis-handling plans promoting CPES self-healing capabilities. These classes will provide tailor-made strategies to overcome emergencies, depending on the current state of the CPES and the under-investigation scenario characteristics. For example, during a transmission system contingency, the corresponding class would provide alternative ways to dispatch power overcoming this issue and potential predicaments. These dynamic re-configurations and self-healing CPES capabilities will stimulate the design of future secure and resilient systems and prove invaluable tools for system operators.

\end{itemize}